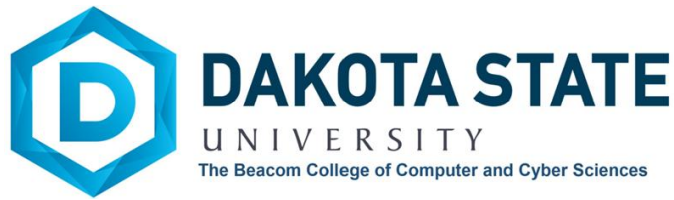

# PRESCRIPTIVE ZERO TRUST: ASSESSING THE IMPACT OF ZERO TRUST ON CYBER ATTACK PREVENTION

A dissertation submitted to Dakota State University for the degree of

Doctor of Philosophy

in

Cyber Defense

December 9, 2024

By

Samuel T. Aiello

Dissertation Committee:

Dr. Yong Wang (Chair)

Dr. Varghese Vaidyan

Dr. Mary Francis

# DISSERTATION APPROVAL FORM

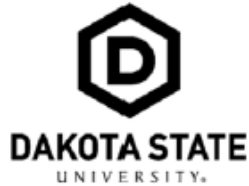

## DISSERTATION APPROVAL FORM

This dissertation is approved as a credible and independent investigation by a candidate for the Doctor of Philosophy degree and is acceptable for meeting the dissertation requirements for this degree. Acceptance of this dissertation does not imply that the conclusions reached by the candidate are necessarily the conclusions of the major department or university.

Student Name: Sam Aiello

Dissertation Title: Prescriptive Zero Trust: Assessing the Impact of Zero Trust on Cyber Attack Prevention

Graduate Office Verification: Brianna Mae Feldhaus　　　Date: 12/03/2024

Dissertation Chair/Co-Chair: *Yong Wang*　　　Date: 12/03/2024
Print Name: Yong Wang

Dissertation Chair/Co-Chair: ___________　　　Date: ___________
Print Name: ___________

Committee Member: *Varghese Vaidyan*　　　Date: 12/04/2024
Print Name: Varghese Vaidyan

Committee Member: *Mary Francis*　　　Date: 12/03/2024
Print Name: Mary Francis

Committee Member: ___________　　　Date: ___________
Print Name: ___________

Committee Member: ___________　　　Date: ___________
Print Name: ___________

# ACKNOWLEDGMENT

First and foremost, I extend my deepest gratitude to my professor and chair, Dr. Yong Wang. His expertise, mentorship, and unwavering support were instrumental in the completion of this thesis.

Gratitude is also due to Dr. Varghese Mathew Vaidyan, whose insights and constructive criticism significantly enhanced the quality of my work.

Special thanks to Dr. Mark Hawkes and Dr. Pat Engebretson for the extensive research opportunities provided by the Ph.D. CD program, and to Dr. Mary Francis for her instrumental support.

My family, especially my wife Linda, and friends' unwavering support and encouragement motivated me during challenging times. Their understanding and patience allowed me to dedicate time and effort to this research.

Colleagues, managers, and classmates provided valuable feedback and engaged in insightful discussions that enriched my understanding and refined my arguments. Their willingness to share their knowledge was invaluable.

In conclusion, this thesis's successful completion and publication would not have been possible without the support and contributions of these individuals and groups. Sincere gratitude goes to my professors, administrators, family, friends, managers, and colleagues for their unwavering support throughout this journey.

# ABSTRACT

Increasingly sophisticated and varied cyber threats necessitate ever-improving enterprise security postures. For many organizations today, those postures have a foundation in the Zero Trust Architecture (ZTA). This strategy sees trust as something an enterprise must not give lightly or assume too broadly. Understanding the ZTA and its numerous controls-centered around the idea of not trusting anything inside or outside the network without verification, will allow organizations to comprehend and leverage this increasingly common paradigm. The ZTA, unlike many other regulatory frameworks, is not tightly defined.

The research assesses the likelihood of quantifiable guidelines that measure cybersecurity maturity for an enterprise organization in relation to ZTA implementation. This is a new, data-driven methodology for quantifying cyber resilience enabled by the adoption of Zero Trust principles to pragmatically address the critical need of organizations. It also looks at the practical aspects ZTA has on capabilities in deterring cyberattacks on a network. Coupled with quantitative statistical methods, the ZTA maturity approach provides guidance on how an organization can objectively gauge its cybersecurity posture.

The outcomes of this research define a prescriptive set of key technical controls across identity verification, microsegmentation, data encryption, analytics, and orchestration that characterize the comprehensive ZTA deployment. By evaluating the depth of integration for each control component and aligning to industry best practices, the study's results help assess an organization's ZTA maturity level on a scale from Initial to Optimized adoption. The research's resultant four-tier model demarcates phases for an organization on its security transformation journey, with each tier adding to the capability of the last. This structured approach will help organizations improve their respective security postures without systematically compromising operational effectiveness, thereby improving risk management and threat response capabilities. This model does much more than just provide security. It helps an organization optimize resources, make focused investments, and measure progress along its Zero Trust journey in quantifiable terms.

# DECLARATION

I hereby certify that this dissertation constitutes my own product, that where the language of others is set forth, quotation marks so indicate, and that appropriate credit is given where I have used the language, ideas, expressions, or writings of another.

I declare that the dissertation describes original work that has not previously been presented for the award of any other degree of any institution.

Signed,

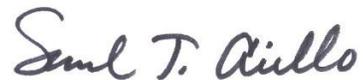

**Samuel T. Aiello**

# TABLE OF CONTENTS

# TABLE OF FIGURES

# TABLE OF TABLES

# CHAPTER 1:  INTRODUCTION

## 1.1     Background

A Zero Trust Architecture is a coordinated strategy and cybersecurity philosophy that secures an organization by eliminating implicit trust and continuously validating every stage of a digital interaction. It is intrinsic in the concept of "never trust, always verify." With Zero Trust, all data and services are treated as resources; thus, it protects them from unauthorized access. A general principle on how to protect modern environments and enable digital transformation includes strong authentication, segmentation of the network, prevention of lateral movement, Layer 7 threat prevention, and granular "least access" policy management.

Zero Trust's architecture was developed on the premise that most traditional security models operate based on an outdated assumption that everything inside the network of an organization should implicitly be trusted. Implicit trust means once on the network, users, whether threat actors or malicious insiders, can laterally move freely around and access or exfiltrate sensitive data because of a lack of granular security controls.

Succinctly, Zero Trust has never been more of an imperative for organizations than today. A properly implemented Zero Trust Architecture creates higher overall levels of security posture but also reduces its complexity and operational overhead.

## 1.2     Statement of the Problem With Motivation

Prescriptive Zero Trust Architecture measures have grown as an important yet sparsely researched area in view of cybersecurity, as organizations find it rather difficult to deploy and study this innovative security method. The absence of explicit prescriptive guidance coupled with the relative newness of the ZTA has resulted in a substantial gap in the understanding of its effectiveness that leaves an enterprise without any secure basis on which to assess its deployments of ZTA (Ghasemshirazi et al., 2023).

This research addresses this critical gap by investigating factors driving the need for comprehensive guidelines to assess end-to-end ZTA deployments. The lack of prescriptive guidance is a major challenge in implementing and assessing ZTA. It is also due to the lack of any well-accepted, prototypical set of security controls that are representative, comprehensive, and usable for assessing enterprise deployments of ZTA.

While risk modeling has evolved over the last couple of years, it has not kept pace with the changing landscape of ZTA; due to this, there exists an abyss between current risk-modeling approaches and what is actually required to support a good ZTA assessment (Rose et al., 2020).

There would also be no established basis to determine the efficacy of the implementation of the elements of a ZTA using standard metrics without clear, prescriptive guidelines for ZTA. In other words, without benchmarks or guidelines, it falls to the enterprises to figure out the intricate landscape of ZTA assessment, which increases the difficulty in quantifying the impact and success of the enterprise's ZTA initiatives.

This further underscores the urgent need for more holistic yet accessible resources that will guide enterprises in assessing the suitability of their implementation and the effectiveness of their ZTA deployments.

## 1.3     Zero Trust Adoption

Organizations are at great risk due to cyberattacks that are increasingly sophisticated in nature. While these are extremely disrupting and affect the entire organization, they compromise sensitive data. The current state of cybersecurity remains, quite frankly, not good enough. The Zero Trust model is a recent move for many organizations. This security framework is based on the simple but revolutionary premise that no one, whether inside or outside the organization, can be trusted to access resources without verification (Tillson, 2024).

A survey of professionals (Unisys, 2023) revealed that nearly half (43%) already have zero trust solutions in place at their companies. A bigger group (46%) is moving towards a zero-trust environment. Just a few (11%) are at the start of cloud security and have not begun using zero trust yet, but they aim to do it later on.

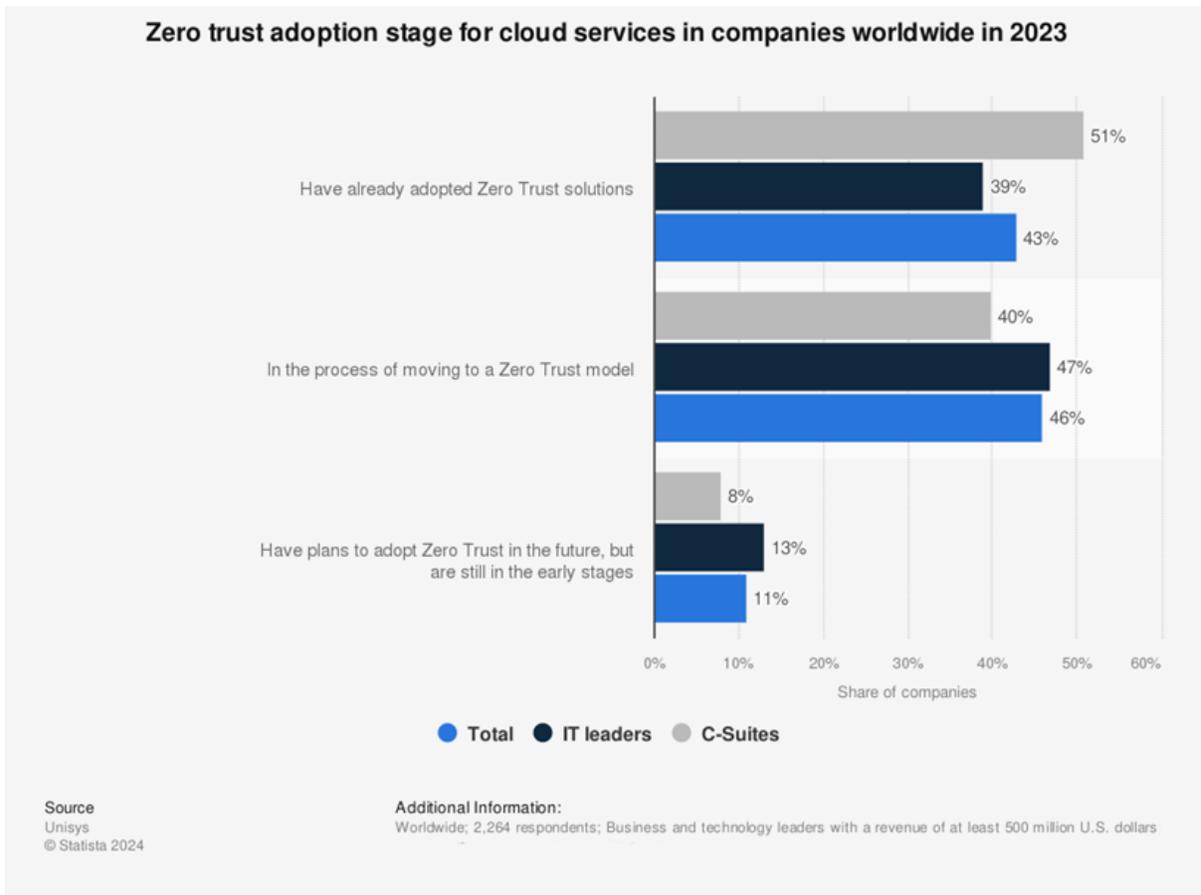

*Figure 1- Worldwide Zero Trust Adoption*

One of the key drivers to consider Zero Trust is due to a growing complexity in the IT environment, with an increase in cloud computing, IoT, OT, mobile devices, and remote work. Traditional security measures based on perimeter defenses no longer safeguard from cyber threats. With Zero Trust in place, an organization will be able to hold a significant edge over the usual policy of "trust-but-verify," as it verifies who is trying to get access to its systems and the security posture of the user and device constantly.

Casildo and Corey's study (2021) shows that organizations that have adopted models of Zero Trust are reporting huge improvements in overall security posture. Organizations should adopt at least privileges and segmentation- a microsegmentation approach - to limit the reach of a security breach in their networks and be able to prevent lateral movement. The potential for a more structured approach to security can, in fact, provide threat detection and response and minimize the danger of data breaches.

Accelerated movement to remote workforces brought on by the COVID-19 pandemic accelerated Zero Trust movements. Organizations realize that with users accessing a corporate network from various disseminated locations and with diverse devices, there is more awareness of a strong security system to adapt to this new reality. Some core factors behind zero trust - such as continuous authentication and encryption - help organizations maintain remote access and protect sensitive data from unauthorized access.

A recent global survey (Cybersecurity Insiders: Zscaler, 2023) shows that many organizations prioritize Zero Trust security. More than half of the e-survey respondents felt this was very important or even more than that, while another 32% said it was important. Zero Trust becomes focused on many users working remotely. Put simply, Zero Trust ensures that whenever and in whatever instance, access to systems and resources is granted to the right user.

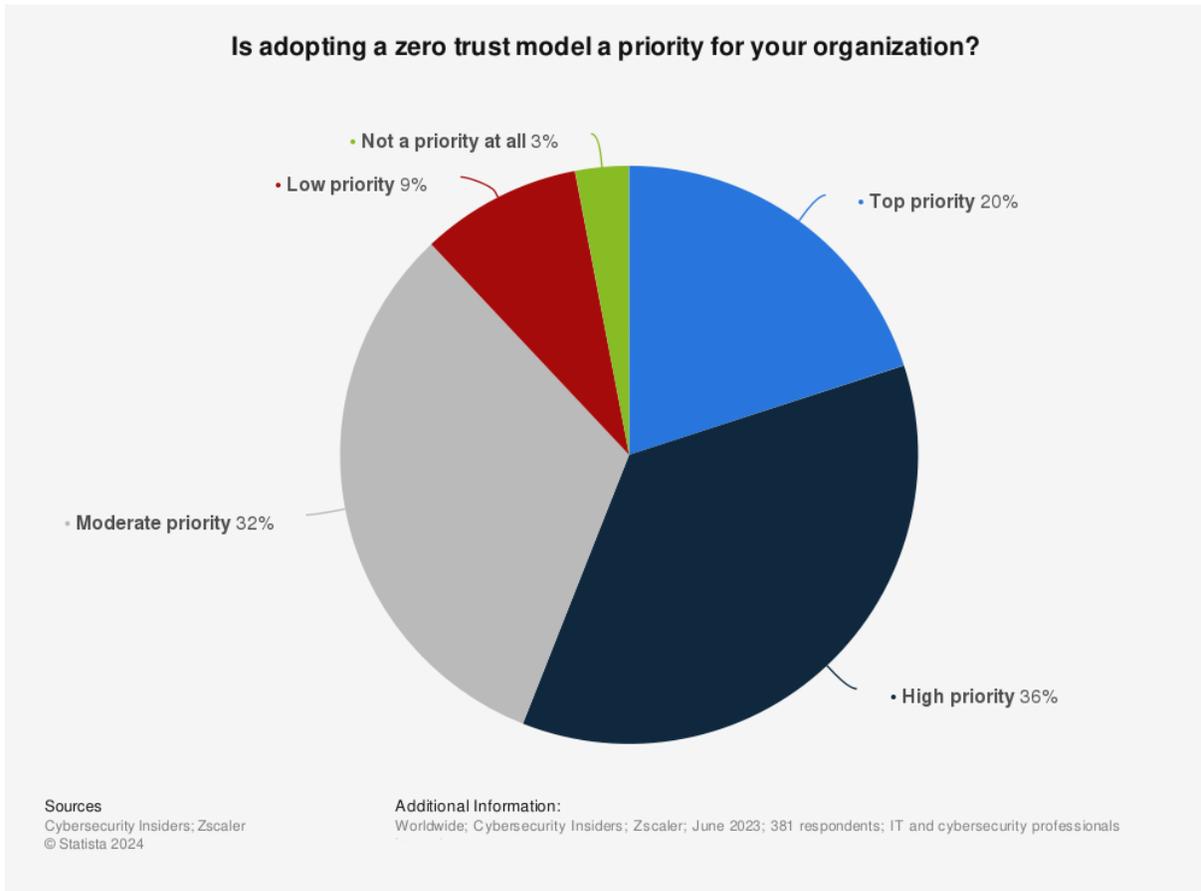

*Figure 2. Zero Trust Adoption Priority*

  The increasing intricacy of IT systems is a major driver propelling many organizations toward Zero Trust. They no longer rely solely on perimeter security to keep out unwanted intruders. With the advent of cloud computing, the Internet of Things, and a mobile workforce that requires access to data from almost anywhere, traditional security approaches have become hard to manage and even harder to enforce. They also have become much less effective. Despite all the effort and expense, there has been an uptick in obvious breaches and many news conferences following "events" that were not prevented. Organizations that were supposed to be defenders against cyber threats had their cyber defenses breached.

  The remote work transition of the COVID-19 pandemic has thrown fuel onto the fire of the adoption of a Zero Trust Architecture. With their employees accessing the corporate network from all over the place and on all sorts of devices, many organizations have come to realize that they need a far stronger security framework than they had before. Zero Trust is not a product but rather a set of guiding principles to follow. Although the "Zero Trust" nomenclature might be trendy, the idea is not new-just ask anyone who has spent any significant amount of time in information security over the past couple of decades. Still, it is a pretty good bet that many organizations will be adopting the style in the coming months and years.

Even with the benefits that are associated with Zero Trust models, some organizations may be reserved when it comes to taking on this new approach. Complexity and cost are the two big reasons why some organizations choose not to adopt a Zero Trust cybersecurity framework. Creating a Zero Trust environment requires a meticulous type of planning that needs to involve all the different departments across an organization that handles information. That is a tall order for organizations that are resource-constrained. To fund the implementation of new security technologies and the staff training that is required to help employees understand the new model are two more reasons why some organizations may be hesitant.

Despite the obstacles associated with putting into place the Zero Trust principles, there remains the possible upside of enhanced security and mitigated risk, making it a pretty tempting option for organizations intent on fortifying their defenses. Following a cybersecurity event, organizations naturally focus on threat and risk management as their foremost concerns and try to prevent business disruption from happening again in the future. But when it comes to securing network architecture, the trend has moved away from perimeter-based security (Tyler & Viana, 2021). Instead, the emerging focus has squarely fallen on the Zero Trust Architecture, which is a security framework that aims to "verify and validate every access request, regardless of where in the world they originate (Kindervag, 2016) or what type of network they are using to connect." (World Economic Forum. 2022)

For large-scale enterprises, implementing a prescriptive Zero Trust framework is a critical step to improving cybersecurity (Barik et al., 2023). Adopting fundamental measures of a Zero Trust approach specifically, identity and access management, network segmentation, continuous monitoring and analytics, and encryption and data protection can effectively reduce opportunities for cyberattacks and improve overall security posture (Casildo & Corey, 2021).

Understanding that the Zero Trust model must be more than just an infrequent process is crucial. The current literature emphasizes that the implementation of this model must be repeated: evaluation and improvement cannot be side-stepped (Kang et al., 2023). Following the principles of Zero Trust better equips organizations to guard their critical assets against today's advanced cyber attackers, who already have proven themselves capable of using unprecedented means and methods to get around traditional security measures (CEA Report, 2018).

Not universal to all organizations, the Zero Trust model is a bit different from one organization to the next (Turner et al, 2021). This makes it hard for outside parties to get a clear picture of how an enterprise organization has set up its cybersecurity mechanisms. In fact, lack of any standardized diagnostic tool or set of standardized procedures makes it difficult to assess any organization's security posture. Still, organizations seem to be increasingly adopting the ZTA in one form or another (Bush & Mashatan, 2022).

Another hurdle in implementing the ZTA is the need for standardization among its many parts. The pivotal policy decision point (PDP) is where ZTA "performs its most significant and impactful work" (Teerakanok et al., 2021). The PDP is the point at which ZTA synchronizes the activity of different components; in effect, the PDP is where ZTA becomes ZT. Different organizations have their own unique ways of implementing Policy Decision Points (PDPs), which can significantly affect how well ZTA works across different systems. This variability in implementation can make it challenging to assess ZTA's overall effectiveness.

In simpler terms, for ZTA to be truly effective, the systems using it need to be able to communicate with each other, whether under specific conditions or more freely. For ZTA to be useful in a conversation about the security posture of a given organization, that organization must be using ZTA in a way that is both conditionally and unconditionally secure.

When organizations quantify the financial effects of ZTA, they can determine whether investing in this security model is justified. ZTA aligns with several industry regulations, thereby reducing the cost of compliance and demonstrating a commitment to protect sensitive data. Most importantly, ZTA establishes a framework that can reduce the risk of non-compliance. With regard to cyber insurance, this compliance aspect is vital because lack of adherence to a standard or a framework can cut off access to low-cost policies with high coverage amounts. Today's cyber threat landscape is growing and more costly by the day. As a result, governance, risk, and compliance (GRC) have never been so critical for your ZTA (Zero Trust Architecture) investments. With a ZTA in place, even if a cybercriminal somehow gets inside, there are minimal ways for them to move around and do damage.

Thus, sound governance, necessary risk management, and even compliance with certain mandates help keep your ZTA within reasonable threat and operational thresholds. Effective processes for securing and managing crucial data are essential to both risk and governance. (Yassine et al., 2021). Models have been proposed to estimate the maturity of an organization's cybersecurity processes. Some of the well-known models can be used to uncover key factors that help assess an organization's level of cybersecurity. Either path taken must cover the essential areas of incident response management, assessment of resource effectiveness, data identification and classification, governance functions, threat and risk management, cybersecurity monitoring, management of compliance and continuity, security testing and auditing, and, finally, assurance of third-party and cloud cybersecurity.

The importance of incident response teams cannot be overstated. These teams do not just secure and fix systems; they also communicate with people (Dube & Mohanty, 2020). When an incident occurs, the public and regulators want to know what happened and why. Resource effectiveness is another critical topic a cybersecurity framework should address (Ghaffari & Arabsorkhi, 2018). Frameworks should include measures to assess whether the resources expended on cybersecurity yield effective results. A more straightforward way of saying this is, is there a good bang for the buck when investing in cybersecurity? To answer these questions, they can adopt a comprehensive framework for cybersecurity maturity assessment.

Drawing from the provided sources, it is possible to propose a quantitative framework for evaluating an enterprise organization's cybersecurity maturity a framework whose only truly missing component seems to be the use of the high-level binary decisions found in the Capability Maturity Model (CMM). In this case, the CMM has been repurposed, and its decisions play an important role in defining the maturity stages of the 10 components found in the framework (Paulik et al, 1993).

Resources that, where possible, support the assumption that the training and development of emergency response teams are indispensable in detection and mitigation, communication, and control collaboration. Additionally, such a system should build a framework that offers good returns to cybersecurity investments and incentivizes stakeholders by rewarding them for the management respect and maintenance of the security infrastructure.

With a comprehensive cybersecurity assessment, organizations can evaluate current security practices and identify areas for improvement (Ghaffari and Abusorkhi, 2018). This assessment helps organizations understand their cybersecurity controls and identify weaknesses that cyberattacks can exploit. With the capability model, a quantitative method could be developed based on analyzing cybersecurity data to determine an organization's capability in preventing cyberattacks.

The analysis must reflect the key elements to be performed in incident management, infrastructure, data analyses and distribution, cybersecurity services, threat management primary and risk, cybersecurity assessment, compliance, and governance, as suggested by Dube and Mohanty (2020). Looking ahead, perform security analysis and management from a cloud network security perspective.

Understanding these key concepts may lead to the development of exposure to cybersecurity risks and threats from management and culture and the creation of appropriate management information to reduce cybersecurity risks and threats. The exercise develops and judges the growth of cybersecurity organizations. Through this, organizations will identify resources and areas for improvement related to cybersecurity management. It therefore entails finding the readiness of the organization for the implementation of a cybersecurity strategy through preliminary assessment of the current state of cybersecurity practices, policies, and resourcing. The assessment should enable the organization to be able to respond in case of incidents; effectiveness in the management of security should be measured; sources of information should be identified and classified. In addition, it needs to ensure this source is managed appropriately. Zero Trust Maturity Model, 2023.

## 1.4    Components

The Zero Trust paradigm is developed on some basic tenets that come together to form a comprehensive security structure. Zero Trust also emphasizes the use of the Unified Policy Engine, which is the administrative center for all security policy creation and enforcement activities in the organization's networks. The Unified Policy Engine enables companies to create granular policies based on user identity, device security posture, and other contextual factors, ensuring that access to resources is granted only to authorized users and devices.

API security constitutes another part of the core components making up the Zero Trust paradigm; APIs Help services and applications communicate and transmit information. In addition, such security protocols on APIs help block attacks that may lead to unauthorized data retrieval as well as keep the API messages secure. (API Security Market Size, Review: Share Projections for 2024-2031).

Microsegmentation and Network Segmentation are two aspects of Zero Trust that entail compartmentalizing a wider area of a network into zones in order to reduce the risk of threat movement. Besides, network breaches can be restricted by virtue of network segmentation and access privileges granting whereby lateral movement of attackers on the network is prohibited.

Endpoint Security is one of the elements within the Zero Trust Architecture, which is important considering that most attacks are focused on endpoints such as laptops, desktop computers, and mobile devices. Recent trends in Endpoint Security focus on implementing antimalware software, deploying Endpoint Detection and Response (EDR) solutions, and device encryption, which arm an organization against malware, ransomware, and other threats.

Within the Zero Trust model, encryption serves as a fundamental tenet by protecting the privacy and authenticity of information whether being moved or stored. Organizations can block unauthorized individuals from overhearing conversations or data and altering or obtaining sensitive information by encoding messages and virtualization.

Analytics and visibility are an integral part of Zero Trust, as they help organizations understand the volume of traffic in their networks, user activities within their environment, and the security status of their networks. Based on the traffic or user activities, organizations can monitor progress and address security incidents.

Orchestration is another key component of Zero Trust that automates security processes and responses to security incidents. By orchestrating security controls and responses, organizations can reduce the time to detect and respond to threats, minimizing the impact of security incidents.

Zero Trust logically leads to the necessity of access control because it pertains to authenticating and validating user and device access to resources. By implementing strong access control, based on the least privilege concept, organizations can effectively reduce the possible exposure of sensitive data and any other valuables to threats.

## 1.5     Principles

Zero Trust is implemented and operated on very few guiding principles or values that direct its usage. The predictable elements of Zero Trust include "Verify Explicitly," which entails that the user, device, and application seeking access to resources should be reviewed regardless of their location or network. Organizations should use strict policies regarding identity and device security before allowing access to them in order to minimize possible security infiltrations.

Another principle of Zero Trust is "Assume Breach." Very simply explained, this principle considers that any institution can be a target of malicious external forces, and such institutions' data cannot be completely safe. Organizations can assume the network is compromised ahead of time and look for more compromise in the center of such an attack as quickly as possible. Therefore, security incidents can be managed more responsively than would be required.

As the term suggests, "Always Authenticate" is another tenet of Zero Trust. In a Zero-Trust environment, this means using strong authentication systems like multi-factor and Biometric Authentication. Organizations that require users to authenticate using several factors will lessen the threat of illicit access and credential harvesting.

"Encrypt Communications" is one of the Trust principles in Zero Trust, where security management emphasizes the need to protect data in movement against any in-transit or data interception. Communication can be made safe and ensure data confidentiality and integrity during transport by using high-securing mechanisms like TLS and IP sec.

## 1.6     Definitions

"No Implicit Trust" is the core principle of Zero Trust that challenges the traditional notion of implicit trust within the network. Instead of assuming that users and devices inside the network perimeter are inherently trustworthy, organizations should verify their identity and security posture before granting access to resources, regardless of their location or network connection.

"Least Privilege Access" is another key principle of Zero Trust that advocates granting users and devices the minimum level of access required to perform their tasks. By implementing least privilege access controls, organizations can limit the exposure of sensitive data and resources to potential threats, reducing the risk of data breaches and insider threats.

"Continuous Monitoring" is an essential principle of Zero Trust that involves monitoring network traffic, user activity, and security events in real time. By continuously monitoring for signs of compromise and security incidents, organizations can detect and respond to threats more effectively, minimizing the dwell time of attackers and reducing the impact of security breaches.

"Evolving Needs" is a core principle of Zero Trust that recognizes the dynamic nature of cyber threats and the evolving security landscape. By adapting to changing threats and technologies, organizations can ensure that their security controls and policies remain effective and relevant in the face of emerging threats and vulnerabilities.

In the context of the Zero Trust paradigm, several key definitions are essential to understanding its principles and components. "Unified Policy Engine" refers to a centralized platform that allows organizations to define, manage, and enforce security policies across the network. By using a unified policy engine, organizations can create consistent security policies that apply across all network segments and resources.

"API Security" involves implementing security measures to protect APIs from unauthorized access, data breaches, and other security threats. By securing APIs with authentication, authorization, encryption, and other controls, organizations can ensure the integrity and confidentiality of API communications.

"Microsegmentation" is a network security strategy that involves dividing the network into smaller segments or zones to limit the lateral movement of threats. By applying access controls between network segments, organizations can contain breaches and prevent attackers from moving laterally across the network.

"Network Segmentation" is a security practice that involves dividing the network into separate segments or zones to control the flow of traffic and limit the exposure of sensitive data and resources. By segmenting the network, organizations can reduce the attack surface and contain breaches more effectively (Hnatiw, 2023).

"Endpoint Security" refers to the practice of securing endpoints such as laptops, desktops, and mobile devices from malware, ransomware, and other cyber threats. By implementing endpoint security measures such as antivirus software, EDR solutions, and device encryption, organizations can protect their endpoints from security breaches.

"Encryption" is a security measure that involves encoding data to prevent unauthorized access and ensure the confidentiality and integrity of information. By encrypting data in transit and at rest, organizations can protect sensitive information from eavesdropping, data tampering, and unauthorized access.

"Analytics & Visibility" involves monitoring network traffic, user activity, and security events to detect and respond to security incidents in real time. By analyzing network data and user behavior, organizations can gain insights into their security posture and identify potential threats before they escalate.

"Orchestration" refers to the automation of security processes and responses to security incidents. By orchestrating security controls and responses, organizations can reduce the time to detect and respond to threats, minimizing the impact of security incidents on their operations.

"Access Control" includes activities that enable the authentication of users as well as the security status of devices before access is allowed to the resources. Applying these controls by recognizing the least privilege principle helps organizations to minimize the risk of inappropriate access to or disclosure of consents and resources.

The concept of Zero Trust is a new direction for organizations in the field of cyber security, which shifts their focus from prevention of external attacks to integrated and flexible security. With the adoption of core components, principles and definitions of Zero Trust, organizations improve their security cross posture and are able to defend their data and resources against continuous cyber threats as well as address the challenges posed by the dynamic environment.

Running head: Aiello Dissertation: Prescriptive Zero Trust       Page **19** of **232**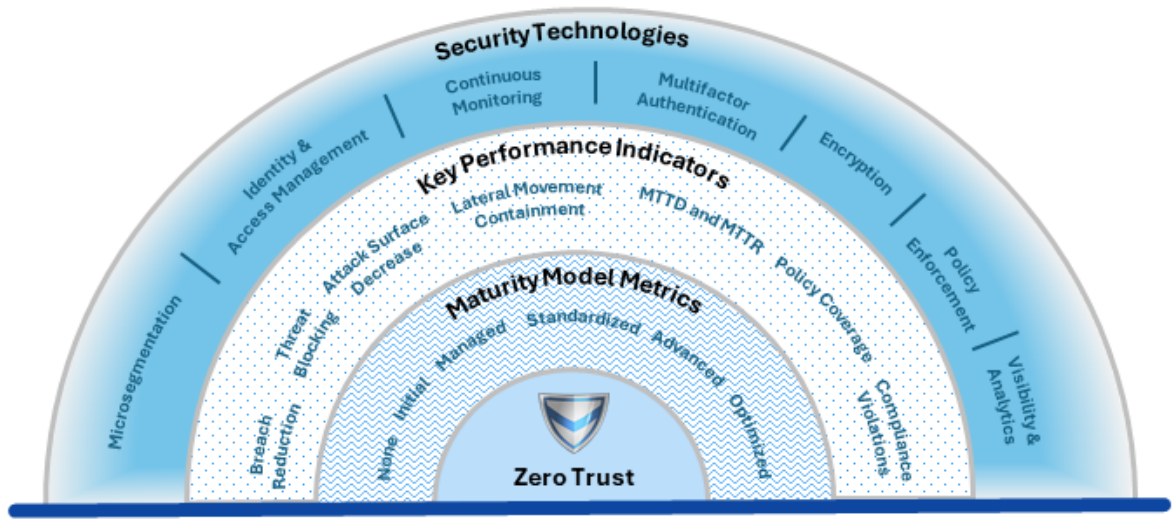

*Figure 3. System Design Diagram*

Companies continue to experience breaches because the antiquated outer defense type of control has its limitations. As a result, a new approach known as Zero Trust paradigm has been adopted as a better way of protecting networks, data, and systems. Based on the findings from the literature review, this figure represents a Zero Trust security framework visualization, showing different layers and components of a comprehensive Zero Trust security architecture. It is impossible to trust a user or device inside or outside one's network perimeter without confirming who they are and how safe they are.

A Zero Trust security posture can be established on a number of principles. Least Privilege Access Control ensures that users and systems are granted access only at the level that is essential to the completion of their duties. With this approach LPAC minimizes the possible damage that can happen in case an account is compromised.

Zero Trust is broader than the single-instance authentication and authorization approach. At every moment of a session, a user identity and the trustworthiness of their devices are verified. There are other risk management procedures where access can be revoked if any suspicious action appears. The strategy of microsegmentation is a way of creating smaller barriers in a more traditional networked structure. This tactic limits the lateral

Page **19** of **232**

movement within the network, thus preventing the quick spread of any security breach to other important components.

Data encryption protects information from unauthorized access and breaches both in storage and in retrieval. Data Loss Prevention (DLP) measures are implemented to add security concerns on unauthorized data export for the organization enhancing further protection layer. Great importance for ZT is laid on efficient Identity and Access Management (IAM) policies.

Multi-factor Authentication helps secure user logins by requiring the user to provide at least one more verification besides the provided username and password.

There are plenty of advantages in looking at security through the lens of Zero Trust. ZT reduces the possibilities associated with compromised accounts or insider attacks. Security is improved as ZT constantly checks who is being allowed access and to what extent.

ZT is in line with privacy regulations including GDPR and CCPA which are inclined towards data access control. Companies can prove they are committed to protecting sensitive information and follow the law at the same time by adopting a Zero Trust approach. The ZT model makes it easy to safeguard remote access to off-site cloud resources and benefits the company with a less deskbound staff. ZT is more concerned about identity than the connection within the network and allows opt-in on any authenticated device.

ZT reduces the extent of loss that may arise in the event of a successful cyber intrusion. It uses access control and network segmentation to restrict the aggressor's lateral movement and shield more systems from being compromised.

While the arguments in favor of Zero Trust are convincing, it can also be noted that there are challenges associated with this strategy, which stands out as the main weakness. Implementing a Zero Trust security framework may be complex as it requires some configurations of the network structure, security mechanisms, and user processes. Both strategic planning and tactical execution over time are critical to successful Zero Trust assimilation.

The use of ZT may equally imply the supplementation of the existing security policies with new security gadgets and technologies in addition to training the users on the new security policies. ZT might include additional policies for those who are accustomed to the traditional way of usage.

To ensure the smooth integration of Zero Trust, it is essential to implement effective communication, training and user support to address any potential scope for resistance. ZT is a huge change in the realm of cybersecurity. Trusting the organization rather than the individual teleport makes it possible for organizations to enhance the security posture.

Despite the presence of obstacles, the advantages of Zero Trust in terms of enhanced security, compliance, and agility justify its value as a valuable investment for organizations of any magnitude. In the digital age, it is becoming more important to adopt a Zero Trust approach in order to protect critical data and systems, as cyber threats continue to develop and change.

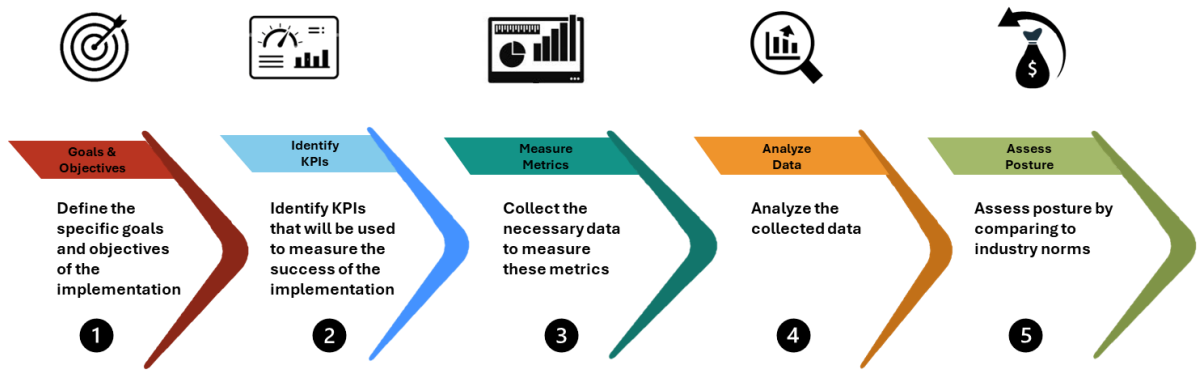

*Figure 4. Posture Assessment Process*

Organizations are constantly implementing new initiatives and programs to improve efficiency, achieve specific goals, or address emerging challenges. However, measuring the true success of these implementations can be difficult without a structured approach. This is where a posture assessment comes into play.

This process (Figure 4) provides a systematic method for assessing the effectiveness of an initiative or implementation by focusing on quantifiable outcomes. Based on the findings from Deming Cycle (Plan-Do-Check-Act) (Deming, 1986), this figure shows the logical progression from setting initial goals through to final assessment, with each step building upon the previous one. The process unfolds in five key stages.

The first step involves clearly defining the specific goals and objectives the implementation aims to achieve. This initial step establishes a baseline for success measurement. With these goals in mind, relevant Key Performance Indicators can be identified in the second stage. KPIs act as measurable metrics that track progress towards the defined objectives. Choosing the right KPIs is crucial, as they should be quantifiable and directly reflect the initiative's impact.

Once KPIs are established, the framework moves into data collection. This stage involves gathering the necessary data to measure the chosen KPIs. Data collection can occur before, during, and after the implementation to capture a holistic picture. The next step involves analyzing the data collected. This analysis helps assess how well the implementation is performing by comparing data against baselines and identifying trends. By analyzing the data, organizations can gain insights into areas where the initiative is succeeding and areas that might require adjustments.

The final stage of the framework focuses on calculating the Return on Investment (ROI). ROI provides a financial perspective on the implementation's success. Here, the costs associated with implementing and maintaining the framework are compared against the benefits achieved. This financial analysis helps organizations determine if the initiative is delivering a positive return on their investment.

This posture assessment offers a valuable tool for organizations. By following this structured approach, organizations can move beyond subjective assessments and gain a data-driven understanding of their initiatives' effectiveness. This allows for informed decision-making about future investments and ensures that resources are allocated towards initiatives that demonstrably achieve their intended goals.

## 1.7     Research Questions

Building on the gap outlined above, this study proposes the following three research questions:

1. What are the key technical controls of a Zero Trust Architecture in organizations?

2. What is the impact of ZTA on cyber attack prevention in organizations?

3. What are the industry best practices for implementing a ZTA?

The first research question addresses which Zero Trust Architecture technical controls are commonly deployed in enterprise organizations. Organizations can self-assess their environments by establishing a baseline to ascertain which technical controls they should deploy. The focus of this question is on which key technical controls make up a Zero Trust Architecture and why they are essential in an enterprise environment.

It is important to understand these components and their relationship to the overall architecture. Without that understanding, organizations could find themselves implementing a "zoned" architecture instead of Zero Trust. More importantly, insight into which components to concentrate on first provides guidance toward significantly strengthening an enterprise's security posture and moving the enterprise closer to the adoption of a Zero Trust environment in the next few years.

By understanding these elements, the organizations will be able to fortify their security posture and adapt to the ever-evolving developments of the threat landscape. By providing insight as to which of the key technical controls to focus on the implementation of, an organization can significantly improve its security posture and move closer to successful adoption of a Zero Trust Architecture to mitigate the risks related to modern cybersecurity.

The second research question seeks to provide insights into how ZTA prevents cyberattacks and safeguards critical assets within enterprise environments. A quantitative research approach is utilized to address this question. This approach explores the effectiveness and implications of implementing ZTA in enterprise organizations.

The reason for this choice is straightforward: to assess the degree to which the implementation of a Zero Trust Architecture influences an organization's security compared to its previous traditional, perimeter-based security model. The numerical data concerning key security metrics was collected from 138 enterprise-level organizations that have adopted ZTA. These organizations have transitioned from more traditional security models to ZTA, and they supplied the data to indicate how security metrics have improved since that transition.

The third research question strives to compile a list of industry best practices for implementing a Zero-Trust Architecture. The process begins with thoroughly assessing the organization. This involves identifying sensitive data, assets, applications, and services that need to be protected. It is also essential to determine the attack surface, which includes all potential entry points for cyber threats, such as network connections, user accounts, and software vulnerabilities.

## 1.8     Project Feasibility

A broad approach is necessary in understanding if it would be viable to use essential technical controls in evaluating the cybersecurity posture within an enterprise organization. This dissertation includes a review of the literature that provides a deep-driven understanding of the state of research, identification of best practices, and practical applications concerning the assessment of cybersecurity posture. The review of the literature touches on different aspects, such as key technical controls identification, assessment criteria, and formal frameworks review, case studies and real-world applications analyses, and investigation of emerging trends and challenges.

Following a thorough literature review, the next step is to devise an appropriate methodology to conduct the research. This involves the intentional selection of the most significant and relevant technical controls to be assessed in this research study, guided by the outcome of the literature review. In respect of the technical controls selected, clear detailed assessment criteria should be stipulated that details how these technical controls will be assessed as effective or otherwise in ensuring cybersecurity posture. This should then be followed by identifying enterprise organizations across industries and sizes that could participate in the study to ensure the sample obtained is representative.

It is based on the fact that data collection methods must be defined-whether it involves surveying, interviewing, on-site assessment, or a combination of these techniques-to thoroughly obtain information dealing with the cybersecurity posture of the organization. This would cover comprehensive information on the implementation, configuration, and performance of the key technical controls within the participating enterprise organizations.

The research also discusses in detail the specific constraints that organizations have in implementing and operating such technical controls effectively, as well as resources and expertise needed for their effective implementation. The investigation makes a critical evaluation based on a well-structured research methodology of the use of principal technical controls to assess and improve the cybersecurity posture of an enterprise organization using the literature review methodology. Results from this study could help support best practice formulation, guidelines, and decision-making frameworks that may guide organizations in proactively improving cybersecurity defenses.

The controls are of little value if they cannot be implemented well. And certainly, expertise combined with resources is necessary for any successful implementation. Selecting the most significant and relevant technical controls from the study is then used to address the problem under interest. Those controls must then be evaluated using clear and comprehensive criteria that allow the enterprise to judge how well each one works to improve its unique cybersecurity environment.

## 1.9     Dissertation Outline

The rest of the dissertation is organized as follows: Chapter 2 presents a thorough literature review on Zero Trust Architecture. Chapter 3 introduces the research methodology used in

Page **25** of **232**

this study, followed by the research design and implementation in Chapter 4. Chapter 5 presents our findings for the three research questions. Chapter 6 introduces the novel Four-Tiered Technical Controls Model. Chapter 7 concludes the dissertation.

# CHAPTER 2: LITERATURE REVIEW

Organizational enterprises are confronted by an upward spiral of cyber threats in this increasingly connected digital world. These cyberattacks disrupt operations, infringe on confidentiality, and erode stakeholder trust. Therefore, it has become important for any organization to proactively review its cybersecurity posture in safeguarding against potential breaches. So how can organizations gauge the effectiveness of their cybersecurity today? More important, how can they do it better than they did in the past? Sun et al. (2022) point out that it is crucial for organizations to stay ahead of the game by regularly checking their cybersecurity measures to prevent any potential breaches.

## 2.1     PRISMA 2020 Methodology

This chapter provides a comprehensive review of the existing literature on the impact of Zero Trust on cyberattack prevention. The review is conducted following the Preferred Reporting Items for Systematic Reviews and Meta-Analyses (PRISMA) methodology (Haddaway et al., 2022). PRISMA employs a transparent and methodical approach to guarantee an impartial selection and evaluation of the papers, allowing for a thorough and reproducible review (Page et al., 2021).

The initial phase involved searching the IEEE Xplore, ACM DL, and Google Scholar databases using the combination of the following keywords ("Zero Trust Architecture" AND Implementation) OR ("Zero Trust Architecture" AND Approach) OR ("Zero Trust Architecture" AND Methodology). Additional focus is on studies published in English from 2020 to 2024. Initially, 1740 research papers were found: Google Scholar returned 1,612 results, ACM DL and IEEE Xplorer returned 128. Of the total 1740 research papers recorded, 402 duplicate papers were eliminated before evaluation, and an additional 1034 papers were excluded for various other reasons. 304 papers were screened, assessed, and pared down to 98 studies included for review.

While conducting the literature review, various technical papers were eliminated for several other reasons. Some papers were excluded due to outdated information, which did not reflect current advancements and trends. The review of the literature was confined to those studies that contained a rigorous methodological approach and had data that was sufficient to support their conclusions. The selection was also made based on the credibility of the studies, which were mostly peer-reviewed. Excluded were a few papers because they were redundant and did not offer any new insights or differ significantly from what was already had. The paper selection process used PRISMA as depicted in Figure 5.

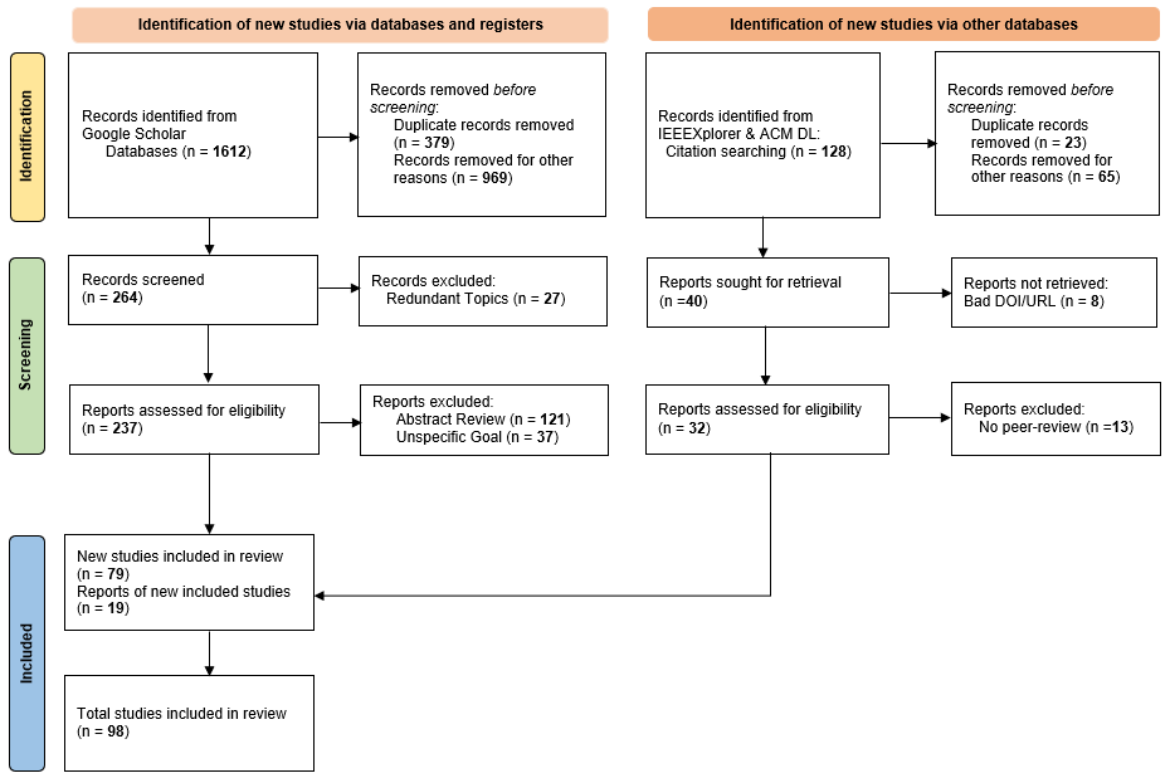

*Figure 5. Selection of Papers for Multivocal Literature Review Using PRISMA*

## 2.2      Litmaps Literature Review Assistant

Litmaps revolutionized the research through their advanced algorithmic way of discovering the academic literature. With Litmaps, users are able to locate the latest scholarly articles in the shortest time possible, hence guaranteeing comprehensive knowledge on the subject matter. Further, it goes beyond traditional database searches and enables complete visualizations and systematic analyses of academic literature networks. Figure 6 shows an example of Litmap visualization carried out for part of this research.

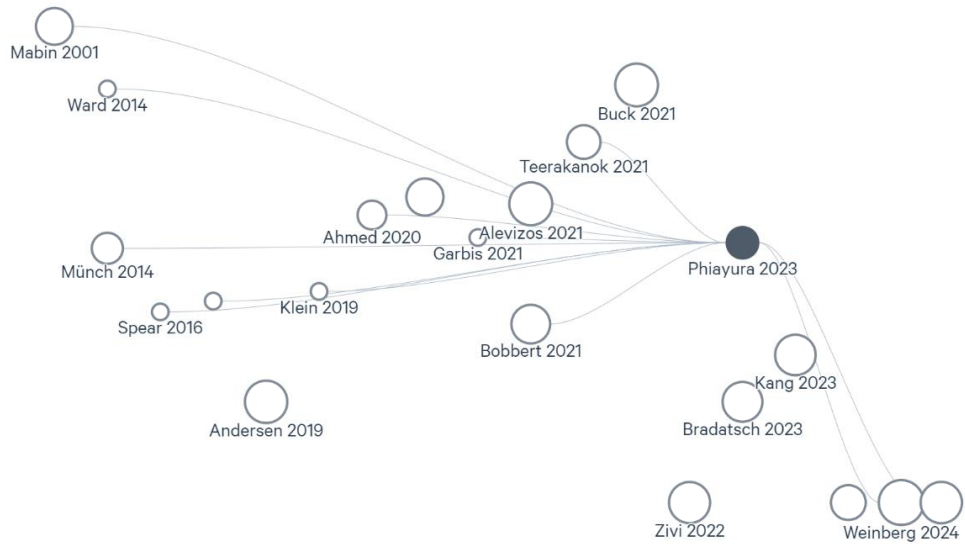

*Figure 6. Phiayura 2024 Litmap Visualization*

Carefully selected seed papers lay as the basis of the systematic discovery process, with their choice based either on citation impact or relevance to the research questions. Each of the research questions resulted in divergent paths within searches, therefore building different citation networks. The final approach was bidirectional in fact-many times both backward and forward citation mapping-enabled the description of foundational works and emerging research. Cross-referencing between various streams of search identified intersecting themes and concepts.

The enhanced visualization capabilities of the platform proved instrumental in understanding complex research relationships. Interactive network maps showed intricate relationships of citations, color-coded clustering showed thematic groupings in literature, and temporal visualization features presented how topics of research evolve over time. Connection

Running head: Aiello Dissertation: Prescriptive Zero Trust                    Page **30** of **232**strength indicators effectively highlighted the works which have the most influence inside each domain of research.

The iterative refinement strategy then moved forward in three clear stages: The primary search broke down research questions into their core conceptual parts, using optimized search precision via Boolean operators. Thus, subsequent filtering of results was done with caution regarding relevance and citation metrics. The citation network analysis utilized bidirectional citation tracking for seminal works and co-citation analysis to outline the intellectual communities in this specialized field. Reference coupling also spawned other cognate studies that were methodologically similar, while the citation paths illuminate how key concepts evolve in these literatures.

In the thematic organization step, the literature was organized categorically by developing a custom tagging system. Collections were made for each of the major research themes identified while subtopic mapping revealed the knowledge gaps associated with the field. Comprehensive network analysis showed cross-theme relationships and provided detailed insights into how different research areas interlink.

The automated suggestions for citation considerably widened the base of literature. The pattern recognition in the network of citations showed emerging research trends, while gap analysis showed less-explored research areas. Such a methodology made integration with several theoretical perspectives possible and pointed out relations between topics that otherwise would appear quite disparate. Such an exhaustive approach led to coverage of the relevant literature being complete, identification of new emerging research directions being possible, proper understanding of theoretical grounds being gained, methodological patterns being recognized, and discoveries related to across-discipline connections taking place.

## 2.3   Implementation Guidance

As a response to the dynamic threat landscape and increased use of cloud Services, ZTA is gaining momentum as a paradigm around escalating cybersecurity issues. One institution is the National Institute of Standards and Technology (NIST), which has provided

Page **30** of **232**

instructions and guidelines on principles of Zero Trust controls. There are, however, not enough implementation guides for these principles. This is one of the most difficult problems facing organizations regarding implementation, while greater adoption and focus could provide the inducement to justify the efforts.

The trusted legacy networks of yesteryear need to go. It is thought that all devices and users in the network "belong" there. This change to a focus on identity changes how we think about security. It has big effects on access rules and how we check users (Chandramouli & Butcher, 2023). NIST Special Publication 800-207 lays out what Zero Trust means. But it does not give enough details for organizations wanting to put this into action.

The NIST publications on Zero Trust are a good start. Zero Trust means focusing more on context, the "What" or "Why" and not so much on the "How." ZTA is currently the best choice for keeping today's dynamically-changing networks safe.

Some studies look at Zero Trust more broadly than just the NIST framework. While some of these studies praise the "flexibility and growth" of Zero Trust, they do not consider how businesses will actually use this method. The main issues focus on reluctance to change (Phiayura & Teerakanok, 2023) and cultural obstacles (Damaraju, 2024) to moving forward, as well as the shift from old systems (Syed et al., 2022) to Zero Trust designs.

Differences between techniques creates difficulties (Tsai et al., 2024) and problems with the mixing of different technologies (Phiayura & Teerakanok, 2023) within a business can create significant obstacles. It is so, particularly when even just one part of the architecture, a necessary access security control, could clash with some of the other components of the Zero Trust model.

When organizations seek to implement ZTA, they frequently encounter a couple of substantial hurdles. To begin with, it is the technical complexity of the ZTA that it is. Then comes the necessity of making all the ZTA's various units cooperate, and this problem of integration is quite a serious one. ZTA is a good concept and a must-have one, however, the path to it is not easy and is filled with challenges.

Zero Trust Architecture model proposed for access control in cloud-native applications (Teerakanok, Uehara & Inomata, 2021) is a notable contribution to addressing these challenges. Chandramouli and Butcher (2023) really stress the need for a solid platform that brings together API gateways, sidecar proxies, and application identity infrastructures. Picture it like a secure gated community where every individual house (or service) is kept safe no matter where it is located. This makes sure that security practices are preserved all the time, even if the services are really close to each other or halfway around the world.

The National Institute of Standards and Technology guidance on Zero Trust Architecture provides a basic framework sufficient for the NIST guidance on Zero Trust Architecture, but still, the organizations need to implement certain solutions that would address the real problems. The existing studies, while important, point out the need for further research and creativity in this area to completely achieve the goals of Zero Trust paradigms and to improve the security posture of modern, dynamic network environments.

Although important, the existing work is only scratching the surface of what a Zero Trust design could be like in its entirety and we have vast room for further exploration and contribution, hence with significant search opportunities within this space. There is already some great research around Zero Trust, but there is still much work to do. This is where a million things in the way of cloud computing, IoT and edge work its wizardry on our now much larger digital boundaries; as well as keeping an eye open for new ideas across Zero Trust but maybe even more critical than that, new innovations. But how can it be made sure that Zero Trust does not have the same theoretical holes, which were continuously exploited over years to just wait for security measures are in place? How do we guarantee that the next version not only reinforces defenses but expands their permeation across every piece of an organization's digital and physical property?

There are countless opportunities to explore in this field. For one, develop industry standards and create sound methods for integrating various technology stacks to work together seamlessly as if part of a single entity. For another, study ZTA to understand the range of its possible implementations across various industries and the performance of those implementations. An architecture's implementation is not its performance, after all. And consider how machine learning (ML) and artificial intelligence (AI) could enforce supremely adaptable security policies in ZTA environments during real time.

This continued and ongoing research helps to push the field forward, so that organizations might have what is necessary in order implement, enforce (and more importantly) real Zero Trust Architectures within ever-more convoluted digital ecosystems. We have a Zero Trust model that needs to be built from the ground up, but if the pieces do not come together properly and create access security then it is half-baked.

## 2.4    Examination of Zero Trust

Weinberg and Cohen (2024) present an extensive examination of the adoption of the Zero Trust (ZT) security framework during the period of emerging technologies. The basic tenets of Zero Trust provide a good foundational understanding of what it is and how it works. ZT is based on the principle of "never trust, always verify. Continuously verifying user identities, application workloads, and the security health of devices (by administrative policy enforcement) is what ZT is all about. ZT uses a variety of techniques to do this, including multi-factor authentication (MFA), and it manages to maintain overall "trust" in its system by keeping tabs on compliance (again, more or less maintaining security by means of policies). Trust in ZT is "conditional," and the conditions are good ones.

The ZT security framework represents a fundamental change from traditional stair-step security models and access control models like ABAC, RBAC, and FGAC (Karatas & Akbulut, 2018). ZT moves away from paradigms that partition information (the essence of both Bell-LaPadula and Biba) to a model based on continual authentication and authorization.

The essence of Zero Trust is not just that you should not trust anyone inside or outside your information system. It is also that you should continually verify everyone and everything trying to access your system. The ZT model embraces a multilayer access control system that makes it very hard for an intruder to get to the valuable assets inside your information system.

AI and ML are very important in a Zero Trust Architecture. For instance, Support Vector Machines are good to use for anomaly detection, classifying network traffic as normal or suspicious. Decision Trees are good for making real-time access decisions based on rules we have predefined. Complex problems, such as forecasting security breaches, are where Artificial Neural Networks come into their own. They find patterns in large volumes of data and make valuable predictions from those patterns. In the architecture of Zero Trust, even advanced predictive capabilities can only be utilized if the access control system is working first so that each and every access request is able to be matched against the architecture's security criteria. The saying "Trust but verify" has never been more appropriate than with access requests, and the even more demanding model of ZT requires complete vetting, carefully checking security controls while giving access to the only needed user and resource.

However, they are not the only methods employed. Social beacons, which are part of the Internet of Things, provide ZT with auditory, visual, and infrared signals about who is where, doing what. Connected through the cloud, these beacons help ZT carry out three disciplines: access control, ongoing monitoring, and threat detection.

The approach towards Zero Trust needs to be systematic and incremental in nature, starting with a small test of proof, followed by thorough asset determination, clear planning, integration of automated procedures, risk-based prioritization, 'change' management, proper governance structure, and continuous monitoring and improvement. Large potential is lying in the ZT architecture when the latter can be integrated with other security frameworks. It works particularly well with Extended Detection and Response - XDR, Identity and Access Management - IAM, Endpoint Detection and Response - EDR, Security Orchestration, Automation and Response - SOAR, and Security Information and Event Management - SIEM systems.

This architecture creates a synergistic outcome: an overall enhanced security posture. The upshot is not only more comprehensive threat detection but also better access control, improved endpoint security, far more efficient automation, and sophisticated log analysis. In other words, we get a more responsive, nuanced, and effective security monitoring system because of this architecture.

## 2.5	Key ZTA Technical Controls

A complete Zero Trust Architecture is composed of various fundamental capabilities that must be used in conjunction to deliver the primary functions needed by a Zero Trust model (Kumar et al., 2022). The first of these is a consolidated policy engine. By centralizing the administration of your access policies, this engine then ensures that your enterprise is following those very same policies through an automated enforcement mechanism. The second is network microsegmentation. This allows the architecture to resemble several small networks, preventing unauthorized lateral movement. User Identity Service This service should work properly because it is applied to logical access control as a part of the first tier security in this architecture (Noel et al., 2021). The final ability is endpoint security. This protects both physical and virtual devices prior to use. The risk treatment builds on all previous capabilities and is driven by the policy engine (Park et al., 2023).

Data encryption protects sensitive information while in transit and at rest, preventing unauthorized data exposure (Thabit et al., 2023). Analytics and visibility tools continuously monitor and analyze activity, user behavior, and system events to detect potential threats and validate trust levels dynamically (Al-Mhiqani et al., 2020). Orchestration capabilities automate the provisioning of resources, and the adaptation of access policies based on situational context and risk postures, enabling the Zero Trust Architecture to respond and adapt to evolving threats and business needs (He et al., 2022).

A resilient Zero Trust Architecture implementation makes coordinated use of multiple core capabilities (Stewart, 2020). The components comprise a centralized policy engine designed to enforce automated access control, network microsegmentation implemented to isolate resources, endpoint security compliance checks utilizing multi-factor authentication to validate identities, data encryption, analytics, and visibility systems to facilitate continuous monitoring and trust validation, and orchestration systems to automate provisioning and dynamic policy adaptations (Marelli, 2022).

Real-time, continuous monitoring and analysis of network traffic and user behavior are foundational to the Zero Trust approach. These are just a few of the mechanisms that comprise a Zero Trust Architecture. The advantage of this architecture is building in an effectual, even if some of the methods used are somewhat labor-intensive, way of ensuring that access attempts are valid, that resources are well protected, and that bad actors are kept out. Violations are not only detected but also reported in real-time.(Kott & Arnold, 2015). Every access attempt is validated. If the attempt is invalid, it is guaranteed that a threat actor made the access attempt.

## 2.6    Framework Adoption and Alignment

To assess an organization's posture regarding ZTA deployment, an extensive evaluation must be carried out in various areas (Sarkar et al., 2022). One basic method of assessing advancement is the number of the most essential ZTA controls adopted, including how many have identity and access management, controls on least privilege, and monitoring capabilities (Currey et al., 2020). This is a good estimate of the degree of integration of the overall architecture. However, it is not sufficient to think about the number of deployed components; in addition to that, they should measure how efficiently every control has been put into practice.

For instance, is the access control implementation granular and context-enforcing, or is it a simple one? The in-depth analysis by Kayes et al. (2018) stipulates that this would analyze the comprehensiveness of the integration of the ZTA. The alignment with industry standards and best practices- for example, those demonstrated in NIST SP 800-207 (Rose et al., 2020)- is also important in ensuring that compliance is observed with recognized frameworks.

Cao et al. (2022) highlight several critical aspects for evaluating ZTA posture. These include the number of core components deployed, the extent of integration for each component, conformity with industry best practices, and the recognition of policy exceptions. Monitoring these metrics against maturity targets over time provides a clear, data-driven insight into the ZTA's advancement and highlights any gaps. Comparing these metrics with trends in risk and breaches demonstrates the guidance's effectiveness.

## 2.7     Assessing the Impact of a Zero Trust Architecture

When trying to establish the impact of a ZTA deployment, its practical application in the organization should be considered. User perception of the implemented ZTA controls, especially in the areas of usability and influence on productivity, is also an important factor to consider (Sarkar et al., 2022). There is a significant danger that such employees will instead reject such measures or compliance simply because the security measures undertaken are too intrusive into their normal workflow (Alevizos et al., 2022).

One of the things that need to be monitored when moving to the implementation of ZTA is how it affects your overall IT infrastructure. This is because the increase in number authentication checks, verification steps, this time not only on the user level but also on the network level may sometimes cause a slight lag and in turn may have an adverse effect on the application and the end user. It is a question of where that ideal line is for any organization where extra security measures do not compromise operational productivity. On the other extreme, if the ZTA is competent it might as well be embedded within the infrastructure without causing any negative impacts however mild it may portray. So the worth of ZTA operational efficiency assessment is at least as much as ZTA security features assessment (Baee et al., 2021).

Zero Trust Architecture's successful deployment is contingent upon multiple factors you must consider, including operational efficiency and what we might term "security accessibility." In other words, can you find a way to deploy Zero Trust that does not bog down otherwise legitimate requests for access to your systems? This is a particularly pressing question in the supply chain context because the sheer number of system-to-system interactions in a connected supply chain makes slowness and fail-to-access scenarios hugely troublesome from a business standpoint (Levine & Tucker, 2023; Park et al., 2023; Collier & Sarkis, 2021).

To facilitate consistent and objective evaluation of an organization's ZTA implementation, it is beneficial to sample peer organizations to identify common ZTA elements deployed (Phiayura & Teerakanok, 2023). Then, a benchmark can be established to compare various organizations against the norm. Employing an archetype as a model is tantamount to having a pliant blueprint for an organization's security infrastructure. It sets a good benchmark without suggesting that a single framework should be followed, which could detract from the efficacy of security setups tailored to an organization's unique needs. In the context of organizational development and improvement, reference points highlighting key areas needing work are invaluable. Using the Balanced Scorecard framework for this purpose within the context of Security Operations Centers is one notable way to do this (Polat et al., 2022).

NIST offers a self-assessment guide in its "Cybersecurity Framework." This guide helps organizations gauge their current state within the standard and chart their improvement initiatives alongside the recognized frameworks to align with their technical controls. The frameworks are well-known in the industry, making them easier to align with and use as a reference. They also simplify the auditing process (NIST, 2022).

Despite the well-established frameworks and associated self-assessment initiatives, the posture of many American enterprises still needs to improve. Therefore, this study seeks to explain to the reader some of the key elements of a Cybersecurity Framework necessary for self-assessment initiatives and "posture improvement" endeavors.

A Multilayered Security approach ensures that the system is protected at every level. The enterprise has varied layers, and each layer has different infrastructure components. For MEC, covering each layer with a security component requires minimal resources and is almost effortless to implement compared to working with the infrastructure at just one layer. This part of the article outlines the steps that should be undertaken to achieve a minimum security posture (MSP) with minimal resources. Coverage with a Multilayered Security approach is the main focus.

One factor that organizations should consider is the human element. Even with robust technical controls, human error is a significant cybersecurity risk. Organizations can mitigate this risk by investing in comprehensive employee training and awareness programs that cover cybersecurity best practices (Aldawood & Skinner, 2018). Staff members should learn to identify phishing attempts, use strong passwords, and recognize social engineering and other contemporary tactics that attackers employ to dupe people into compromising an organization's security.

## 2.8     Strengthening an Organization's "Human Firewall"

Issues to consider that would help further strengthen an organization's "human firewall." Safety and security, rather than just meeting compliance, should be part of the focus. One easy way to effectively do this is through an awareness-centered information security compliance model. This model places much emphasis on awareness in attaining compliance and proposes that good leadership and employee trust are the enablers of awareness. Workers are far more likely to adhere to security policies when they realize their vulnerabilities and their self-efficacy is enhanced. Pérez-González et al. (2019); Koohang et al. (2019). Metrics should be fitted to the organization like a fine suit, aligned lock and step with the industry vertical, regulatory requirements and risk appetite, along with outputs against strategic goals of that very organization.

The latter might include, for example, a financial institution that prioritizes customer records and may become overly fixated on metrics related to the encryption of that data and the access controls protecting those records. A manufacturing firm, on the other hand, would be more concerned with safeguarding its control systems and IP and may pay more attention to the type of metrics which would disclose whether either was secure.

Using published standards or frameworks as a basis may yield an insufficient understanding. To get a clearer view of the effectiveness of the Zero Trust Architecture, one can combine the not-yet-available standardized benchmark test results with the qualitative assessments of security SMEs and citizen judges (Zaber & Nair, 2020). This combination will provide a clearer understanding of how the ZTA behaves under a variety of (problem) conditions, including its operational impacts.

## 2.9	Barriers and Challenges in Implementing Zero Trust

The threat landscape in cybersecurity is fluid, with "diverse threat actors" using a constantly changing landscape of tactics, techniques, and procedures. As Kerman et al. affirms, ZTA itself is an adaptive risk-based approach, calling for fine-grained, context-aware access controls since organizations need to always verify user and endpoint trustworthiness continuously. Even ZTA, however, requires a means by which its effectiveness could be measured. Current best practices are for the organizations that have adopted a Zero Trust Architecture to evolve appropriate, relevant, and context-aware metrics that can do that measuring. (Zero Trust Core Principles, 2021).

Today's digital world has given tremendous opportunities to cybersecurity attackers to target enterprise organizations. The sophistication of today's cyber threats requires more than just a few traditional security measures. Fortunately, Zero Trust can be turned into a cybersecurity framework to increase the organization's level of protection for sensitive data and systems. Mitra offers a practical interpretation that it shall be considered a narrative guide through which cybersecurity frameworks are translated into protection postures. The legend is that only legacy infrastructure stands in the way of implementing Zero Trust within enterprise organizations.

The advancement of new technologies and the changing of the existing ones can, on the other hand, be time and resource-consuming when it comes to implementation. For instance, implementing a Zero Trust cybersecurity framework can bring the same challenges. Implementation of Zero Trust may also follow similar patterns whereby several technologies and "pieces" may need to be employed that work together. One needs (or gains) a certain level of savvy on how to orchestrate all of these many technologies in order to be able to pull it off. And the way that this system of many parts works with the aggregate capabilities of the organization's environment is what gives the institution further heights of security than what was obtained earlier. (Richter & Sinha, 2020; Clements & Horton, 2024)

In the Zero Trust approach, the user experience is more important than in typical practices. A Zero Trust perspective generally emphasizes that with the right individuals and devices, the right accessibility can be enjoyed (Ghosh et al., 2021). Concerning the Zero Trust paradigm, a user can reasonably expect to be authenticated several times and through varied methods prior to gaining very limited access (Buck et al., 2021).

Implementing Zero Trust requires specialized skills and expertise. Organizations need security professionals who understand the principles and technologies underlying Zero Trust and can effectively design, implement, and manage the framework. However, there is a shortage of skilled cybersecurity professionals in the industry, making it challenging for organizations to find and retain the talent necessary for successful Zero Trust implementation (Haleliuk, 2023).

## 2.10    Additional Challenges and Their Impact

Yet, legacy technologies can also inhibit Zero Trust implementations. Notably, this points to the problem of broader coverage: the inability to support modern protocols, controls, and architectures. The move to Zero Trust represents a major shift in culture and demands extensive change management to prompt user adoption. Failure to secure buy-in will undermine security policies, as cultural opposition can arise, as it has on occasions regarding virus protection and other security controls. These require significant new tooling, systems re-architecting, and ongoing operational complexity, adding substantial cost. The skill sets required for Zero Trust- from Cloud Security to microsegmentation- are rare skills. Mis-configurations due to lack of expertise can also happen with them.

### 2.11 Zero Trust Timeline Evolution

Zero Trust developed as the cybersecurity framework that evolved over time, particularly with the growing risks and vulnerabilities of network security. But the origin can also be traced to 1994 when Stephen Paul Marsh provided a model that removes an assumption of trust-thus implementing secure information flow within organizations. It has undergone many changes in different stages of development since the beginning when the concept was put forth. Marsh was among the first to come out with the concept of de-perimeterization. It is a conceptual approach and challenges the conventionally adopted idea of secure perimeter relied upon for network security. Instead, Marsh promoted the philosophy of "trust nothing, verify everything" for organizations where, by default, no trust should be placed in any user or device, and its authenticity should always be checked and verified.

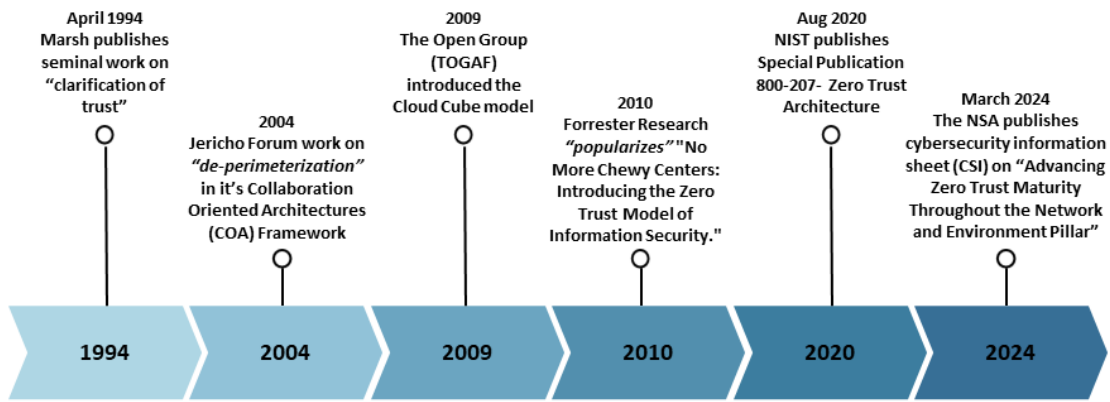

*Figure 7. Zero Trust Timeline Evolution*

The concept formed the basis of what has become known as Zero Trust, which has since been expanded and honed by several groups and personas. This concept, by Marsh, was further developed on by security professionals in a think-tank known as the Jericho Forum in 2004, which then coined the term "de-perimeterization" to refer to the emergent necessity for enterprises to quit a traditional approach to security based on such a perimeter model. They focused their call on an alternative strategy for network security that was dynamic and flexible to shift the emphasis away from boundary defense to asset protection. Also in the same year of 2009, The Open Group introduced the Cloud Cube model to further develop

Zero Trust. This model gave organizations a means of ascertaining a degree of trust for different cloud deployment models and a way of deciding on the various strategies for securing the clouds.

Further elaborating on this idea in his research paper titled "No More Chewy Centers: Introducing the Zero Trust Model of Information Security", it has been John Kindervag, a principal analyst at Forrester Research, in 2010. Kindervag (2016) For a traditional perimeter-based security model, he claimed that this enforces a "chewy center" wherein an attacker, once able to breach the perimeter, can easily move laterally within the network and freely access critical assets. In response, Kindervag introduced the Zero Trust model to address the problem. Under this model, controlled authentication and verification should be continuously practiced at all points in a network.

His approach emphasizes the principle of not over-relying on the network's boundary for security, but truly implementing strong access controls, microsegmentation, and the least privilege principles within an organization. In August 2020, the National Institute of Standards and Technology published NIST Special Publication 800-207, by Rose et al. (2020), describing the guidelines for an implementation of a Zero Trust Architecture. This document puts forth basic tenets and best practices for the implementation of Zero Trust, such as multi-factor authentication, encryption, automation of security access rules, and treating all applications as if they were directly and openly exposed to the internet. It also recommends several other best practices for security: continuous monitoring, data protection, and secure configuration management.

To this end, the NSA issued the Cybersecurity Information Sheet (CSI) in 2024 to transfer knowledge pertaining to the network and environment pillar that centers on perimeter defense and the security controls closer to resources and data in the ZT security model. This pillar concerns data flow mapping and network segmentation by applying strong access controls that prevent lateral movement. This allows for host isolation, network segmentation, encryption, and visibility of the enterprise. The better the internal network control matures, the better the defense in depth to allow organizations to limit a network intrusion to only a small section of the entire network.

## 2.12     Summary

The literature review examined the development, implementation, and impact of Zero Trust Architecture in modern cybersecurity. Applying the PRISMA 2020 methodology, supported by the literature review assistant Litmaps, the research reviewed close to 2,000 academic papers and focused on 98 key studies that shed light on how ZTA was developed and how effective the concept has become. The literature review discusses how ZTA originated from Stephen Paul Marsh's 1994 trust model and then evolved through the efforts of the Jericho Forum, John Kindervag's foundational work, and recent NIST frameworks to transform what was once a purely theoretical conception into one of today's major security paradigms.

The findings note that against an expanding cyber threat, combined with the complexity of today's IT environment, organizations are embracing Zero Trust Architecture, citing the latest surveys where 43 percent of firms have already implemented zero trust solutions and 46 percent are on course to do the same. The literature review reiterated some of the key technical controls being considered integral to the implementation of ZTA, including consolidated policy engines, network microsegmentation, user identity services, and endpoint security. It also, however, underlines significant challenges to the adoption of ZTA: technical complexity, integration, resource constraints, and a requirement for specialized resources within an already skills-constrained profession.

The literature review further underscores that ZTA's effectiveness needs to be evaluated through quantitative metrics. It, therefore, calls on organizations to develop context-sensitive measures that meet the needs of their industries and the appetites for risks. This chapter concludes with correlation analysis to understand the relationship between cybersecurity implementation variables. This helps bring into light the importance of considering linear and nonlinear relationships in assessing the security posture and how effective these postures are in preventing cyberattacks.

    The ultimate objective of this analytical approach is to help make the assessment of security posture more effective. Organizations would be able to make more prudent decisions based on the relations among the various security measures and their outcomes concerning their investments in, and strategies for, the sphere of security. An understanding of such nature would help indicate which combination of security controls provides the most effective protection, thereby helping the organizations optimize their security infrastructure. This would give organizations a more wholistic and, hence, effective approach to cybersecurity based on the data-driven insights relating to how the different elements work in concert to prevent and mitigate cyber threats.

# CHAPTER 3: RESEARCH METHODOLOGY

The study uses a nine (9) step quantitative method research approach (Kumar, 2011), including a survey. The survey is conducted primarily among cybersecurity professionals from around the world. These professionals are reached via the internet. The survey is available in English. The participants use a web-based program called QualtricsXM to record their responses.

The survey questions are grouped into three parts. Part one asks about the value of Zero Trust. Specifically, the first three questions ask about where these professionals see value in the Zero Trust concept. (McMillan, 2021)

This study intends to use a quantitative research methodology and techniques. The first half of the study employs a mostly quantitative framework to help answer the guiding research questions. Following the first half study, the second half mostly uses techniques to help better understand the significant findings that emerged from the first half.

The research begins by creating a set of foundational questions that lead to fulfilling the measurement objectives tied to evaluating the maturity of a prescriptive ZTA framework. These questions are straightforward but crucial to the subsequent step of defining what serves as the actual metrics for the evaluation. When we think about what it means to evaluate something, we typically think about using a set of rubrics or metrics to gauge how well something functions at a basic level, how well it operates on a prescriptively stated level, or how mature it is (or is not) as an architecture or framework.

This quantitative research approach derives its real power from being not just a one-off undertaking but a necessity-driven cyclic process- of focusing first, then refining last. This research is about making things better. Better in a way that allows what's being researched to fit and work in today's rapidly changing world, of ever-evolving and sometimes unstable conditions. Such conditions make the cybersecurity fortification problem a slippery one.

A quantitative approach was used in this evaluation. The quantitative method provides a well-rounded assessment of an enterprise's cybersecurity maturity and of the applicability of the findings derived during the evaluation. This well-rounded assessment is further enhanced by the respondents' insights contributed by a number of cybersecurity professionals, who applied their years of experience and expertise to the evaluation. In essence, the use of this approach allows for the evaluation to make better judgment calls with regard to the applicability and accuracy of the evaluation's findings.

The approach for conducting the research using the quantitative method research is based on the findings from Kumar (2011). Figure 8 shows the nine (9) steps in the cycle.

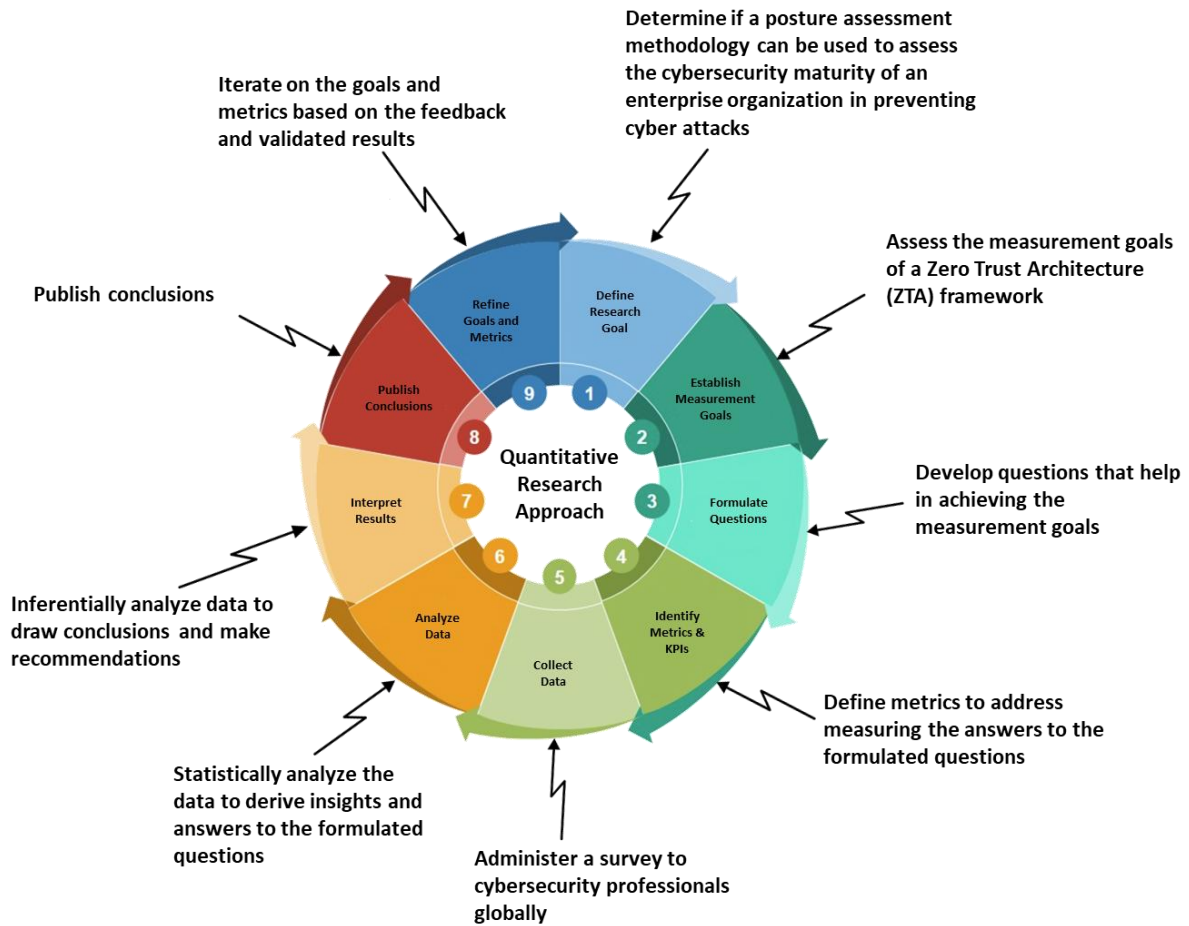

*Figure 8. Proposed Research Approach*

## 3.1    Define Research Goal

Organizations across all industries face a substantial risk from cyber threats. Many are embracing a "Zero Trust" security approach to protect against rising threats. The basic premise of the Zero Trust paradigm is to "never trust, always verify." The security advantages of Zero Trust are apparent, but organizations need to consider many other elements when looking at the various implementation options and techniques.

This effort attempts to lay the foundation for coming up with an approach that enables the assessment of an organization's cybersecurity posture in terms of its ability to prevent cyber threats. The focus of this effort is on making the Zero Trust security model not just a theoretical concept, but one that can be translated into actual practice, where organizations can implement it and get some value from it and measure that value.

The value proposed is to be measured using a four-tier model. The tiered model is easy to understand, and the names of the tiers have a clear and direct relationship to the level of value an organization could be expected to get from a Zero Trust implementation.

Each level signifies a successive step in ZT adoption, with the higher levels representing a more thorough and effective security posture.

## 3.2    Establish Measurement Goals

Incorporating Zero Trust into the complex infrastructures that support identities, devices, networks, and data demands an intricate melding of many components of the IT architecture. Even though the various frameworks that have been published provide useful guiding principles, not much is out there that helps leadership know what kind of progress is being made when Zero Trust is being implemented. This is really the first step in such an assessment.

Posture indicators will offer coverage percentages for each fundamental aspect such as identity management, network microsegmentation, multi-factor authentication, analytics, and user/device posture checks. Weightings can be used to adjust for varying priorities within an organization. An aggregate ZTA score might indicate progress made at specific phases. Comparing performance to industry norms is another aspect to consider. A quantified maturity score helps firms pinpoint capacity gaps to inform strategic Zero Trust roadmaps and investments.

### 3.3    Formulate Questions

Increasingly, enterprise firms are adopting ZTA models as a strategic imperative to push back against sophisticated cyber threats. ZTA is not a transformation of current security measures but rather an improvement- consistent validation of all network connections and access to data, for one. Yet, what is ZTA, really? And how can its level of advancement be evaluated quantitatively? In a prescriptive sense, ZTA can be thought of as having requisite elements; and in a more progressive sense, as having maturing "stages" (it can be advanced or not) and "scoring" (it can be well-implemented or not). This chapter discusses not just the "what" and "how" of ZTA but also the "why," offering context for why this framework is important and timely.

While enhanced security certainly promotes the transition to ZTA, the impetus is very much tied to economic considerations. Organizations want to be assured that the move from legacy systems to Zero Trust networks will pay off in terms of their budgets. Thus, on a common accounting basis, enterprises would like to understand how and in what specific ways ZTA impacts their bottom line, both positively and negatively. Clearly, no organization wants to find itself incurring costs to an architecture that, say, has a very minimal effect in cutting down the probability of a cyberattack. And given the substantial upfront investment that Zero Trust requires, it is all the more critical that an organization understands precisely where that money is going and what it is buying in terms of security. Ultimately, understanding ZTA's effectiveness economically is fundamental to justifying the architecture's implementation in the first place.

### 3.4    Identify Metrics, KRIs and KPIs

To effectively appraise the maturity of a ZTA implementation, one needs to examine measurable key performance indicators. The most obvious place to start is coverage. This is quite similar to the deployment of a security system such as an intrusion detection system (IDS). It involves assessing how many resources have policies enforced in an automated manner. From conversations with early adopters, that number tends to be a bit lower than most would prefer.

Some of the most crucial metrics involve the number of breaches and threats thwarted each year. When implementing something like ZTA, the goal is to see a reduction in successful attacks. To get a clear picture, it is necessary to compare annual breach statistics to the annual number of blocked malware or phishing attempts before and after adopting ZTA. By doing this, one can determine if there is a clear trend toward fewer successful attacks and more thwarted threats.

Another key step is reducing the attack surface with microsegmentation, and access based on the principle of least privilege, and that is especially important because the next vital containment step is limiting lateral movement across the surface. If a determined attacker manages to penetrate the system, they should not be able to move around much. If they can, the situation becomes critical, and that's what Zero Trust aims to prevent.

By tracking these KPIs across the various pillars of the Zero Trust Architecture model, a quantitative analysis of the maturity and impact of cyber attack prevention capabilities is enabled. Executives can use the data to make strategic and investment decisions.

Implementation of a zero-trust architecture requires continuous expansion and enforcement of policies for infrastructure and large-scale environments. Policy coverage is expressed as a percentage of assets under ZTA's automatic policy support that provides an equal amount of maturity. Another issue is the depth of microsegmentation, measured by the number of network levels and zones of instability established to prevent external movement.

Encrypted data transfer and storage with various devices complying with security standards shows maturity.

Additional metrics should be calculated in technologies and methods that support other unpublished reliability pillars. For example, the percentage of identification that requires multiple factor analysis indicates the maturity of a robust identification system. The final level of achievement includes qualified hardware management. Mobile security measures the flexibility of visibility required to monitor behavior and risk across infrastructure.

Performance metrics indicate a competent ZTA implementation. The percentage of ZTA's automated and manually managed forecasting services affects compliance. The frequency of facility inspections affects the ability to detect structural changes or emerging weaknesses. Progressing the maturity map through these milestones will help guide the guidelines and level-level definition of Zero Trust Architecture progress.

## 3.5      Structured Questionnaire Approach

The structured questionnaire is a research instrument that contains a series of questions that are already written and that contain options for answers that are already decided. It is among the most common research tools used in survey research to obtain a large amount of data from a large sample size in a systematic and standardized way.

What we usually call a "questionnaire" can be mostly classified as a structured questionnaire because it consists mainly of "closed-ended" questions that prompt the respondents to select one of the many provided answer choices. This characteristic enables highly standardized and highly quantifiable data collection, which makes structured questionnaires almost ideal for any kind of research that requires statistical analysis.

Gathering information from a varied group of participants can be done in a few different ways. One of the best methods for achieving consistent results from a lot of people is the good old questionnaire. As far as I know, the first use of a questionnaire dates back to the early 1900s. Since then, this method has been modified to make it even better for a wide range of different situations. Today, we use structured questionnaires more or less by default. The main reason why we do this is because the consistent results we get from using them are quite trustworthy. And when we follow the instructions that come with them, they are also quite easy to analyze.

Simplifying the tasks of inputting and analyzing data, these forms often reduce what was once a manual, labor-intensive, procedure to entering and receiving a statistic- in a few cases a "number"- that conveys an essential point about a pattern, trend, or relationship within the data. Of course, researchers can and do use more sophisticated methods to explore these features of their data (and will when necessary). But many of us would probably be a bit disoriented were we to sail all the way back to the manual procedures of the 1980s. And when these "sailors" return, they often state how much they appreciate the more "user-friendly" aspect of entering data and receiving results that are comparable to what they once received with prompts in a manual "open up to page X, don't forget to carry [something important] forward" solution.

When developing structured questionnaires, researchers must carefully consider the wording and structure of the questionnaire to ensure clarity and clarity. The questionnaire should be tested on a small group of participants to determine if there are any errors or problems with the survey instrument. Researchers may also consider using established scales or validated methods to increase the reliability and validity of questionnaires.

It is important to note that structured questionnaires are not without limitations. The predetermined response options may limit respondents' ability to fully express their thoughts or opinions on a topic. In some cases, the response categories provided may not align with the respondent's true feelings or experiences, leading to response error. Researchers must also be mindful of potential biases that may arise from the wording or framing of the questions.

Structured questionnaires are a valuable tool in survey research for collecting standardized and quantifiable data from a large sample size. By using closed-ended questions with predetermined response options, researchers can streamline data collection and analysis processes. While structured questionnaires have limitations, when designed and implemented thoughtfully, they can provide valuable insights into a wide range of research topics.

## 3.6     Participant Engagement Strategies and Pilot Survey Details

The most integral part of any research study or survey project that one undertakes is the engagement by participants. In their absence, active participation by the target audience will lead to a set of data that is not representative or reliable. But in general, effective participant engagement is attained by the researcher through use of several strategies and tools. This section outlines some of the effective participant engagement strategies and provides details relating to the pilot survey conducted using Microsoft Forms and Qualtrics XM.

One of the most successful ways to engage participants is through personalized email campaigns. Personalized emails sent to a potential participant may raise awareness and be indicative that the researcher values them. Personalized emails can include, but are not limited to, a greeting by the name of the recipient, the purpose of the survey, and why their participation is valuable. Such methods could bring in higher response rates and overall engagement.

Other areas included the social media forums, which could be used to reach and engage in a dialogue with members of a community or group, such as groups on LinkedIn. Various groups posted relevant to security, such as the Cloud Security Alliance CSA, Information Systems Security Association ISSA, and Information Systems Audit and Control Association ISACA, would provide the researchers with the opportunity to connect with those people that have a greater interest in the topic at hand in the survey. Announcements and postings in these groups will aid the researcher in obtaining cognizant and interested participants on the topic.

Second to the outreach strategy, a focus on perfecting the research design will also engage the participants. Two successive refinements and tunings were performed in making the questions clear, relevant, and interesting. Typographical corrections and enhancement in format also form a very important part of making for a better overall participant experience.

The pilot survey was initially rolled out using Microsoft Forms to a small group of participants. This pilot phase allowed for testing the survey instrument, identifying any technical issues, and gathering feedback on the overall user experience. Based on the feedback received, the survey was revised and refined before the "go live" phase on Qualtrics XM.

Qualtrics XM offers advanced features for survey design, distribution, and data analysis. By conducting a pilot before the official launch on Qualtrics XM, researchers can ensure that the survey is optimized for maximum participant engagement and data quality. This iterative approach to survey design and implementation can lead to more meaningful insights and a higher level of participant satisfaction.

By implementing strategies such as personalized email campaigns, social media outreach, survey refinement, and pilot testing, researchers can increase participant engagement and collect high-quality data for analysis. The use of tools like Microsoft Forms and Qualtrics XM can further enhance the survey experience for participants and researchers alike.

### 3.7    Implementation

QualtricsXM was used to conduct the online survey. QualtricsXM is mainly used to generate and take online surveys and questionnaires. The section explained an eclectic set of issues for the carrying out of the online survey using QualtricsXM: sampling bias, methods for data exploration, justification of the study, feedback sessions with experts, design of the questionnaire, response rate, data privacy, nonresponse bias, question order effect, validation of the questionnaire, sample selection, transparency, leading questions, and response scales. An example of what the survey looks like can be seen in Figure 10.

**Informed Consent.**
**Assessing the Impact of Zero Trust on Cyber Attack Prevention**

**Introduction:** This survey is being conducted by a student researcher from Dakota State University. In this study, I will assess the impact of Zero Trust on cyber attack prevention. Your participation is completely voluntary.

**Please read this consent agreement carefully before you decide to participate in the study.**

**Consent Form Key Information:** Respond to a 30-question online survey about cybersecurity and Zero Trust Architecture (ZTA).

**Purpose of the research study**: The purpose of the study is to develop a comprehensive framework for assessing the impact of implementing a Zero Trust Architecture in enterprise organizations for cyber attack prevention.

**What you will do in the study**: The study is composed of multiple-choice and ranking questions broken into five distinct sections.

**Time required:** The study will require less than 15 minutes during a single session.

**Risks:** There are no physical, psychological, legal, or otherwise risks anticipated with the survey.

**Benefits:** There are no direct benefits to you for participating in this research study. The study may help to understand whether implementing a Zero Trust Architecture improves an organization's posture in defending against cyber attacks.

**Confidentiality:** The information that you give in the study will be anonymous. The information that you provide in the study will be handled confidentially. Your information will be assigned a code number.

**Voluntary participation:** Your participation in the study is completely voluntary.

**Right to withdraw from the study:** You have the right to withdraw from the study at any time without penalty.

**How to withdraw from the study:** An opt-out is provided here at the end to abort the responses. If you want to withdraw from the study, the OPT OUT option is clearly identified in the survey. There is no penalty for withdrawing.

**Payment:** You will receive no payment for participating in the study.

**Using data beyond this study**: After the conclusion of the study, the data will not be used for any other purpose.

The study researchers may use a software application or an Artificial Intelligence (AI) program, like Chat GPT, to analyze data for this study, including your data. The researcher will remove any directly identifying information (such as your name, contact information, etc.) connected to the information you provide. This AI program may store your data outside of DSU for future use.

The researcher will keep the data you provide for this study in a secure manner for three years before destroying it.

**Please contact the researcher on the study team listed below to obtain more information or to ask a question about the study.**

Samuel Aiello
The Beacom College of Computer and Cyber Sciences, 820 N Washington Ave.
Dakota State University, Madison, SD 57042.
DSU email address: sam.aiello@trojans.dsu.edu

**You may also report a concern about a study or ask questions about your rights as a research subject by contacting the Institutional Review Board listed below.**

Dakota State University, 820 N. Washington
Madison, SD 57042
Telephone: (605) 256-5100
Email: irb@dsu.edu

DSU IRB-Approval #20240612

If you would like to retain contact information, please make a copy of the informed consent before moving on to the survey

○ I consent, begin the study

○ I do not consent, I do not wish to participate

>> Next Question

*Figure 9. QualtricsXM Survey Screenshot*

*Sampling Bias*

Using QualtricsXM, the respondent`s sampling could show a sampling bias if the sample itself did not appropriate the population. Such a situation leads to inaccuracies and wrong conclusions. Sampling bias could be reasonable in certain cases. For example, in other studies, the investigators purposely targeted one sub-region for their research in order to gather in-depth information about that particular segment. The researcher may have been selective in issuing out questionnaires to particular sub-groups within the surveyed population even where random sampling was adopted. Considerations on the bias that may arise in survey designs have to be made and reported along with any implications that such bias may have on study outcomes.

The use of QualtricsXM to put forward online surveys and questionnaires formed an efficient way of gathering data, analyzing patterns and acquiring information concerning different areas of research. Sampling bias, data exploration methods, questionnaire design and structure, response rates and data privacy and other relevant strategies/ and or methods were addressed to improve on the quality and credibility of the research outcome. Considering the overall planning that was conducted, the design that was carefully chosen and ethical principles that were adhered to, this is what online surveys are used for; to expand knowledge, aid in making decisions and facilitate change in various fields.

*Mitigating Sample Bias*

The approach to counteract sample bias employed probability sampling techniques on a foundation of random sampling from the target population. The strategy for distribution was all-inclusive to begin with, using dual platforms for wide coverage across its demographic boundaries. First, the platform Microsoft Forms for the piloting phase and Qualtrics XM for the "go live" phase. This was, however, suffering from limitations since the options for offline response were not included; this might have avoided possible participation by respondents who may have had limited access to the internet. This in particular might have furthered a poor response rate and a lack of demographic representation.

Technically, while implementing this survey, various points of strengths and weaknesses were identified that must be focused on for further improvements: Compatibility testing across different browsers and various devices was done, which created a friendly environment for diversified participation. There was no multilingual support which may have improved response participation.

The follow-up protocol was systematic: for non-respondents, there was email communication, and posts on social media advising the survey progress and remaining timeframe window for participation. Transparency about the purpose of the survey and information related to the estimated time it would take to complete provided a base of trust that supported response rates in spite of the lack of incentives.

*Strategies for Understanding the Data*

In this context, descriptive statistics is a way of exploring data collected through an online survey using QualtricsXM to analyze it. Exploring data in this way gives a complete understanding of the research results and enables meaningful decisions.

Analyze the collected data using IBM SPSS®. The SPSS software package offers a wide range of statistical analysis methods. Basic indicators include methods such as calculating new variables, generating multi-item scaling indicators such as frequency distributions and descriptive statistics, and visual examination of data distributions using methods such as quantitative QQ plots.

After performing preliminary procedures, groups such as cross-tabulation of single variables, t-tests to compare means between two groups or variables, univariate analysis of variance to compare means between multiple groups, and ANCOVA for control variables were conducted. Subsequent results can be extrapolated using more advanced ANOVA methods, including two-way ANOVA, repeated-measures ANOVA, MANOVA for multiple dependent variables, and t-tests with comparison within the ANOVA model.

*Rationale for Study to Enhance Zero Trust*

Computers and the Internet provide a mechanism to research how to improve Zero Trust security measures. Organizations are receptive to solutions to strengthen their security measures to counter the increase in cyberattacks and data breaches. As it is possible to obtain data through online surveys with the help of QualtricsXM, insights on the current ZT practices, the current status, and how better to secure the ZT methodology can all be obtained.

*Response Rate*

The response rate is among the key indicators of the level of participation and engagement in online surveys. In this respect, reminders were issued to respondents, progress updates posted on social media, and multi-device accessibility allowed to enable participants to complete the survey. A high response rate acts as a way to achieve representatively reliable data for analysis.

*Confidentiality of Data*

Data privacy is a key concern when conducting online research. Steps were taken to ensure that participant information was stored securely, encrypted and accessed in a secure manner and in accordance with data protection legislation. The confidentiality and integrity of the data collected was ensured by applying confidentiality measures such as anonymous responses and obtaining informed consent. An Institutional Review Board on Human Subjects Evaluation (IRB) was submitted and approved.

*Nonresponse Bias*

Nonresponse bias refers to a problem where those who have not responded differ significantly from those who have, and the findings may be biased. Of course, this can be an even greater concern for those surveys that were started but then abandoned. Both Microsoft Forms and Qualtrics XM keep statistics on abandonment. The pilot was helpful in determining that the survey, originally 60 questions, was to long and hence encouraged abandonment.

The abandonment of the survey may be due to various reasons, such as lack of time, disinterest, or unease with the questions. This might differ from those who responded to the survey, hence generating a bias in the final results. The consequence of this could be over- or under-representation of certain groups, hence distorting the overall validity and reliability of the findings. If the results are to accurately represent the target population, there is an important need to address nonresponse bias. In all cases, nonresponses were eliminated from the dataset.

*Question Order Effects*

Question order effects also describe how the sequence of questions influences the responses of participants. In the course of web based survey development while utilizing QualtricsXM, the sequence of questions presentation was particularly addressed so as to eliminate bias focus and enhance data accuracy. The elements of question order effects were addressed by randomization of the question order, usage of skip logic, and pre-testing the questionnaire on a pilot group before the study.

*Sample Selection*

For this research, convenience sampling was employed. It is also referred to as availability sampling or accidental sampling. The terms all describe a research method where there is data collection from those participants who are readily available and who are willing to take part, rather than any form of random sampling.

This method is particularly useful in cases when the target population has some peculiar characteristic or expertise, as is the case with cybersecurity professionals possessing ZTA experience, or if the probabilistic sampling is highly impractical or too expensive. The term "accidental" comes from the fact that participation occurs through chance availability and willingness to participate at the time of data collection rather than through systematic random selection. Convenience sampling is, however, the more generally used and accepted term in research methodology.

This can be inferred from several key aspects of the research design and implementation. The online survey, designed in QualtricsXM, was open to cybersecurity professionals of various ranks from around the world with practical experience in Zero Trust Architecture implementations. Self-selection was inherent in the nature of participation, given the opt-out question about experience with ZTA at the very outset, whereby those with no relevant experience were allowed to opt out of the survey immediately.

Convenience sampling also finds evidence in the demographic distribution of participants, as responses are accumulated from a total of 138 participants, the majority being from North America at 78.3%. The geographical concentration suggests that the sample was based mostly on convenience rather than systematic random selection. Additionally, the online targeting of Cybersecurity professionals as a sampling approach also depicted convenience in sampling the participants to meet a convenient quota which may be available and willing to contribute to the research.

This approach will, therefore, enable the survey to collect data from professionals with relevant experience in the implementation of Zero Trust Architecture, although it limits the generalization of the findings in the broader population of the organizations that are implementing ZTA.

*Transparency*

Participants of online studies need to be informed of the information contained in the studies because failure to do so will erode the level of trust that the participants had towards the research. Participants were informed about the study's aim, the procedures for collecting the information, the assurance of confidentiality of the information, and the risks and the benefits of being in the study. Because the rationale behind the survey implementation is elucidated, proper functioning in ethics is cultivated alongside the willingness of the participants.

*Leading Questions*

Efforts have been made to eliminate any form of leading questions which are designed to bias a respondent in one direction and compromise the integrity of the survey when designing surveys in QualtricsXM. Rather, neutral and objective language was employed in formulating the questions in order to provoke as accurate and bias free responses from the participants as possible.

## 3.8    Collect Data

Researchers can simplify the processes of gathering and analyzing data by using closed-ended questions with preordained response options. Though structured questionnaires have their drawbacks, when they are designed and used with care, they can yield large amounts of valuable information across a variety of research subjects. One bastion of the implementation evaluation of ZTA can be found in two indicators, which track across time the two key performance indicators (KPIs) and three maturity model metrics that frame the ZTA implementation.

Since this implementation involves both complicated technical structures and the processes of human actors, it is of utmost importance to gather the insights of experts working in these domains. Primarily an online survey was used to collect data from an extremely global set of ZTA practitioners. These cybersecurity professionals were best able to comment about the complex, layered implementation of ZTA. The survey was offered in English on the QualtricsXM platform. Conducting the survey in English allowed for reaching the largest possible number of ZTA practitioners and also ensured a standardized set of responses for analysis.

Including participants from a range of industry sectors and organizational sizes allowed the research to draw on a diverse set of experiences from which to craft the benchmark. It yielded a composite score across the key metrics of policy coverage, microsegmentation, and automation- three areas in which Cybersecurity Excellence is critically judged.

This comprehensive survey approach yields actual data from ZTA implementations in production settings instead of data from test cases or proof-of-concept projects. Those real-world scenarios give us the data to reliably span our ZTA progression measures from the earliest stages of implementation through the fully adaptive condition that characterizes the maturity endpoint.

Trends extrapolated from this worldwide data source would help in the construction of models to track the maturity and effectiveness of the Zero Trust Architecture. They would also serve as the groundwork for a series of benchmarks to help operationalize and optimize an organization's cybersecurity strategy.

## 3.9     Understanding Numerical Variable Relationships

In the field of statistics, understanding the relationships between variables is crucial for making informed decisions and predictions. Two common methods used to analyze these relationships are correlation analysis and regression analysis. While both techniques involve numerical variables, they serve different purposes and provide distinct insights into the data. This dissertation focuses on the importance of correlation analysis in measuring the strength of the association between numerical variables, as opposed to the predictive nature of the simple linear regression model.

Correlation analysis is a statistical technique that measures the strength and direction of a relationship between two numerical variables. It provides a numerical value, known as the correlation coefficient, which indicates how closely the variables are related. A perfect negative relationship is denoted by a correlation coefficient of -1, no relationship is indicated by a correlation coefficient of zero, and a perfect positive relationship is denoted by a correlation coefficient of 1. By calculating the correlation coefficient, researchers can determine the degree to which changes in one variable are associated with changes in another.

According to renowned statistician Sir Francis Galton, "Correlation measures the strength and direction of a linear relationship between two variables, allowing researchers to quantify the extent to which changes in one variable are associated with changes in another" (Galton, 1888). This quote highlights the fundamental purpose of correlation analysis in understanding the relationships between numerical variables.

Unlike regression analysis, which aims to predict the value of a dependent variable based on one or more independent variables, correlation analysis does not involve prediction. Instead, its primary objective is to assess the strength of the association or covariation that exists between two variables. This distinction is important because it emphasizes the exploratory nature of correlation analysis, focusing on understanding the relationship itself rather than using it for predictive purposes.

In his book "Statistical Methods for Research Workers," Sir Ronald Fisher emphasized the significance of correlation analysis in research, stating that "Correlation analysis allows researchers to uncover patterns and associations between variables, providing valuable insights into the underlying relationships within the data" (Fisher, 1925). Fisher's words underscore the importance of correlation analysis as a tool for exploring and understanding the complex interplay between numerical variables.

One of the key advantages of correlation analysis is its ability to detect both linear and nonlinear relationships between variables. While the Pearson correlation coefficient is commonly used to measure linear relationships, other correlation measures such as the Spearman rank correlation coefficient can capture monotonic relationships that may not be strictly linear. This flexibility allows researchers to assess a wide range of relationships and uncover hidden patterns in the data.

In their seminal work on correlation analysis, (Yule, 1897) highlighted the versatility of correlation measures in capturing diverse relationships, stating that "Correlation analysis provides a comprehensive framework for examining the associations between variables, accommodating both linear and nonlinear patterns with precision and accuracy." This acknowledgment of correlation analysis's adaptability underscores its importance in exploring complex relationships in data.

Furthermore, correlation analysis plays a crucial role in hypothesis testing and model building. By examining the strength of the relationship between variables, researchers can determine whether the observed associations are statistically significant or occur by chance. This statistical rigor is essential for drawing valid conclusions from data and establishing robust models that accurately represent the underlying relationships.

In their influential paper on hypothesis testing in correlation analysis, (Working & Hotelling,1929) emphasized the importance of statistical significance in interpreting correlation results, stating that "Significance testing allows researchers to assess the reliability of the observed relationships between variables, ensuring that conclusions are based on solid evidence and not random fluctuations." This emphasis on statistical rigor highlights the critical role of correlation analysis in hypothesis testing and model validation.

The correlation coefficient is our standard tool for statistically measuring how well two variables associate with one another. From a Bayesian viewpoint, the correlation coefficient takes on a more nuanced meaning. It is not merely a measure of association between two variables. Working with the coefficient is more about the story you are telling with your model and how much you get to believe in that story. The more you get to believe in it, the higher the correlation coefficient tends to be. Indeed, deceitful relationships can be correlated. For this reason, the correlation coefficient can also be viewed as measuring the "goodness of fit" of a model that tells a story about the relationship between two variables.

## 3.10    Analyze Data

Surveys of cybersecurity professionals yield sample data rich in meaning. These data were collected to gain better insight into the impact of ZTA on performance. Hypothesis testing methods, including t-tests and ANOVA, were used to determine whether the observed differences in performance between environments secured by ZTA and those using traditional models were statistically significant. For the most part, they were. Regression analysis was also used to relate specific ZTA mechanisms to performance impacts, primarily cost-related and risk/recovery-based, that were hypothesized. Finally, the accuracy and reliability of the predictive regression models were examined. In each case, ZTA environments were found to be superior.

Going beyond basic statistical methods, machine-learning algorithms can identify patterns among data elements with non-intuitive relationships that impact both security and economics. For example, using clustering analysis, unexpected groupings of technical controls might be found among the many measures employed by organizations with high ZTA maturity scores. Additionally, neural networks could better predict cost and risk by processing the complex combinations of device posture, network traffic, access patterns, and policy configurations that ZTA relies upon. The bottom line is that using these advanced machine-learning tools provides much more confidence in the predictions made about the enhanced security ZTAs can bring to organizations.

The analysis carried out using several different methods enhances the credibility of the story being told about the results that organizations of various maturity levels are experiencing as they work their way toward the implementation of a Zero Trust Architecture. The analysis provided by the author within the report is largely focused on IT operational efficiency that organizations are seeing as a result of moving to a ZTA, which is something that is not often heard much about. Most of the time, when stories about ZTA implementations are shared, they are focused on how secure the end result is. However, this report provides some good insights into how ZTA can be a secure and less exhaustive process alternative to the traditional approach.

The organization's recommendations would be bolstered by these quantitative validations. They would serve to reinforce the recommendations to upgrade legacy authentication protocols, consolidate networks into software-defined microsegments, and invest in centralized policy and posture management platforms.

### 3.11    Interpret Results

One way to discern what the report is saying is to clearly define the architecture being discussed in a model for the uninitiated. After that, seeing how the recommendations aim to achieve a set of goals. And what are those goals? They boil down largely to a robust implementation of Zero Trust principles.

The leadership stakeholders can assert that the model's forecasts of (1) security improvements, (2) operational efficiency benefits, and (3) a path to successful Zero Trust implementation are indeed valid and represent what can be achieved in their organizations.

## 3.12    Refine Goals and Metrics

To ensure that the statistical analysis and conclusions fully reflect the implementation of Zero Trust Architecture in various enterprises worldwide, the study includes validation from representative groups. Cybersecurity experts from multiple industry organizations will review maturity levels, risk mitigation strategies and cost reduction strategies. They provide dynamic solutions based on challenges and goals.

An iterative enhancement involves modifying the existing measurements' objectives to include additional factors that are relevant to influencing the level of efficiency of the deployments and operations of Zero Trust Architecture. This broadens the analytical framework of the costs and benefits to include productivity impacts, technology stability risks and marginal costs of complexity rather than just including reductions of cyber risk. Organizations may use the findings to make trade-offs between security concerns and productivity disruption while mapping out changeover periods from legacy systems to Zero Trust networks, based on the feedback given by the respondents. Introducing an efficiency ratio and its metrics to the model adds practical applicability to real situations.

The maturation of the research also incorporates the perspectives, which were emphasized by the quantitative elements in the engagement. The expansion of dimensional analysis likely leads to diversification in adoption in various sectors.

## 3.13    Summary

The research methodology presents a nine-step detailed quantitative research plan focused on the implementation of ZTA in organizations whether it is being properly implemented and how effective it has been. The design of the study is based on the survey of willing cybersecurity professionals worldwide, conducted through the web-based Qualtrics$^{XM}$ platform. Questions asked are deemed appropriate for the measurement of the value and maturity of a Zero Trust implementation.

This chapter outlines the coverage percentage basis of its key performance indicators on key aspects, such as identity management, network microsegmentation, multi-factor authentication, and analytics, aside from designing proper measurement goals and metrics that tell whether ZTA is effective or not in implementation. It also examines practical outcomes, including how many breaches are impeded, reduced successful attacks, and effectiveness in lateral movement across networks. The research design involves the administration of structured questionnaires to a large sample to elicit standardized and quantifiable data.

The analytical phase utilizes various statistical methods, such as hypothesis formulation and testing, regression analysis, and machine learning algorithms, which may indicate patterns and correlations in the data. In general, the study tries to prove that the efficiency of ZTA implementations is valid regarding a range of security improvements, benefits related to operational efficiency, or successful implementation pathways.

The methodology concludes with the refinement phase, which involves iteration for measurement improvements and includes other factors that affect the efficiency of deploying ZTA. The methodology can then provide an organization with direction for making informed decisions about its security investments and strategies.

## CHAPTER 4: RESEARCH DESIGN AND IMPLEMENTATION

This chapter provides details of the research design used in the study.

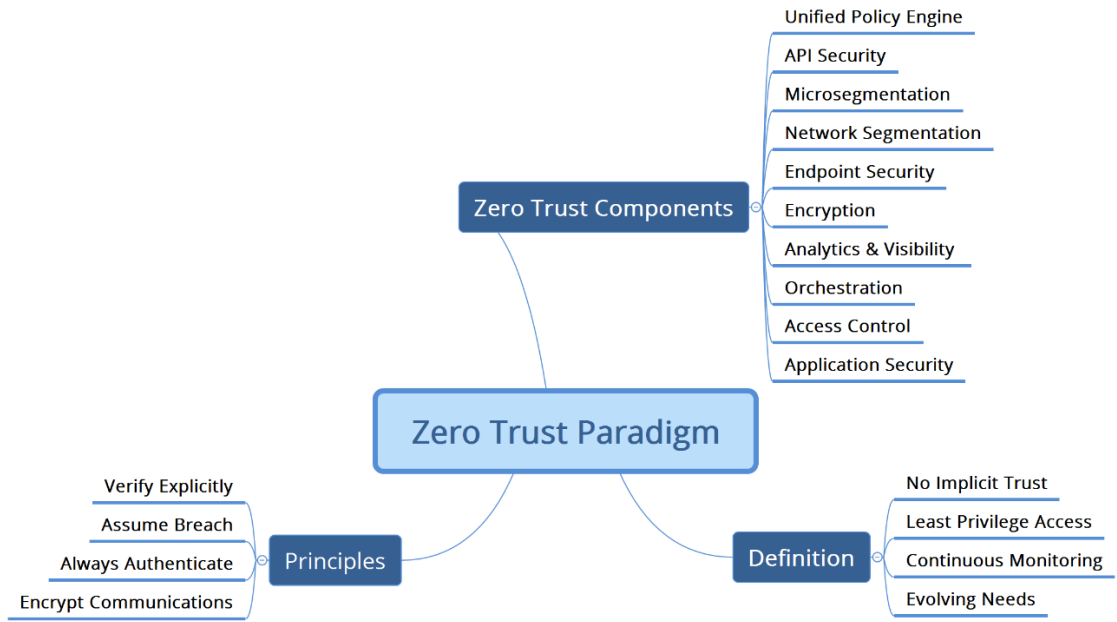

*Figure 10. Zero Trust Paradigm*

The zero-trust paradigm and its subtopics, components, principles, and definitions are described in Figure 10. Based on the findings from NIST 800-53, this figure shows the relationships of the Zero Trust Paradigm. This section details the components, principles, and definitions of the Zero Trust paradigm, including the consensus engine, API security, subsegments, network segments, endpoint security, encryption, analytics and transparency, orchestration, access control, application security, key emergence. Concepts include authentication, presumption of breach, persistent authentication, encrypted messages, absolute trust, restricted access, continuous monitoring, and variable requirements.

### 4.1    Validation

*Questionnaire Validation*

Questionnaire validation is a process of assessing the reliability and validity of survey instruments. Online surveys created using QualtricsXM could be validated by conducting pilot tests, analyzing internal consistency, and comparing results with established measures. By validating the questionnaire, the accuracy of the survey in capturing the intended constructs and producing reliable data for analysis could be ensured.

*Questionnaire Design*

The design of the questionnaire plays a central role in the success of the online survey. When using QualtricsXM, the wording of the questions, the order of the questions, the response options provided, and the overall layout of the survey were carefully considered. Clear and concise questions, logical flow, and user-friendly design help to maximize response rates and ensure the quality of the data collected.

*Sample Size Using G\*Power to Achieve Statistically Significant Results*

G\*Power provides a detailed and thorough method to figure out the required sample size for a linear multiple regression model. The basic idea behind this software is that it serves the user by calculating how the unspecified model could operate under given fixed conditions. In other words, G\*Power can help you see in a "what if" framework and enable you to understand how strong effects have to be in order for your model to function reliably and validly. The core of this framework centers on the effect size you set. In our case, the effect size is set at $f^2 = 0.15$.

Figure 11 depicts the setup parameters used to assess the required number of respondents

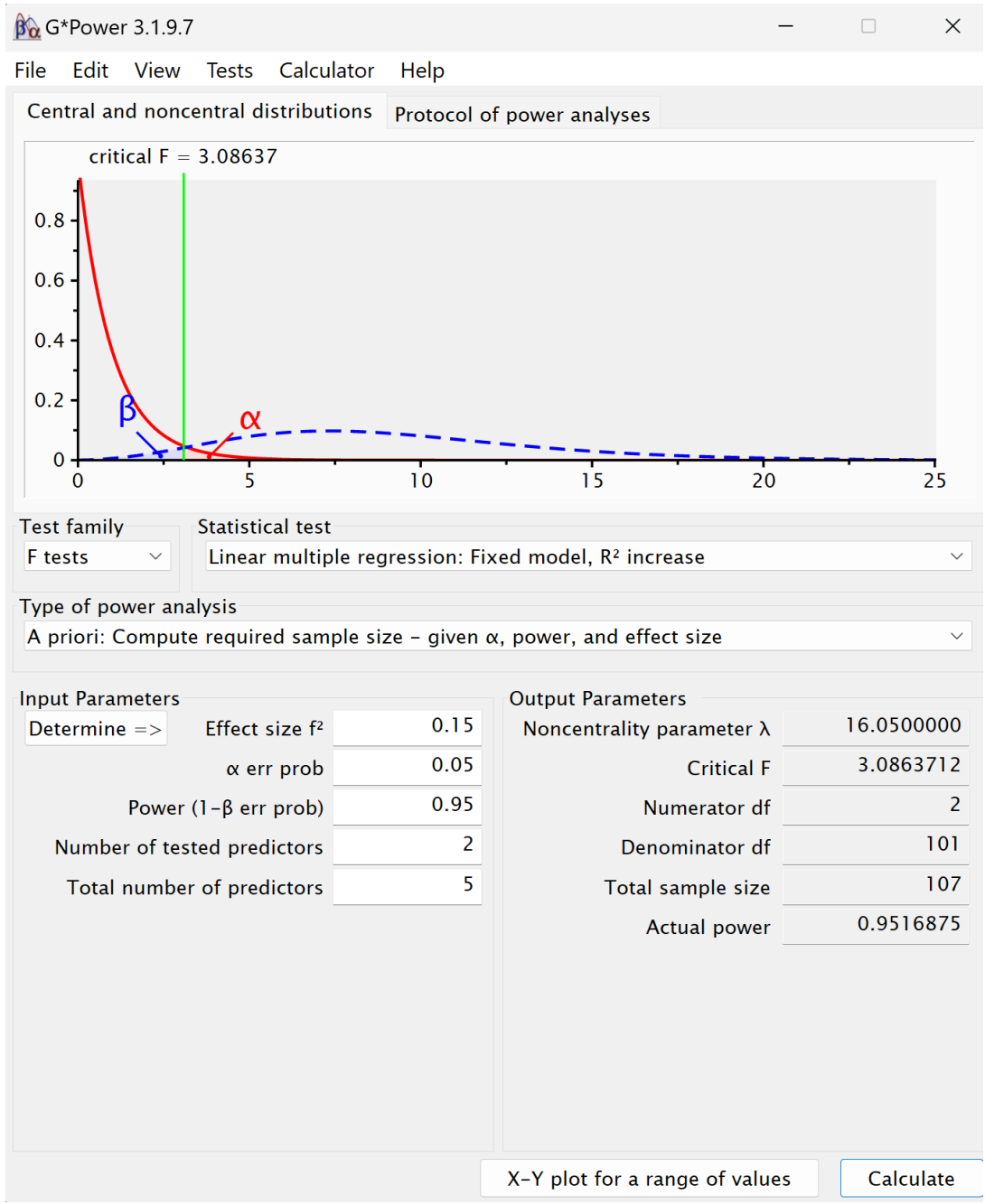

*Figure 11. G*Power Linear Multiple Regression Test*

　　　　An f² value of 0.15 is considered a medium effect size according to Cohen's conventions. This means the independent variables (IVs) explain a fair amount of the variance in the dependent variable (DV) in the model being tested. An f² value of 0.15 for the IVs' relative covariance shows that the relationship between the IVs and DV is significant. This is not merely an observed relationship that could result from a Type I error, where the null hypothesis (Ho) is incorrectly rejected. The model is valuable as it provides reasonably good predictions of the DV based on the IVs. The alpha level is set at 0.05, a commonly used threshold.

This level of importance is almost universally accepted because it knows no border in good scholarship. It is not too lenient, and it is not too stringent. The power settings are the entire purpose for utilizing the tool. Power is the probability of correctly rejecting a false null hypothesis. The power desired is 0.95. This means our chance of correctly detecting a true effect when we observe one is 95%. It is recommended that our reliability estimate not be below 0.90; otherwise, there could be concern about overfitting the problem.

The decision to use such a high-power level stem from a desire to ensure that the study's findings are robust and reliable, to allow one to trust the results. In this particular analysis, two specific predictors are being looked at within a model that contains a total of 5 predictors. What this means is that the model accounts for potentially influential other variables, but our main interest is in understanding the unique contribution of the two predictors of primary interest.

The noncentrality parameter ($\lambda = 16.05$) sheds light on the distribution of the test statistic under alternative hypothesis. This basically takes into consideration both the effect size and the sample size to give an index that conveys the degree of the overall effect in terms of the sample. Accordingly, higher values of the non-centrality parameter usually lead to a higher probability of the study obtaining a significant result, obtaining a "true" effect with overwhelming bet ($1-\beta = 0.95$). The threshold F value (3.0863712) is the critical value of the test statistic whereby the study understudied was found to be significant due to the surpassing of the threshold.

In case the regression analysis is performed for the first time, and to further analyze the primary coefficient of determination F, if the calculated F value is greater than this critical point, the conclusion can be drawn that the set of predictors is able to explain the variation of the criterion well. Other Predictors that are not included in the regression model, could not find their way into the regression; hence, they must not be very applicable to the outcome variable in question. We may assume with some degree of confidence, that the two pairs of words, which we have stated and tested in the model do not result in proportional differences in the dependent variable. And it is safe to assume that the regression prediction of variances does not bring about any significant difference in the outcomes due to the predictive model constructed.

In order to achieve the desired power level under the effect and alpha levels given, the total sample size should comprise of a minimum of 107 participants. This sample size results in a study which is clearly overpowered enough, not to mention reliable and with generalizations made being more applicable. While the authors note in the analysis report that the power of the study is 0.9516875, i.e. the study is robust enough to find out if any "true" effect occurs when required detecting such, most likely moderate alpha levels were intended by the authors of the report.

Finally, the GPower software was used to establish how large a sample would be needed, with a focus on attainment of sufficient statistical power. Given the medium effect size ( $f^2 = 0.15$), this value was used in all subsequent calculations and is termed as effectiveness expected. In searching for ways to make the tool most useful others often also use the effect size ( $f^2 = 0.15$ ) in GPower because it is what is mostly considered average. Therefore, insert 'significance level of 5%' useful at this stage and 'power of 95%'. With these values, G*Power indicated that the study should target at least 107 participants. This section would typically include such calculations in the methods section of the dissertation to convince the readers and the reviewers that the design of our study was not underpowered.

## 4.2     Statistical Analysis Discussion

*Regression Model to Evaluate the Quantified Effect of Various Factors on Breaches*

In the realm of data analysis and statistical modeling, regression analysis plays a significant role in understanding the relationships between variables and predicting outcomes.

One particular type of regression model that is commonly used in research is logistic regression. Logistic regression is especially useful when the outcome of interest is categorical, such as whether a breach occurred or not. By quantifying the impact of different factors on the likelihood of a breach, organizations can better understand and mitigate risks. In this essay, we explore how a logistic regression model can be used to evaluate and rank the factors that have the most significant impact on breaches.

To begin with, it is essential to understand the basic principles of logistic regression. Unlike linear regression, which is used when the outcome variable is continuous, logistic regression is employed when the outcome is binary. In the context of breaches, the outcome variable would be whether a breach occurred (1) or did not occur (0). The goal of logistic regression is to model the probability of a breach as a function of one or more independent variables, also known as predictors.

When setting up a logistic regression model to evaluate the impact of different factors on breaches, it is essential to carefully select the independent variables that are likely to influence the likelihood of a breach. These independent variables can include a wide range of factors, such as the organization's security measures, employee training programs, technological infrastructure, and external threat landscape. Additionally, net promoter scores, which measure customer satisfaction and loyalty, can also be included as independent variables in the model. By incorporating these diverse factors, the logistic regression model can provide a comprehensive analysis of the drivers of breaches.

Once the independent variables have been identified and included in the logistic regression model, the next step is to analyze the results and quantify the effect of each factor on the probability of a breach. In logistic regression, the coefficients of the independent variables indicate how much the log-odds of the outcome variable change with a one-unit increase in the predictor variable, holding all other variables constant. By exponentiating these coefficients, researchers can calculate the odds ratios, which represent the factor by which the odds of a breach increase or decrease for each unit change in the predictor variable.

After estimating the odds ratios for each independent variable, researchers can rank the factors based on their impact on breaches. Factors with higher odds ratios indicate a more substantial influence on the likelihood of a breach, while factors with odds ratios close to 1 have a minimal effect. By ranking the factors in descending order of their odds ratios, organizations can prioritize their efforts to address the most significant risk factors and strengthen their security posture effectively.

In addition to quantifying the impact of different factors on breaches, logistic regression models can also provide valuable insights into the relationships between variables and the overall predictive power of the model. By examining the significance levels of the coefficients, researchers can determine which factors are statistically significant in predicting breaches and which may not have a meaningful impact. Furthermore, measures such as the concordance statistic (C-statistic) can assess the predictive accuracy of the model and its ability to distinguish between breaches and non-breaches.

Overall, logistic regression is a powerful tool for evaluating and quantifying the effect of various factors on breaches. By setting up a logistic regression model with appropriate independent variables, researchers can identify the key drivers of breaches, rank them based on their impact, and prioritize risk mitigation strategies accordingly. Through careful analysis of the odds ratios, significance levels, and predictive performance of the model, organizations can enhance their cybersecurity defenses and protect against potential threats effectively.

### *Relationship Between Regression and Correlation Analysis*

Regression and correlation analysis are two essential statistical techniques used to examine the relationships between two or more numerical variables. These methods provide valuable insights into how changes in one variable may affect another, allowing researchers to make predictions and draw conclusions based on data. While both regression and correlation analysis involve numerical variables, they serve slightly different purposes and are applied in various scenarios to uncover patterns and associations within the data.

Regression analysis is a statistical technique that is primarily used for prediction. The main objective of regression analysis is to develop a mathematical model that can be used to predict the values of a dependent or response variable based on the values of one or more independent or explanatory variables. Regression analysis includes different techniques such as simple linear regression, multiple linear regression, and logistic regression, each tailored to specific data types and research inquiries. This predictive capability makes regression analysis a powerful tool in various fields, including economics, finance, psychology, and epidemiology.

One of the key concepts in regression analysis is the regression equation, which represents the relationship between the dependent and independent variables.

**The equation is, therefore, written in the form $Y = a + bX$.**

where Y is the dependent variable, X is the independent variable, a is the intercept (the value of Y when X is zero), and b is the slope (the change in Y for a one-unit change in X). By estimating the values of a and b from the data, researchers can create a model that best fits the observed relationship between the variables.

Regression analysis includes different techniques such as simple linear regression, multiple linear regression, and logistic regression, each tailored to specific data types and research objectives. Simple linear regression is used when there is a single independent variable, while multiple linear regression is employed when there are two or more independent variables. Logistic regression, on the other hand, is used when the dependent variable is binary or categorical in nature.

In contrast to regression analysis, correlation analysis focuses on measuring the strength and direction of the relationship between two or more variables. Correlation coefficients, such as Pearson's r or Spearman's rho, are used to quantify the degree to which variables are related. Strong positive relationships are indicated by correlation coefficients near +1, whereas strong negative relationships are suggested by coefficients near -1. Strong positive relationships are indicated by correlation coefficients near +1, whereas strong negative relationships are suggested by coefficients near -1. A coefficient near zero indicates little to no relationship between the variables.

Correlation analysis is valuable for identifying associations between variables and determining the extent to which changes in one variable are associated with changes in another. However, it is important to note that correlation does not imply causation. Just because two variables are correlated does not mean that changes in one variable cause changes in the other; there may be other factors at play that influence the relationship.

In practice, both regression and correlation analysis are often used together to gain a comprehensive understanding of the relationships between variables. Regression analysis is a statistical method that allows us to make predictions about the value of a dependent variable using one or more independent variables. Correlation analysis is a statistical technique that helps us evaluate the intensity and direction of the relationship between variables. By combining these techniques, researchers can uncover patterns, make predictions, and evaluate hypotheses based on empirical data.

Overall, regression and correlation analysis are powerful tools in the field of statistics, allowing researchers to explore the relationships between numerical variables and make informed decisions based on data. Whether predicting future outcomes, identifying trends, or testing hypotheses, these techniques play a crucial role in advancing knowledge and understanding in various disciplines.

### *Binary Logistic Regression and Mediation Analysis in Predicting Breaches*

In statistical analysis, binary logistic regression stands out as a powerful tool used to predict the probability of a binary outcome. This outcome typically involves a yes or no, true or false, or 0 or 1 scenario. For instance, binary logistic regression can be applied to predict whether a customer will churn, whether a patient has a particular disease, or whether a loan will be repaid. This method is particularly useful in situations where the dependent variable is dichotomous, and the goal is to understand the relationship between the independent variables and the probability of a specific outcome occurring.

Binary logistic regression is a valuable technique in the field of data analysis, as it allows researchers to model the relationship between a binary outcome and one or more independent variables. By estimating the probability of the outcome occurring based on the values of the independent variables, binary logistic regression provides insights into the factors that influence the likelihood of a particular event taking place. This predictive capability is especially beneficial in various fields, including business, healthcare, and finance, where understanding and forecasting binary outcomes are necessary for decision-making processes.

One area where binary logistic regression can be particularly useful is in predicting breaches in architectural components. In the context of cybersecurity and information technology, the identification and prevention of breaches in architectural components are of paramount importance to safeguard sensitive data and ensure the integrity of systems. By applying binary logistic regression and mediation analysis techniques, researchers and practitioners can gain valuable insights into the factors that contribute to breaches in architectural components and prioritize their mitigation efforts effectively.

Mediation analysis aims to extend the concepts of mediation for investigating the boundaries of mediation and complement series of binary logistic regression for predicting influencing factors in occurrence of breaches in architectural components. This mediation analysis seeks to illuminate a greater understanding of the mechanisms at work in the dependent and the independent variable, for instance, the system vulnerability and occurrence of breaches. Such clarity on how exactly conditions leading to such breaches come about would enable policies to be formulated towards such specific areas thereby improving the security of the architectural components.

When conducting binary logistic regression and mediation analysis to predict breaches in architectural components, identifying statistically significant variables is key. These variables play a key role in determining the probability of breaches occurring and can aid in ranking the architectural components based on their vulnerability levels.

By assessing the impact of each variable on the likelihood of breaches and determining which variables have the greatest or least effect on the potential for breaches, organizations can prioritize their resources and focus on mitigating the most significant risk factors.

In the context of binary logistic regression, statistically significant variables are those that have a notable impact on the probability of the binary outcome. These variables are characterized by their ability to influence the likelihood of breaches in architectural components and can be either positively or negatively associated with the occurrence of breaches. A positive relationship indicates that an increase in the value of the variable is associated with a higher probability of breaches, while a negative relationship suggests the opposite – a decrease in the variable's value is linked to a higher likelihood of breaches.

By analyzing the coefficients and odds ratios of the statistically significant variables identified through binary logistic regression, researchers can quantify the strength and direction of the relationships between the independent variables and the probability of breaches in architectural components. This quantitative assessment enables organizations to rank the architectural components based on the impact of each variable and prioritize their security measures accordingly. Variables with the greatest effect on the potential for breaches can be targeted for immediate attention, while those with lesser influence may receive less emphasis in the risk mitigation strategy.

In the process of ranking the architectural components based on statistically significant variables, organizations can gain valuable insights into the vulnerabilities present in their systems and infrastructure. By understanding which variables have the most significant impact on the likelihood of breaches, decision-makers can allocate resources effectively and implement targeted security measures to strengthen the overall resilience of their architectural components. This proactive approach to risk management can help organizations mitigate potential threats and enhance their cybersecurity posture in an increasingly complex and dynamic threat landscape.

Binary logistic regression and mediation analysis are powerful tools that can be leveraged to predict breaches in architectural components and enhance cybersecurity practices. By identifying statistically significant variables and assessing their impact on the probability of breaches, organizations can rank the architectural components based on their vulnerability levels and prioritize their security efforts accordingly. This data-driven approach to risk management enables organizations to proactively address potential threats and safeguard their systems against malicious attacks. By integrating binary logistic regression, mediation analysis, and statistical significance testing into their risk assessment processes, organizations can strengthen their defenses and mitigate the risks associated with breaches in architectural components effectively.

### *Multicollinearity in Multinomial Regression: Understanding the Impact on Statistical Inferences*

Multicollinearity is a statistical concept that arises when two or more independent variables in a regression model are highly correlated with each other. In the context of multinomial regression, multicollinearity can have a significant impact on the reliability of statistical inferences drawn from the model. When independent variables are correlated, it becomes difficult to disentangle their individual effects on the outcome variable, leading to less precise estimates of the coefficients and potentially misleading results.

For specific multicollinearity, two variables are very highly related; technically speaking, their correlation coefficient may either be +1.0 or -1.0. Hence, these two types give additional information to the model, and thus working out unique coefficients for each type becomes impossible, hence the model might be inexact with high sensitivity coefficients on small changes in data.

Since the multinomial regression outcome to be modeled falls into more than two groups, multicollinearity between independent variables can definitely have an effect on interpreting the results. On one hand, the high correlation between independent variables may make it difficult to identify the independent contribution of a variable to the probability of each outcome. This will, in turn, be prone to model bias, and it will be difficult to determine the level of importance of the relationship between the independent and the outcome variables.

One of the major problems with multicollinearity in multivariate regression involves the effects it could have on point estimates of odds ratios for independent variables. Normally, in multivariate regressions, odds ratios are considered in quantifying the association between independent variables and the different outcomes of interest. In cases where there is a problem of multicollinearity, odds ratios may be imprecise and unreliable upon which meaningful conclusions from an analysis may arrive.

To illustrate the impact of multicollinearity on multinomial regression, consider a hypothetical study examining the factors associated with the likelihood of data breaches in a large organization.

Data breach incidence is any outcome variable that could be differentiated into a number of categories based on the severity or type of breach. Independent variables of interest in the data will, among many others, include both continuous variables such as the number of security measures being set up and dichotomous variables such as whether there is a dedicated cybersecurity team.

If multicollinearity is present among the independent variables in this study, the odds ratios estimated from the multinomial regression model may be biased and unreliable. For example, if the number of security measures and the presence of a dedicated cybersecurity team are highly correlated, it may be difficult to determine the unique contribution of each variable to the likelihood of data breaches. As a result, the odds ratios associated with these variables may be inflated or deflated, leading to erroneous conclusions about their impact on the likelihood of breaches.

In addition to the impact on odds ratios, multicollinearity in multinomial regression can also affect other aspects of the model, such as the overall fit and predictive power. When independent variables are correlated, the model may have difficulty distinguishing between the effects of different variables, leading to a poorer fit to the data. This can result in a loss of predictive accuracy, making it harder to use the model to make reliable forecasts or decisions.

Multicollinearity is a common issue in multinomial regression that can have serious implications for the reliability of statistical inferences. When independent variables are correlated, it becomes difficult to estimate the unique effects of each variable on the outcome variable, leading to less precise estimates of coefficients and odds ratios. Researchers conducting multinomial regression analyses should be aware of the potential for multicollinearity and take steps to address it, such as including interaction terms or reducing the number of correlated variables in the model. By addressing multicollinearity, researchers can ensure that their multinomial regression models provide accurate and reliable insights into the relationships between independent variables and multiple outcomes of interest.

## 4.3    Logical Elements

Appendix B: Survey Questions contains the thirty (30) questions that comprised the survey entitled "Mapping Exercise_v8 30 Questions." The subsequent discussion follows the specific question sequence in Section 3- Questions 7-17. Table 1 provides the logical cybersecurity groupings.

*Table 1. Logical Cybersecurity Groupings*

| Survey Question | Logical Group | Security Control Family |
|---|---|---|
| | **Logical Cybersecurity Groupings** | |
| 7 | Identity and Access Management (IAM) | Multi-Factor Authentication, Single Sign-On (SSO), Adaptive Risk-Based Authentication, Biometric Authentication, Identity Federation and Directory Services |

Running head: Aiello Dissertation: Prescriptive Zero Trust		Page **82** of **232**| 8 | Access Control and Endpoint Security | Endpoint Detection and Response (EDR), Role-Based Access Control (RBA), Attribute-Based Access Control (ABA), Privileged Access Management (PAM), User and Account Lifecycle Management |
|---|---|---|
| 9 | Endpoint Security and Management | Unified Endpoint Management (UEM), Mobile Device Management (MDM), Device Posture and Compliance Checks, Secure Boot and Hardware-Based Integrity, Micro-Agent or Agentless Endpoint Security |
| 10 | Network Security and Access Control | Software-defined perimeter (SDP), Zero Trust Network Access (ZTNA), Application Control and Whitelisting, Device Isolation and Quarantine, Network Access Control (NAC) |
| 11 | Network Security and Segmentation | Microsegmentation and Network Isolation, Virtual LANs and Microsegmentation, Network Access Control (NAC), Secure Web Gateways (SWG), Cloud Access Security Brokers (CASB) |
| 12 | Data Protection and Network Security | Data Classification and Labeling, API Gateways and Web Application Firewalls, Network Traffic Analysis and Anomaly Detection, Secure remote access (VPN, VDI, RDP), Data loss prevention (DLP) |
| 13 | Data Protection and Information Security | Encryption of Data at Rest and in Transit, Digital Rights Management (DRM), Data Access Control and Granular Policies, Rights Management Services, Secure File Sharing and Collaboration |
| 14 | Security Monitoring and Incident Response | Data Loss Prevention for Cloud Storage (DLPC), Security Information and Event Management (SIEM), User and entity behavior analytics (UEBA), Security Orchestration, Automation, and Response (SOAR), Data Activity Monitoring and Analytics |
| 15 | Security Monitoring and Threat Detection | Threat Intelligence Platforms, Security Analytics and Machine Learning, Centralized Logging and Auditing, Automated Policy Enforcement and Remediation, Continuous Monitoring and Anomaly Detection |
| 16 | Governance, Risk, and Compliance (GRC) | Policy and Risk Management Frameworks, Compliance Monitoring and Reporting, Integrated Dashboards and Reporting, Vendor and Third-Party Risk Assessments, Automated Compliance Checks and Controls |
| 17 | Operational Security and | Security Awareness and Training Programs, Change Management and Configuration Control, Incident Response and |

Page **82** of **232**

| | Incident Management | Disaster Recovery Planning, Reporting and Executive Dashboards, Device Endpoint Patching and Updates |
|---|---|---|

*Identity and Access Management Controls- Survey Question #7*

The grouping of identity and access management controls follows an underlying logic, with the ordering reflecting a progression from simpler and more widely adopted controls to more advanced and risk-based controls. This progression allows organizations to layer controls based on feasibility and risk tolerance, implementing simpler controls broadly first before layering more sophisticated capabilities over time.

Fundamental Authentication Control: The most basic and effective control that security professionals apply in any process is multi-factor authentication. In addition to real passwords, MFA is a fundamental authentication control that provides further protection for users by requiring a password and an added factor controlled by a mobile device or computer. MFA is a very practical base control that provides authentication due to the existent password and also requires something else such as a key code that is sent to the user's device.

Risk-Based Authentication: Building upon MFA, the next control is an adaptive system referred to as risk-based authentication. This control permits modification of authentication needs so that they are in tune with the situation on the ground and also known as the risk level. People's authentication is sometimes based on the extent of the additional information provided such as where the user is logged in from, the device he or she is using, or even how the user operates on the system.

Centralized Identity Management: The third control, Identity Federation and Directory Services, pushes the envelope further by making identity management as well as cross-domain and cross-system integration feasible. This control reduces the complexity of managing users and their access privileges by providing user identity and access management capabilities in a single interface rather than in multiple applications and services.

Improved Access Convenience: One of the controls brings onto the discussion Single Sign-On (SSO), which is the ability of the user to sign in once and gain access to numerous applications and/or services. This control improves the usability and convenience for the users since there are no longer burdens of remembering and entering password for each application or service.

Advanced Authentication Methods: Finally, the control discussed is cited as Biometric Authentication, which is a higher form of Authentication that uses biological characteristics such as a person's fingerprints, facial characteristics, or even iris. Biometric Authentication is associated with improved security and helps principally when there is a need for physical presence or transactions of high worth.

Alignment with Zero Trust Principles: As a whole, the identity and access management controls captured in these groups are arranged in ascending degrees of authentication. This is in conjunction with the Zero Trust core goals. These controls can be implemented in phases, allowing managers to improve their organization's security in stages without overbearing the constraints and risks involved. It allows the gradual progression of moving to the next level of control as the organization becomes more secure in both the levels and requirements.

*Access Control and Endpoint Security - Survey Question #8*

The grouping of Access Control and Endpoint Security follows an underlying logic in how those endpoint protection and access control elements are grouped together in column N. The grouping follows a progression from foundational endpoint security controls to more granular access management capabilities.

The ordering starts with the core endpoint threat detection capability, Endpoint Detection and Response. This fundamental endpoint security solution provides visibility into threats and suspicious activities on endpoints, establishing a secure foundation for the subsequent access control measures.

Building on Secure Endpoints, Role-Based Access Control (RBAC) allows access privileges to be managed based on defined roles and responsibilities. This coarse-grained access control model serves as the initial layer of access management.

User and Account Lifecycle Management supports the provisioning, management, and de-provisioning of user accounts throughout their lifecycle. This capability ensures that access controls remain effective by responsibly managing identities from creation to termination.

Granular Access Management: Privileged Access Management (PAM) focuses specifically on controlling and monitoring access to sensitive accounts and systems with elevated privileges. This layer of access control adds an extra level of security for high-risk resources.

Attribute-Based Access Control (ABAC) represents a more advanced and dynamic access model based on factors beyond just roles. This fine-grained access control mechanism aligns with Zero Trust principles, enabling highly granular access decisions based on various attributes.

Aligned with Zero Trust: The grouping follows a logical progression of first establishing endpoint security visibility, followed by implementing increasing levels of access control granularity aligned with Zero Trust principles. It starts with the core endpoint threat detection capability, then layers access control models from coarse role-based to highly granular attribute-based access.

The user and account lifecycle piece enables responsibly managing identities throughout their full lifecycle, ensuring that these access controls remain effective and up-to-date.

By building on strong endpoint security foundations before layering more robust identity-centric access management capabilities, this grouping allows for a comprehensive and well-structured approach to endpoint protection and access control.

*Endpoint Security and Management- Survey Question #9*

The Endpoint Security and Management grouping follows a progression from broad device management controls to more granular security validation and enforcement at the endpoint level.

Unified Endpoint Management (UEM) provides a centralized way to manage and secure all types of endpoints (desktops, laptops, mobile devices, etc.) across their lifecycle. Building on UEM, Mobile Device Management (MDM) focuses specifically on managing mobile devices like smartphones and tablets which have different requirements. The grouping then shifts towards validating and enforcing security posture on the endpoints.

Device Posture and Compliance Checks allow validating the security posture and policy compliance of endpoints before granting access. Secure Boot and Hardware-Based Integrity involve utilizing secure hardware capabilities to establish endpoint root of trust and integrity validation.

Micro-Agent or Agentless Endpoint Security represents different ways to deploy security sensors/controls directly on the endpoints themselves. The ordering starts with broad endpoint management capabilities, then narrows to specific requirements around mobile devices, and focuses on validating and enforcing security posture on the endpoints through compliance checks, secure boot processes, and deploying endpoint security controls.

Aligning with Zero Trust Principles allows organizations to first establish broad device management, then layer on more granular security hardening and enforcement at the endpoint level. The grouping progresses from management controls to security validation to endpoint enforcement in a structured way to reduce the risk of compromise.

*Network Security and Access Control Elements Survey Question #10*

The network security and access control elements grouping follows a progression from implementing foundational Zero Trust Network Access Controls to more granular application/device level enforcement.

Establishing Zero Trust Network Access: It starts with establishing the overarching Zero Trust Network Access (ZTNA) approach, enabled by software-defined perimeter (SDP) technology. ZTNA establishes the core zero trust model for secure remote access to applications/services without traditional VPNs, while SDP is a key enabling technology to implement the ZTNA zero trust access model.

Enforcing Access Controls: The progression then extends to broad Network Access Control policies allowing for enforcement of policies to validate user, device, and application criteria prior to network access. It then narrows down to application level targeting and control and enforcement through application control/whitelisting which allows only the execution of approved/trusted applications.

Responding to Threats: The grouping includes attending to known threats through Device Isolation and Quarantine capabilities in which risky devices are isolated. This makes it possible to know and remove or restrict access to the affected or lawless devices from the system.

Aligning with Zero Trust Principles: The ordering also subscribes to a methodology that starts with implementing Zero Trust Network Access, then enforcing access control at a degree of a granular nature, and finally responding to whatever threats that have been identified, which is in itself core to zero-trust principles. The grouping makes it possible to build a ZTNA base, use multilayers of access control granularity, and have the enforcement and response for policy offenses.

### *Network Security and Segmentation- Question #11*

The network security and segregation elements rotation proceeds from basic and such foundational elements as network segmentation advanced to such capabilities related to control of access in cloud and web environments. It starts with the core capability to segment and isolate different parts of the network securely through Microsegmentation and Network Isolation.

Microsegmentation in networks can be achieved through Virtual Local Area Networks. Network Access Control allows implementing restriction and segmentation of the network for the devices or users that connect within its domains.

Cloud Access Security Brokers extends the access rights, control and monitoring over the use of the cloud services' traffic.

Microsegmentation capabilities partition the network into securely defined regions. It then bound and explains policies of protection namely Networking Access Control Along with VLANs. From there, it addressed more advanced aspects, such as web and cloud security – critical external traffic flows within the Zero Trust model. The advancement was logical and propagational i.e. first create the physical separation of the network, implement enforcement, then control the access to the websites and to the cloud services.

Adhering to the tenets of Zero Trust, this organization makes it easier to minimize environmental risk via layered network demarcation, verification, and secure online/cloud operations, which are key elements of Zero Trust. The first step in zero trust is network segmentation, where apart from access controls and policies, there is a focus on volumetric traffic flow control, which in terms of zero trust is essential.

*Data Protection and Network Security- Question #12*

Grouping the elements of data protection and network security together shows an underlying logic, moving from data security foundational controls to enabling secure access while implementing monitoring and prevention controls. This also logically flows into key data protection tenets and gives a structured approach to holistic data risk reduction.

It starts with the very basics, the identification and categorization of sensitive data assets through Data Classification and Labeling. This gives the basic level of recognition and categorization of sensitive data that needs protection. Expanding from this, DLP controls introduce ways in which the implementation of monitoring and preventing unauthorized data exfiltration can be assured.

Enabling Secure Remote Access Capabilities: Essential for accessing data and applications in a Zero Trust model. Technologies such as VPNs, virtual desktop infrastructure (VDI), and remote desktop protocol (RDP) facilitate secure remote access to sensitive resources.

Ensuring Data Accessibility through APIs and Web Interfaces Secure access via API interfaces and/or web interfaces by including API Gateways and Web Application Firewalls that protect and manage access to data and applications exposed via these vectors.

Network traffic pattern analysis and the detection of anomalous behaviors: In general, network traffic analysis provides visibility into network events on a continuous basis and offers an additional layer of protection against exposure risks to sensitive data.

Data Risk Reduction Approach: This logical flow aligns with first discovering and categorizing data assets, implementing preventive controls, enabling secure access capabilities, hardening exposure vectors, and then implementing continuous data-centric monitoring – all key data protection tenets. The grouping provides a structured approach to holistic data risk reduction through progressive classification, prevention, access, hardening, and monitoring controls.

Aligning with Zero Trust Principles: Organization can minimize environmental risk via layered network demarcation, verification, and secure online/cloud operations, which are key elements of Zero Trust. The first step in zero trust is network segmentation, where apart from access controls and policies, there is a focus on volumetric traffic flow control, which in terms of zero trust is essential.

*Data Protection and Information Security- Question #13*

The data protection and information security elements grouping follows a progression from foundational data encryption controls to more granular access management and rights enforcement for data.

The process begins with the fundamental data encryption capabilities as a basic protection measure, establishing the baseline of protecting data confidentiality. Building on encryption, it then covers implementing granular access control policies to validate and authorize data access requests based on defined conditions.

Encryption of Data at Rest and in Transit: whether the data is stored or transmitted across networks.

Digital Rights Management (DRM) allows organizations to control not only who can access the data but also how the data can be used, edited, printed, or shared, even after it has been accessed.

Rights Management Services extends access control by enabling persistent protection and enforcement of data usage rights and restrictions.

Secure File Sharing and Collaboration Capabilities provides the ability to share files and data securely while enforcing the defined access rights and usage restrictions during collaboration workflows. This ensures that data remains protected even when shared or collaborated upon, upholding the established access policies and usage rights.

This logical flow aligns with a progression of encrypting data, validating access, persistently enforcing data rights, and then enabling controlled sharing and collaboration-implementing progressively stronger data-centric controls. The grouping allows organizations to layer data protection capabilities in a structured way, progressively reducing data risks through encryption, access policies, usage restrictions, rights management, and secure sharing.

Aligning with Zero Trust Principles: Organizations can establish a comprehensive data protection strategy that starts with encrypting data, implements granular access controls, enforces persistent data usage rights, and enables secure collaboration while upholding those rights. This structured approach helps organizations effectively manage and mitigate data risks, ensuring the confidentiality, integrity, and availability of their sensitive information assets throughout the data lifecycle.

*Security Monitoring and Incident Response- Question # 14*

The security monitoring and incident response elements grouping follows a progression from foundational security monitoring and analytics capabilities to more advanced automation and orchestrated response.

The ordering begins with implementing a Security Information and Event Management (SIEM) platform. The progression then covers specific monitoring of data activities and extends that monitoring to cloud storage environments.

Security Information and Event Management: the central security monitoring and event correlation engine. This establishes the core capability to collect, analyze, and correlate security event data from multiple sources, providing a comprehensive view of an organization's security posture.

User and Entity Behavior Analytics (UEBA) provides advanced analytics focused specifically on detecting anomalous user and entity behaviors that could indicate potential threats. UEBA enhances the SIEM's capabilities by leveraging machine learning and advanced analytics techniques to identify deviations from normal behavior patterns, enabling more effective threat detection.

Data Activity Monitoring and Analytics allows organizations to monitor and analyze data access and usage activities, enabling the detection of potential data loss or misuse incidents.

Data Loss Prevention for Cloud Storage extends this monitoring capability to cover cloud data risks, ensuring comprehensive visibility and protection across on-premises and cloud environments.

Security Orchestration, Automation, and Response (SOAR) enables automating and orchestrating response actions based on the detected threat intelligence. SOAR solutions leverage the insights gathered from the SIEM, UEBA, and data monitoring components to automate and streamline incident response workflows, accelerating the time to respond and mitigate threats.

This logical progression aligns with first establishing monitoring and analytics foundations, then enhancing with advanced behavioral and data analytics, and finally automating response actions- implementing progressively intelligent detection and automated response. The grouping allows organizations to build robust security monitoring, enhance with targeted analytics for user and data risks, extend visibility into cloud environments, and ultimately leverage automation to accelerate and streamline incident response processes.

Aligning with Zero Trust Principles: By following this structured approach, organizations can effectively detect and respond to a wide range of security threats, from traditional cyber attacks to insider threats and data breaches, while leveraging automation to improve efficiency and reduce the risk of human error in incident response.

*Security Monitoring and Threat Detection- Question #15*

The underlying logic behind how security monitoring and threat detection elements are grouped follows a progression from foundational security monitoring and logging capabilities to advanced analytics, machine learning, threat intelligence integration, and automated enforcement.

Centralized Logging and Auditing provides a core capability for collecting and recording, for monitoring and auditing purposes, centralized security-related events and activities. Centralized logging is source data, the basis for further developments.

Continuous monitoring with anomaly detection can always monitor the data arriving continuously to find deviations from normal behavior patterns that might allow a quick identification of threats or security incidents.

Security Intelligence and Machine Learning: Advanced analytics is utilized, with the inclusion of machine learning models, thus allowing for deep analytics in monitoring data for complex threats and patterns not easily detectable by traditional means.

Threat Intelligence Platforms unify global threat intelligence, enabling organizations to share monitoring data fed with new indicators and threat intelligence, offering wide context and enabling end-to-end search and response.

Standard policies and remediation help define the outcomes of advanced analysis, while threat intelligence, automation, and optimization together enable timely and consistent responses to any detected threats or breaches, thereby reducing possibilities for human error and hastening the process.

This logical process has to do with the first understanding, via the cutting of the middle trees, by methods in analysis that slow down the process, bringing in global threat intelligence, and lastly encouraging action reinforcement and response due to the information obtained in the previous sections.

Aligning with Zero Trust Principles: This approach enables the organization to create a program of security analytics and solutions that could drive data analytics, machine learning, and threat intelligence toward effective and efficient detection and response against various security threats.

*Governance, Risk, and Compliance- Question #16*

There is an underlying logic behind how governance, risk, and compliance elements are grouped. The grouping follows a progression from establishing foundational GRC frameworks and policies to enabling compliance monitoring, reporting, and automation.

Core Policy and Risk Management Frameworks sets the standards for compliance across the organization. This lays the groundwork by implementing overarching frameworks, policies, and processes to manage security risks and define compliance requirements.

Compliance Monitoring and Reporting Capabilities allows for continuously monitoring the organization's compliance posture against the defined policies and generating reports to provide visibility into the current state of compliance.

Integrated Dashboards and Reporting provides a centralized view with dashboards and reporting capabilities to surface compliance insights and metrics to relevant stakeholders in a consolidated and easily consumable manner.

Automated Compliance Checks and Controls enable automating the validation of compliance requirements and the enforcement of associated controls, reducing the risk of human error and increasing the efficiency of compliance management processes.

Vendor and Third-Party Risk Assessment covers assessing and managing the risks introduced by the use of third-party vendors, suppliers, or service providers, ensuring that compliance requirements are met throughout the organization's extended ecosystem.

This logical flow aligns with first defining compliance targets through the establishment of Policy and Risk Management Frameworks, implementing monitoring and reporting mechanisms to validate compliance, automating compliance validation checks and control enforcement actions based on the defined policies, and extending to incorporate third-party risk assessments- progressively maturing GRC capabilities.

Aligning with Zero Trust Principles: By following this structured approach, organizations can establish a comprehensive GRC program that starts with defining compliance standards, enables continuous monitoring and reporting, leverages automation for efficient compliance validation and enforcement, and extends to manage risks introduced by third-party relationships, ultimately fostering a culture of accountability and risk-aware decision-making.

*Operational Security and Incident Management- Question #17*

There is an underlying logic behind how these operational security and incident management elements are grouped. The grouping follows a progression from establishing foundational security processes and controls to enabling effective incident response and reporting.

Device Endpoint Patching and Updates are fundamental practices of keeping devices and endpoints updated with the latest security patches and software versions, reducing vulnerabilities. This lays the groundwork by implementing processes for device and endpoint patching and updates, establishing a secure baseline for the organization's IT infrastructure.

Change Management and Configuration Control processes ensure that changes to systems and configurations are responsibly managed and controlled, maintaining the organization's security posture and preventing unintended vulnerabilities or misconfigurations.

Security Awareness and Training Programs educate employees on security best practices and reinforce the importance of operational security controls, fostering a security-conscious culture within the organization.

Incident Response and Disaster Recovery Planning includes having well-defined plans and processes for responding to security incidents and recovering from disasters, ensuring the organization's ability to effectively manage and mitigate the impact of security events.

Reporting and Executive Dashboards: This provides visibility through reporting and dashboards, informing executives and stakeholders about the organization's operational security posture, enabling data-driven decision-making, and facilitating leadership's oversight of operational security initiatives.

This logical progression aligns with first establishing secure baselines through endpoint patching and updates, implementing core operational processes such as Change Management and Configuration Control, educating the workforce through Security Awareness and Training Programs, enabling response readiness with Incident Response and Disaster Recovery Planning, and ultimately providing oversight and visibility through reporting and dashboarding – progressively maturing operational security capabilities.

Aligning with Zero Trust Principles: By following this structured approach, organizations can build a robust operational security program that starts with secure foundations, implements essential processes, cultivates a security-aware workforce, prepares for effective incident response and recovery, and provides leadership with the necessary visibility and insights to make informed decisions and drive continuous improvement in operational security.

## 4.4     Summary

This chapter highlights various design and implementation methodologies from basic to detailed scientific for research on the security paradigm of Zero Trust. It begins by elaborating on the basic configuration for Zero Trust, which includes identity and access management, endpoint security, network segmentation, data protection, and security monitoring. The chapter then provides an in-depth breakdown of how these components fit within an end-to-end expanded framework for security, showing diagrams that support an explanation of each component within the overall security architecture.

The methodology outlines how online surveys were carried out through QualtricsXM and discusses, among other methodological considerations, issues of sampling bias, the approach utilized to explore the data, and how the questionnaire had been validated. This is followed by a detailed discussion of eleven logical groupings of cybersecurity questions, Questions 7-17, inclusive of IAM, Network Security, Data Protection, and GRC.

Each group is then described, from the perspective of its development from foundation controls to the most interesting capabilities and specifically how they relate to Zero Trust principles. Statistical considerations, such as binary logistic regression and multicollinearity in multinomial regression, are mentioned to ensure that the data collected is appropriately analyzed.

# CHAPTER 5: DATA ANALYSIS

## 5.1     Descriptive Analyses

This chapter delves into the pivotal findings stemming from the research on ZTA, a cybersecurity paradigm that challenges the traditional perimeter-based approach. It examines the essential technical controls organizations employ to implement ZTA effectively, unveiling the critical components underpinning this robust security model.

The research investigates the profound impact of ZTA implementation on an organization's ability to prevent and mitigate cybersecurity threats, assessing its effectiveness in fortifying defense mechanisms against malicious actors. It explores the industry-recognized best practices that have emerged as organizations navigate the complexities of ZTA adoption, distilling the most effective methodologies that have proven successful in real-world scenarios.

The research unveils significant insights contributing to the ever-evolving body of knowledge surrounding ZTA by analyzing patterns, correlations, and unexpected nuances within the data. These findings address existing gaps in understanding and offer practical implications for relevant stakeholders, empowering organizations to make informed decisions and implement robust security measures in the face of an increasingly hostile cyber landscape.

## 5.2     Dataset Analysis on Participant Demographics

To conduct the following research, the survey responses were exported from Qualtrics$^{XM}$ into an IBM SPSS file format for analysis. The file was then interrogated primarily using the Analyze => Descriptive Statistics => Frequencies function. The respondent counts were verified for accuracy similarly for variables A2 through A18_5. The individual results were collated and presented in a series of tables that follows.

A snippet of the IBM SPSS code used to perform this analysis is shown in Figure 12.

```
 1  * Encoding: UTF-8.
 2
 3  DATASET ACTIVATE DataSet1.
 4  FREQUENCIES VARIABLES=YOE Job_Role Company_Size Region
 5   /ORDER=ANALYSIS.
 6  OUTPUT MODIFY
 7    /SELECT TABLES
 8    /IF COMMANDS=["Frequencies(LAST)"] SUBTYPES="Frequencies"
 9    /TABLECELLS SELECT=[VALIDPERCENT CUMULATIVEPERCENT] APPLYTO=COLUMN HIDE=YES
10    /TABLECELLS SELECT=[TOTAL] SELECTCONDITION=PARENT(VALID MISSING) APPLYTO=ROW HIDE=YES
11    /TABLECELLS SELECT=[VALID] APPLYTO=ROWHEADER UNGROUP=YES
12    /TABLECELLS SELECT=[PERCENT] SELECTDIMENSION=COLUMNS FORMAT="PCT" APPLYTO=COLUMN
13    /TABLECELLS SELECT=[COUNT] APPLYTO=COLUMNHEADER REPLACE="N"
14    /TABLECELLS SELECT=[PERCENT] APPLYTO=COLUMNHEADER REPLACE="%".
```

*Figure 12. IBM SPSS Code Snippet No 1 Frequencies*

A sample of 138 participants completed the survey between June 2024 and August 2024, all of whom indicated that their organizations have implemented ZTA. The majority of the participants ($n = 58$, 42%) had more than 20 years of experience in cybersecurity and were engineers/architects ($n = 42$, 30.4%). Also, the majority of the participants ($n = 72$, 53.6%) were employed in large organizations with company annual revenue exceeding one billion United States dollars. Most of the organizations where the respondents were employed were predominantly located in North America ($n = 108$, 78.3%). A summary of the participants' demographics are presented in Table 2.

*Table 2. Participants' Demographics*

| Variable | Category | *n* | % |
|---|---|---|---|
| **Years of Experience** | < 5 | 10 | 7.2% |
|  | 6-10 | 21 | 15.2% |
|  | 11-15 | 23 | 16.7% |
|  | 16-20 | 26 | 18.8% |
|  | >20 | 58 | 42.0% |
| **Current Job Function** | Administrative/executive | 13 | 9.4% |
| | Cybersecurity/IT staff | 32 | 23.2% |
| | Engineer/Architect | 42 | 30.4% |
| | Staff/Technology Manager | 9 | 6.5% |
| | Professional Staff | 2 | 1.4% |
| | Academics/professor/faculty member | 4 | 2.9% |
| | Consultant | 35 | 25.4% |
| | FT/PT Graduate Student | 1 | 0.7% |
| **Size of the Company** | < $1M | 20 | 14.5% |
| | 1.1M - $10M | 13 | 9.4% |
| | 10.1M - $50M | 11 | 8.0% |
| | 50.1M - $200M | 9 | 6.5% |
| | 200.1M - $500M | 8 | 5.8% |
| | 500.1M - $1B | 3 | 2.2% |
| | >1B | 74 | 53.6% |

| Geographic Location | Africa: North Africa and Sub-Saharan | 2 | 1.4% |
|---|---|---|---|
| | Asia | 4 | 2.9% |
| | Europe | 20 | 14.5% |
| | Latin America and the Caribbean | 1 | 0.7% |
| | North America | 108 | 78.3% |
| | Oceania | 3 | 2.2% |

## 5.3     Dataset Analysis Supporting RQ1

In the dataset, variables A7_1 through A17_5 were the responses to a series of questions regarding the sentiment toward which technical controls would be perceived to be more or less important. In essence, a stack rank was established for each technical control group by tabulating the "modes" for each. To facilitate handling the variables within the groups the following schema was developed (Table 3).

*Table 3. Technical Control Group Schema*

| Group | Variable | Label |
|---|---|---|
| 1 | A7_All_in_IAM | Tech Ctrls Grp 1- Identity and Access Management |
| 2 | A8_All_in_ACES | Tech Ctrls Grp 2- Access Control and Endpoint Security |
| 3 | A9_All_in_ESM | Tech Ctrls Grp 3- Endpoint Security and Management |
| 4  | A10_All_in_NSAC  | Tech Ctrls Grp 4- Network Security and Access Control |
| 5  | A11_All_in_NSS   | Tech Ctrls Grp 5- Network Security and Segmentation |
| 6  | A12_All_in_DPNS  | Tech Ctrls Grp 6- Data Protection and Network Security |
| 7  | A13_All_in_DPIS  | Tech Ctrls Grp 7- Data Protection and Information Security |
| 8  | A14_All_in_SMIR  | Tech Ctrls Grp 8- Security Monitoring and Incident Response |
| 9  | A15_All_in_SMTD  | Tech Ctrls Grp 9- Security Monitoring and Threat Detection |
| 10 | A16_All_in_GRC   | Tech Ctrls Grp 10- Governance, Risk, and Compliance |
| 11 | A17_All_in_OSIM  | Tech Ctrls Grp 11- Operational Security and Incident Management |

The first research question identified the key technical controls of ZTA based on the rating provided by the respondents concerning how they are prioritized in their organizations. The first category that were listed as key technical controls included: (a) multi-factor authentication; (b) Single Sign-On; (c) Adaptive Risk-Based Authentication; (d) Biometric Authentication; and (e) Identity Federation and Directory Services. One of the ZTA key technical controls was MFA, which was ranked and tied for first place by the plurality of participants: n = 48 or 34.8%. Next in line is Identity Federation and Directory Services, with n = 48 or 34.8%, followed by Adaptive Risk-Based Authentication at n = 28 or 20.3%, and with the smallest number, Biometric Authentication takes the last place with n = 5 or 3.6%. Table 4 provides a breakdown of the ranking of the five technical controls. Based on these results, the top two prioritized key ZTA controls at most organizations are identity federation and directory and MFA.

*Table 4. Ranking of Technical Controls Group 1*

| Ranking | MFA | SSO | Adaptive Risk-Based Authentication | Biometric Authentication | Identity Federation and Directory Services |
|---|---|---|---|---|---|
| First | 48 | 9 | 28 | 5 | 48 |
| Second | 40 | 26 | 39 | 14 | 19 |
| Third | 36 | 27 | 33 | 20 | 22 |
| Fourth | 12 | 47 | 30 | 28 | 21 |
| Fifth | 2 | 29 | 8 | 71 | 28 |

The second category of ZTA key technical controls assessed were Endpoint Detection and Response, Role-Based Access Control (RBA), Attribute-Based Access Control (ABA), privileged access management, and User and Account Lifecycle Management. For the second category, the prominent ZTA key technical control that was rated to be prioritized by the majority of the participants was User and Account Lifecycle Management ($n = 46$, 33.3%). The least prioritized technical control was ABA ($n = 19$, 13.8%) (Table 5).

*Table 5. Ranking of Technical Controls Group 2*

| Ranking | EDR | RBA | ABA | PAM | User and Account Lifecycle Management |
|---|---|---|---|---|---|
| First | 27 | 21 | 19 | 25 | 46 |
| Second | 22 | 38 | 19 | 41 | 18 |
| Third | 29 | 36 | 26 | 28 | 19 |
| Fourth | 25 | 31 | 33 | 26 | 23 |
| **Fifth** | 35 | 12 | 41 | 18 | 32 |

The third category of key ZTA technical controls comprised unified endpoint management, Mobile Device Management, Device Posture and Compliance Checks, Secure Boot and Hardware-Based Integrity, and Micro-Agent or Agentless Endpoint Security. Among these technical controls, Device Posture and Compliance Checks were ranked first by the majority of the respondents ($n = 57$, 41.3%). The least rated technical control was Micro-Agent or Agentless Endpoint Security ($n = 12$, 8.7%) (Table 6).

*Table 6. Ranking of Technical Controls Group 3*

| Ranking | UEM | MDM | Device Posture and Compliance Checks | Secure Boot and Hardware-Based Integrity | Micro-Agent or Agentless Endpoint Security |
|---|---|---|---|---|---|
| **First** | 23 | 24 | 57 | 22 | 12 |
| **Second** | 39 | 24 | 31 | 16 | 28 |
| **Third** | 39 | 39 | 26 | 15 | 19 |
| **Fourth** | 30 | 26 | 14 | 41 | 27 |
| **Fifth** | 7 | 25 | 10 | 44 | 52 |

The fourth category of ZTA technical controls comprised software-defined perimeter, Zero Trust Network Access, Application Control and Whitelisting, Device Isolation and Quarantine, and Network Access Control (NAC). ZTNA was ranked first by the majority of the participants ($n = 96$, 69.6%). The least ranked technical controls were Device Isolation and Quarantine and NAC, with ($n=7$, 5.1%) respondents ranking them first (Table 7).

*Table 7. Ranking of Technical Controls Group 4*

| Ranking | SDP | ZTNA | Application Control and Whitelisting | Device Isolation and Quarantine | NAC |
|---|---|---|---|---|---|
| **First** | 16 | 96 | 12 | 7 | 7 |
| **Second** | 52 | 18 | 25 | 13 | 31 |
| **Third** | 27 | 11 | 43 | 25 | 31 |
| **Fourth** | 21 | 10 | 43 | 41 | 23 |
| **Fifth** | 22 | 3 | 15 | 52 | 46 |

Another category of ZTA key technical control that was compared comprised Microsegmentation and Network Isolation, Virtual LANs and Microsegmentation, Network Access Control, secure web gateways, and Cloud Access Security Brokers (CASB). Microsegmentation and Network Isolation were ranked first among the majority of the respondents ($n = 65$, 47.1%). VLANs and microsegmentation were the lowest ranked (n = 11, 8%) (Table 8).

*Table 8. Ranking of Technical Controls Group 5*

| Ranking | Microsegmentation and Network Isolation | VLANs and microsegmentation | NAC | Secure web gateways | CASB |
|---|---|---|---|---|---|
| **First** | 65 | 11 | 17 | 27 | 18 |
| **Second** | 33 | 38 | 19 | 22 | 26 |
| **Third** | 20 | 43 | 35 | 21 | 19 |
| **Fourth** | 13 | 22 | 28 | 42 | 33 |
| **Fifth** | 7 | 24 | 39 | 26 | 42 |

The sixth category of ZTA key technical controls that were evaluated were Data Classification and Labeling, API Gateways and Web Application Firewalls, Network Traffic Analysis and Anomaly Detection, Secure remote access (VPN, VDI, RDP), and Data loss prevention. Among these controls, Data Classification and Labeling were ranked first by the majority of the respondents ($n = 61$, 44.2%) followed by Secure Remote Access (VPN, VDI, RDP) ($n = 38$, 27.5%). The least ranked technical controls were API Gateways and Web Application Firewalls ($n = 8$, 5.8%) (Table 9).

*Table 9. Ranking of Technical Controls Group 6*

| Ranking | Data Classification and Labeling | API Gateways and Web Application Firewalls | Network Traffic Analysis and Anomaly Detection | Secure remote access (VPN, VDI, RDP) | DLP |
|---|---|---|---|---|---|
| **First** | 61 | 8 | 18 | 38 | 13 |
| **Second** | 34 | 43 | 18 | 19 | 24 |
| **Third** | 11 | 35 | 46 | 22 | 24 |
| **Fourth** | 25 | 30 | 30 | 21 | 32 |
| **Fifth** | 7 | 22 | 26 | 38 | 45 |

The seventh category of technical controls compared were Encryption of Data at Rest and in Transit, Digital Rights Management (DRM), Data Access Control and Granular Policies, Rights Management Services, and Secure File Sharing and Collaboration. Among these controls, Encryption of Data at Rest and in Transit were ranked first by the majority of the participants ($n = 81$, 58.7%) (Table 10).

*Table 10. Ranking of Technical Controls Group 7*

| Ranking | Encryption of Data at Rest and in Transit | DRM | Data Access Control and Granular Policies | Rights Management Services | Secure File Sharing and Collaboration |
|---|---|---|---|---|---|
| First | 81 | 6 | 42 | 3 | 6 |
| Second | 22 | 13 | 58 | 8 | 37 |
| Third | 19 | 32 | 21 | 28 | 38 |
| Fourth | 12 | 55 | 16 | 43 | 12 |
| Fifth | 4 | 32 | 1 | 56 | 45 |

The eighth group of technical controls comprised Data Loss Prevention for Cloud Storage, security information and event management, user and entity behavior analytics (UEBA), Security Orchestration, Automation, and Response (SOAR), and Data Activity Monitoring and Analytics. In this category, the highly rated control was SIEM (n = 50, 36.2%) and the lowest was Data Loss Prevention for Cloud Storage (n = 14, 10.1%) (Table 11).

*Table 11. Ranking of Technical Controls Group 8*

| Ranking | Data Loss Prevention for Cloud Storage | SIEM | UEBA | SOAR | Data Activity Monitoring and Analytics |
|---|---|---|---|---|---|
| First | 14 | 50 | 34 | 24 | 16 |
| Second | 18 | 25 | 35 | 27 | 33 |
| Third | 26 | 22 | 32 | 36 | 22 |

| | Fourth | 34 | 26 | 20 | 24 | 34 |
|---|---|---|---|---|---|---|
| | Fifth | 46 | 15 | 17 | 27 | 33 |

The ninth category comprised Threat Intelligence Platforms (TIPS), Security Analytics and Machine Learning, Centralized Logging and Auditing, Automated Policy Enforcement and Remediation, and Continuous Monitoring and Anomaly Detection. In this category, Continuous Monitoring and Anomaly Detection were ranked first by the majority of the participants (n = 44, 31.9%) (Table 12).

*Table 12. Ranking of Technical Controls Group 9*

| Ranking | Threat Intelligence Platforms | Security Analytics and Machine Learning | Centralized Logging and Auditing | Automated Policy Enforcement and Remediation | Continuous Monitoring and Anomaly Detection |
|---|---|---|---|---|---|
| First | 24 | 3 | 35 | 32 | 44 |
| Second | 23 | 22 | 26 | 30 | 37 |
| Third | 16 | 34 | 23 | 33 | 32 |
| Fourth | 38 | 43 | 22 | 30 | 5 |
| Fifth | 37 | 36 | 32 | 13 | 20 |

Another category of ZTA key technical controls comprised Policy and Risk Management Frameworks, Compliance Monitoring and Reporting, Integrated Dashboards and Reporting, Vendor and Third-Party Risk Assessments, and Automated Compliance Checks and Controls. Among these controls, Policy and Risk Management Frameworks were ranked first by the majority of the participants ($n =78$, 56.5%) (Table 13).

*Table 13. Ranking of Technical Controls Group 10*

| Ranking | Policy and Risk Management Frameworks | Compliance Monitoring and Reporting | Integrated Dashboards and Reporting | Vendor and Third-Party Risk Assessments | Automated Compliance Checks and Controls |
|---|---|---|---|---|---|
| **First** | 78 | 12 | 8 | 10 | 30 |
| **Second** | 26 | 48 | 19 | 17 | 28 |
| **Third** | 15 | 45 | 19 | 35 | 24 |
| **Fourth** | 13 | 15 | 43 | 37 | 30 |
| **Fifth** | 6 | 18 | 49 | 39 | 26 |

The last category of ZTA key technical controls comprised Security Awareness and Training Programs, Change Management and Configuration Control, Incident Response and Disaster Recovery Planning, Reporting and Executive Dashboards, and Device Endpoint Patching and Updates. Among these controls, Device Endpoint Patching and Updates ($n = 44$, 31.9%) and Security Awareness and Training Programs ($n = 43$, 31.2%) were ranked first and second respectively by the majority of the participants (Table 14).

*Table 14. Ranking of Technical Controls Group 11*

| Ranking | Security Awareness and Training Programs | Change Management and Configuration Control | Incident Response and Disaster Recovery Planning | Reporting and Executive Dashboards | Device Endpoint Patching and Updates |
|---|---|---|---|---|---|
| **First** | 43 | 29 | 17 | 5 | 44 |
| **Second** | 13 | 47 | 33 | 4 | 41 |
| **Third** | 25 | 30 | 47 | 8 | 28 |
| **Fourth** | 40 | 25 | 34 | 26 | 13 |
| **Fifth** | 17 | 7 | 7 | 95 | 12 |

## Research Question 1

**What are the key technical controls of a Zero Trust Architecture in organizations?**

## Quantitative Analysis of Cybersecurity Control Maturity

This unique approach needs to be applied to classify the variables more nuanced and meaningfully, considering the nature of the data and the context of the cybersecurity maturity model.

## Related Definitions

1. **Mean**: This is the average value of all the Rank 1 variables. It provides a central point around which the data is distributed.

2. **Standard Deviation**: This measures how spread out the numbers are from the mean. A slight standard deviation means the numbers are close to the mean, while a large standard deviation means they are spread out over a wider range.

3. **Using Mean and Standard Deviation for Classification**:
   - A threshold can be set by using the mean and standard deviation to help classify the data into different categories.
   - For example, if a percentage is much higher than the mean, it might be considered "Optimized (Very High)" because it is significantly better than average.
   - If a percentage is much lower than the mean, it might be considered "Initial (Low)" because it is significantly worse than average.

4. **Outliers**:
   - Outliers are values much higher or lower than the rest of the data.

- Using the mean and standard deviation provides a better way of accounting for these outliers. Instead of letting them skew the understanding of the data, the approach uses the standard deviation to understand how unusual these values are.

- This approach helps make more informed decisions about which values are exceptional (high or low), and which are more typical.

Using the mean and standard deviation allows the classification of the data in a way that considers both the average performance and the variability in the data. This method helps identify which variables perform exceptionally well or poorly, even if the dataset has extreme values.

**Logic and Analysis**

The detailed logic and analysis are outlined below:

1. Calculate the Average Ranking for each respondent
2. Calculate Statistics:
   - Mean of Average_Rank: 2.9994545454545456
   - Standard Deviation of Average_Rank: 0.4076736993207253
3. Define Thresholds:
   - Optimized (Very High): >= Mean + 1.5 * Standard Deviation
   - Advanced (High): >= Mean + 0.5 * Standard Deviation
   - Developing (Moderate): >= Mean - 0.5 * Standard Deviation
   - Initial (Low): < Mean - 0.5 * Standard Deviation

Apply the defined thresholds results using the calculated mean ($\mu$=3.00) and standard deviation ($\sigma$=0.408) for Threshold Calculations:

- Initial (Low): $\mu - \sigma$ = 3.00 - 0.41 = 2.59
- Developing (Moderate): $\mu$ = 3.00
- Advanced (High): $\mu + \sigma$ = 3.00 + 0.41 = 3.41

- Optimized (Very High): µ + 2σ = 3.00 + (2 * 0.41) = 3.81

These calculations show how the breakpoints are derived using the actual values for the mean and standard deviation.

*Table 15. Posture Levels and Percentages*

| Posture Level | Count | % | Cumulative % |
|---|---|---|---|
| Optimized (Very | 11 | 20% | 100% |
| Advanced (High) | 15 | 27.2% | 80% |
| Developing | 21 | 38.2% | 52.7% |
| Initial (Low) | 8 | 14.5% | 14.5% |

**Distribution Insights**

Analyzing the distribution of technical controls among various maturity levels enables concluding the general trend of an organization's cybersecurity approach. The "Developing (Moderate)" category holds the largest group of controls, at 38.18%. This large percentage reveals that many of the technical controls are of average importance or priority in the organization's security framework. Therefore, such a significant middle ground indicates a more balanced practice in implementing security, where most controls have received considerable attention without being flagged as critical.

The "Advanced (High)" category is the second-biggest group, comprising 27.27% of the controls. This large relativist share would indicate that a fair share of controls is highly pertinent to an organization's security posture. Such controls most likely represent more sophisticated or targeted measures to deal with specific security concerns or compliance requirements of higher priority.

The top category in the spectrum is the "Optimized," of which only 20% of the controls are available. The group consists of the most significant or highest-priority technical measures internally adapted as part of the organization's security strategy. The controls that would go into this category would most certainly be of an unusual nature, with heavy reinforcements, frequent changes, and scrupulous supervision. What is termed the 'core' is most potent to the defense posture of the information system.

The narrowest category that might be considered would probably be "Initial," constituting 14.55%. That could mean there are rather few such controls that are not so central or less cumbersome to implement or that information security controls have advanced so much that further developments on the standard controls are not expected.

This categorization distribution indicates that balance and nuance are in order in the approach to cybersecurity. Indeed, there is an emphasis on having an overarching focus on a few key controls, which is observable from the large volume of Optimized and Advanced categories.

It corroborates the hypothesis of the range of importance for controls by the smooth transition of the graph, with no erratic shifts or identifiable marked clusters. This subtle distribution might imply that prioritization for these controls is likely to be context-dependent and could be influenced by specific threats within the organization's environment, industrial standards, or even regulatory requirements. This is, in turn, a progressive process to allow flexible and adaptive security management where management or resources can be distributed across a range of control priorities instead of being condensed into a few high-priority areas alone.

They are the numbers that are most closely related to the real percentages; however, they are not exactly the same. This is normal for the data of the real world. Everything falls under the category of "Optimized" since it is the highest threshold.

This is a very common approach in statistics and data analysis, where thresholds take their basis from the mean and standard deviation. It adds another categorical layer to the data, reflecting the distance of each data point from the mean in terms of standard deviations. The idea behind this comes from the assumption that the underlying data for the most part would follow a normal distribution; this, however, is rarely the case in real life.

For the categorization in the context of the technical controls, the ranking prioritizes controls to be treated on an average rank. A lower rank-upper mean is critical or important. Controls that rank in higher positions relative to the mean are not as critical and/or are lower in priority.

These thresholds create intervals that, in a normal distribution, would correspond to certain percentages of the data:

- About 16% of the data is below μ - σ (Initial)
- About 50% of the data is below μ (Developing)
- About 84% of the data is below μ + σ (Advanced)
- About 97.7% of the data to be below μ + 2σ (Optimized)

4. Classification Results

The mean value (3.00) is significant because it serves as the central point for the posture level thresholds:

1. It is precisely at the boundary between "Developing (Moderate)" and "Advanced (High)" posture levels.
2. It is the reference point for calculating the other thresholds using standard deviations.

The fact that the mean is 3.00 on a scale that seems to range from about 1.89 to 3.79 suggests that the Average_Rank values are somewhat normally distributed around this central point, with a slight skew towards lower values (since 1.89 is further from 3.00 than 3.79 is).

This information indicates that, though different rankings of the technical controls are present, most cluster in the middle ranks, with very few notable exceptions at the higher and lower ends of the spectrum.

Calculated values from the dataset where:

- **n** is the number of technical controls (55)
- **sum$_i$**= $\sum_{i=1}^{55} X_i$  $x_i$ is the sum of all Average_Rank values (164.97)
- **mu** is the calculated mean (3.00)

$$\mu = \frac{1}{n} \sum_{i=1}^{n} x_i$$

$$\mu = \frac{1}{55} \sum_{i=1}^{55} x_i$$

$$\mu = \frac{1}{55}(164.97)$$

$$\mu = 3.00$$

Where:

The sum of all Average_Rank values: 164.97

Number of controls (n): 55

**Observations**

Cybersecurity measures vary from the lowest average rank of 1.89 to a high of 3.79, with an average of about 3.00. High-ranking activities considered to be of maximum priority are "Encryption of Data at Rest and in Transit," and this received the lowest average rank of 1.89, reflecting that the protection of information requires data encryption.

By contrast, "Micro-Agent or Agentless Endpoint Security" tops out at 3.79, which would support they are less often considered key. This distribution shows a captured and varied emphasis on different security strategies: encryption tends to remain core, whereas endpoint security is related but less urgent for most. Other security measures fall in the range between these seamless and separated extremes, reflecting the nature of cybersecurity, which is multi-layered, needing various solutions to face the numerous, different, and ever-evolving threats.

Identity and Access Management Controls: One of the strongest conclusions from the ZTA survey was on the essential need for strong identity and access management controls within the ZTA. Such as multi-factor authentication, privileged access management, Role-Based Access Control and attribute-based access control, Single Sign-On and Biometric

Authentication. These controls have consistently been reported as the topmost priorities in ZTA implementations in the "Developing Posture."

Endpoint Security Controls: Endpoint security controls, such as EDR, UEM, MDM, and device posture compliance checks, are used mainly to protect endpoints and allow only defined secure endpoints to connect to the resources within a ZTA.

Network Access and Segmentation Controls: Access control mechanisms to the network such as Zero Trust Network Access, Software defined perimeter, Network Access Control and microsegmentation are necessary for containment and protection of network resources in a zonal architecture and security absolute defense. These controls are widely used to prevent lateral movement and minimize the attack surface.

Data Protection Controls: Data protection controls like encryption, data loss prevention, data classification, and secure file sharing form the basis for securing sensitive information in a ZTA. These controls always feature prominently in the list of most respondents by survey.

Security Analytics and Monitoring: Security information and event management, user and entity behavior analytics (UEBA), continuous monitoring, and Threat Intelligence Platforms are increasingly gained attention for enabling detection, evaluation, and containment of threats in ZTA.

Cloud Security Controls: Cloud Access Security Brokers, cloud connectivity, and cloud security posture management (CSPM) have been identified as effective metrics towards the protection of resources in the cloud and management of cloud services in a multicloud ZTA ecosystem.

Application Security Controls: Web application firewalls (WAF) and in the case of APIs, API gateway/web application firewalls are commonly used in the protection of enterprise systems and APIs when deploying ZTA in the organization.

These technical controls include a subset of many of the ZTA components which include identity access management, endpoint security, network segmentation, data protection, security analytics, cloud, and application security. The findings of the survey showed that organizations tend to take a more holistic approach towards the implementation of ZTA as they seek to combine some of these technical controls in order to meet the security requirements sought.

## 5.4    Dataset Analysis Supporting RQ2

The main objective of the second research question was to determine the contribution of ZTA architecture to organizations' efforts to prevent cyberattacks. It should be noted that a good number of the participants (n = 61, 44.2%) stated that their organizations experienced instances of unauthorized access to their computer systems or any data for the past 12 months.

A number of respondents (n = 51, 37%) respectively reported that their organizations encountered external concerns regarding the potential for breaches of their data within the last 12 months. Analyzing the reported responses, it can be noted that the prevalence of several cyberattacks within the organizations is quite worrisome, highlighting that these attacks are still being perpetrated despite certain ZTA approaches (Table 16).

*Table 16. Prevalence of Different Cyberattacks on Organizations*

| Question | Response Category | *n* | % |
|---|---|---|---|
| **In the past 12 months, has your organization identified any unauthorized access to your computer systems or data?** | Yes | 61 | 44.2 |
| | No | 77 | 55.8 |
| | Yes | 51 | 37 |
| **Has your organization received any external reports (e.g., from customers or partners) of a potential data breach within the past year?** | No | 87 | 63 |
| **Has your organization been notified by law enforcement or regulatory bodies of a potential data breach in the past 12 months?** | Yes | 43 | 31.2 |
| | No | 95 | 68.8 |
| **Has your organization had to restore any systems or data from backups due to a suspected breach in the past 12 months?** | Yes | 39 | 28.3 |
| | No | 99 | 71.7 |
| **Has your organization issued any public statements or notifications regarding a data breach in the past year?** | Yes | 39 | 28.3 |
| | No | 99 | 71.7 |
| **Has your organization provided any financial compensation or credit monitoring services to customers affected by a breach in the past 12 months?** | Yes | 21 | 15.2 |
| | No | 117 | 84.8 |
| **Has your organization made any significant changes to its cybersecurity policies or procedures in the past year (e.g., increased employee training, enhanced security software)?** | Yes | 120 | 87 |
| | No | 18 | 13 |

Table 17 cross tabulates unauthorized access incidents within an organization with context of the use of Cloud Security Posture Management solutions. It portrays the organizations which have experienced unauthorized access incidents, and the status of the CSPM solutions initiated in the organizations. However, despite being insignificant, it is believed that the employment of Cloud Security Posture Management solutions appears to correlate with the lower level of unauthorized access made/committed internally to your computer systems or data inclusive ($\chi2$ =.940, p =.332) based on Pearson Chi-square test of homogeneity. Since the p-value is higher than 0.05, there is no association between recognizing the unauthorized access incidents and applying CSPM. In other words, using CSPM solutions does not make it more probable that the organization will detect unauthorized access incidents according to the data.

It was determined that organizations having no CSPM solutions had a bit more unauthorized access to data and/or systems during the last 12 months (32 people) areas than 29 among those who have been employing CSPM. This analysis indicates that merely adopting CSPM solutions may not greatly improve the detection of access incidents towards illegal access. For effective management of unauthorized access, there is a possibility that organizations will have to adopt more security measures that are not only limited to CSPM (Table 17).

*Table 15. Cross-Tabulation Unauthorized Access Incidents and Use of CSPM*

|  |  | Have you implemented a Cloud Security Posture Management (CSPM) solution to provide visibility and control overshadow IT and unsanctioned cloud services in your Zero Trust Architecture? | | Total | $\chi^2$ | *p* |
|---|---|---|---|---|---|---|
|  |  |  |  |  | Pearson Chi-Square | Asymptotic Significance (2-sided) |
|  |  | Yes | No |  |  |  |
| **Has your organization identified any unauthorized access to your computer systems or data?** | Yes | 29 | 32 | 61 | .940 | .332 |
|  | No | 43 | 34 | 77 |  |  |
|  | Total | 72 | 66 | 138 |  |  |

Figure 13 shows the IBM SPSS code used to carry out this analysis.

```
213  CROSSTABS
214    /TABLES=A24_Internally_Identified BY A21_Use_CSPM
215    /FORMAT=AVALUE TABLES
216    /STATISTICS=CHISQ CC PHI LAMBDA UC RISK
217    /CELLS=COUNT
218    /COUNT ROUND CELL.
```

*Figure 13. IBM SPSS Code Snippet No 2 Crosstabs*

**Research Question 2**

What is the impact of ZTA on cyberattack prevention in organizations?

**Security Posture and Breach Impact: An Organizational Perspective**

The variables A24_Internally_Identified and A25_Externally_Identified are essential to this research question as they show whether an organization has been breached.

*Table 16. Dichotomous Variables A24_Internally_Identified, A25_Externally_Identified*

| Posture Level | Internally Identified Count | Externally Identified Count |
|---|---|---|
| Initial (Low) | 36 | 32 |
| Developing (Moderate) | 25 | 19 |
| Advanced (High) | 0 | 0 |
| Optimized (Very High) | 0 | 0 |
| Total | 61 | 51 |

**Impact Based on Posture Levels**

This data confirms that organizations with an Initial (Low) posture level experience most internally and externally identified breaches. Organizations with a Developing (Moderate) posture level experience fewer breaches overall, but still a significant number.

Below are summarized the findings related to the relationship between an organization's security posture and the frequency of the number of disclosed breaches experienced.

**Initial (Low) Posture Level**

Organizations classified as possessing an "Initial" or "Low" Security Posture are those that, in all probability, have not put security mechanisms of any sort or the security best practices in place. The statement suggests that these organizations suffer from more breaches than most, especially those detected either within the organization by its security personnel or outside the organization by law enforcement and other agencies. Further, this alludes to the fact that the absence of such security frameworks increases the susceptibility of these organizations to attacks.

**Developing (Moderate) Posture Level**

Organizations belonging to the 'developing' or the 'moderate' security level may have commenced introducing certain security procedures but still have not reached the level of a full-blown security policy. Moreover, this statement can be analogically interpreted from the view of many organizations whose users have fewer breaches compared to others having a low posture. Even when there are few improvements made towards cybersecurity, failures are still experienced since there are still risks that can be exploited by malicious parties to execute successful breaches.

**Implications**

The low-posture companies would have more breaches than others, while the developing-moderate ones would face their risks, but not as many. This underlines the urgent need for advanced security solutions, more training, and broad awareness for better results against breaches. Briefly summarized, the outcomes mean that while poorer security postures are the ones that mostly undergo breach incidents, the moderate ones are those that are somewhat less vulnerable. It is, therefore, essential to raise the level of security to a point where the breaches would be controlled.

*Table 17. Summary of All Dichotomous Variables*

| Variable | Yes Count | No Count | % Yes |
|---|---|---|---|
| A24_Internally_Identified | 61 | 77 | 44.20% |
| A25_Externally_Identified | 51 | 87 | 36.96% |
| A26_Law_Enforcement_Identified | 43 | 95 | 31.16% |
| A27_Needed_to_Restore | 39 | 99 | 28.26% |
| A28_Issued_Notifications | 39 | 99 | 28.26% |
| A29_Financially_Compensated | 21 | 117 | 15.22% |
| A30_Policy_Changes | 120 | 18 | 86.96% |

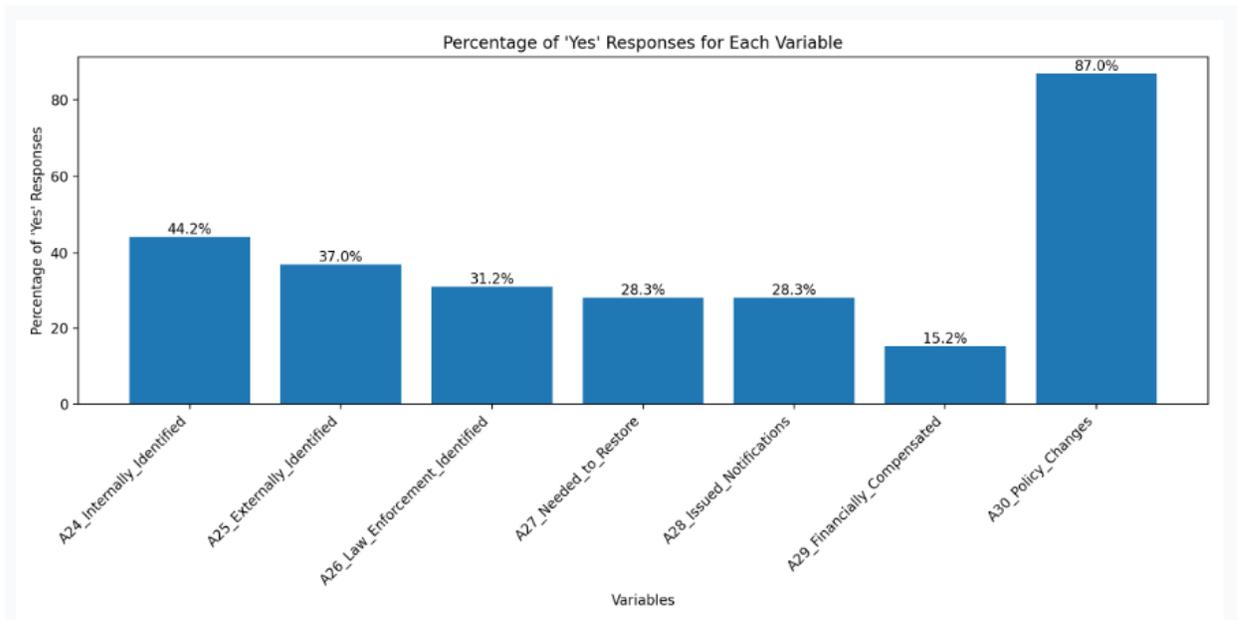

*Figure 14. Percentage of 'Yes' Responses for Each Dichotomous Variable*

**Observations**

Internal identification of breaches (44.20%) is dominant over external identification (36.96%). Though law enforcers themselves identified some of the breaches (31.16%), it is relatively low compared to internal and external identification reasons. Only around 28% of the respondents needed to recover data from backups or issued takedown notifications after a severe breach. Financial compensation was the least common response, with only 15.22% of respondents indicating they had to compensate financially.

Rounding up, policy changes were the single most common reaction upon the pledges with 86.96% of the respondents asserting that policy amendments had been made. After security incidents, it is given that in the majority of cases, organizations do suffer losses, and as a precautionary measure, they change policies to avoid a repeat of the loss incidents.

This indicates an active posture towards security, with institutions admitting that such security measures must evolve with the emergence of the threats used in the breach. Moreover, it indicates that such conduct is a positive trait as organizational maturity regarding cybersecurity is increasing.

**Low Percentages for Other Responses**

The data shows that there are relatively low percentages for responses such as restoration from backups or financial compensation, it could imply a few things:

Severity of Breaches: Sometimes, a breach may not be severe enough to warrant actions such as restoring data from backup. This may mean that such breaches were resolved rapidly or did not involve particularly sensitive data.

Preparedness: Certain organizations seem to prepare themselves so well for breaches that there is hardly any disruption at the institutional level. These organizations may have effective incident response plans to manage breaches.

Responses: Organizations do not respond to some activities or requests because they do not consider their responses or requests rational in view of their status. Thus, for instance, they may concentrate on policy changes and preventive actions instead of financial or recovery interventions.

Overall Implications: The analysis implies that organizations are generally improving their security measures after breaches occur, but the type and effects of those breaches are different.

**Additional Insights**

Anticipating such trends will enable the stakeholders to determine measures to improve the prioritization and performance of incident response activities. It also points out the need for regular practice, education, and funding for cyber security measures to lessen the likelihood of similar breaches happening in the future.

To conclude this section, These findings underline the need to improve the security posture to minimize the probability of a breach. The most urgent need for enhancement in security would, therefore, be organizations with Initial (Low) levels of posture since they tend to bear the highest probability both internally and externally identified for breaches.

Analyzing the survey results, there are some insights on the implementation of ZTA that seeks to enhance cyberattack prevention in organizations.

The study of variables A24 to A30 is most important in terms of appreciating the organization's structures for addressing security breaches. The data orchestrates a sophisticated pattern of cybersecurity and its components' effectiveness.

The higher rate of internal identification (44.20%) compared to external identification (36.96%) suggests that organizations are more effective at detecting breaches through their own monitoring systems.

Law enforcement identification of breaches (31.16%) is less common than both internal and external identification. The low level of such respondents in terms of the law intervention could suggest that organizations are finding sufficient means to respond to many more security attacks by themselves. Nevertheless, it raises the issue of the question concerning the level of the attack that will make an organization think it is necessary to call in the police and whether there are cases which may not have been declared that may be solved through investigation.

The percentage of respondents who experienced the need to draw data from backups was 28.26% and corresponds to the percentage of organizations that informed the clients after a breach.

The low percentages for needing to restore from backups and issuing notifications suggest that many breaches may not have resulted in significant data loss or required extensive communication with stakeholders.

Financial compensation was the least common response, with only 15.22% of respondents indicating they had to be compensated financially. The very low percentage of organizations that had to provide financial compensation indicates that most breaches did not lead to severe financial repercussions.

By far, the most interesting aspect revealed by the research is the percentage of organizations that report policy changes after security breaches. Such changes were reported by 86.96% of the organizations. This overwhelming majority illustrates a strong commitment to learning from breaches and improving security practices, which is a positive sign for overall security posture.

As much as this suggests that the policies are made with an understanding of the rapidly changing environment and the threats it comes with, so has it over the years been evident that there is need for change in the policies conducted. It is noteworthy that the analysis of these factors characterizes the capacity of organization to withstand and respond to cybersecurity issues. It appears that in the context of a particular breach detection effort, internal attempts may be more efficient than external ones; however, there is the implication of complex mechanisms for breach detection and management.

That being said, many organizations have a standard practice, which is after a data breach almost all of them undertake some policy changes. It indicates the existence of a progressive and flexible philosophy in cybersecurity. Such an attitude is significant in times when threat actors change their strategies over time. Instead of treating weakening of one form of security as a temporary setback, awareness of the necessity for considering every security event as an opportunity to strengthen their posture is welcomed.

This study offers a deeper understanding of the reactions of organizations to security breaches and the vulnerability of organizations to breaches. The high percentage of policy changes made after breaches means that the organizations have been able to learn from breaches and have taken steps to level up their security. On the other hand, the percentages for the other responses (for instance, restoration from backups or financial restitution) are quite low, and this may reflect the fact that some breaches are not too critical to warrant these measures being taken, or that corporations have appropriately prepared for management of such breaches with little or no serious effect.

The study reveals several links between the detection and response of the security breach and learning in the organization concerning the methods employed. Even though some aspects, such as the integration of internal and external detection mechanisms, could be improved, the overall trend that may be highlighted is one of increasing sophistication and resilience of organizational cyber defense within the practice. This even goes so far as to say that the high rate of policy modification, in particular, bodes well for the future posture of organization security within an ever-adapting landscape of threats.

Reduced Attack Surface and Limited Lateral Movement: Several respondents in the survey mentioned that ZTA enhances the level of cyber security since both the attack surface and lateral movement within the network are limited. By using techniques such as microsegmentation, network isolation and granular access controls, ZTA reduces the number of lateral movements and access to more resources by attackers.

Continuous Monitoring and Analytics: The survey findings show that the organizations are widely using SIEM, UEBA, continuous monitoring, and Threat Intelligence Platforms as part of ZTA as a measure to prevent cyber threats. These controls provide the possibility of performing monitoring, analysis, and detection of threats in real time enabling the organization to respond to the threats as soon as they arise.

Visibility and Control over User Access: This statement received positive feedback from several respondents given that ZTA improves an organization's ability to avert cyberattacks as it provides better visibility and control over user access. Organizations keep an eye on users' activities over time to identify and help address threats that may result from hacked accounts or internal risky behaviors.

Data Protection and Encryption: The survey results stressed the need and presence of certain data loss prevention measures such as encryption, DLP, and secure file exchange in ZTA suitable environment. These measures aim at reducing the chances of data breaches where unauthorized individuals gain access to sensitive information, thus minimizing the success extent of cyberattacks.

Limiting Spread of Threats: However, some of the respondents observed that ZTA assists in networking by curbing the spread of threats by way of resource segregation and isolation. Even when an attacker succeeds in penetrating the network, ZTA implements segmentation and population of various access controls which restrict movement within the network thereby reducing the full effect of an attack.

Strengthened Security Posture: Some respondents included other factors of reducing risks' concerns by stating that their ZTA strategy has improved their overall security posture due to reduced attack surface and enforced adequate security measures on several domains including ID and A, endpoint and network segmentation, and data protection.

It is clear that ZTA is not the primary custodian against cyberattacks, but is a significant contributor in making successful rather efficient measures aimed at reducing cyber security risks by means of a rich layers of technical controls and security practices.

## 5.5     Dataset Analysis Supporting RQ3

The third research question was aimed at exploring the current best practices in the industry concerning the implementation of ZTA architecture. Most (n =119, 86.2%) of the participants reported using password-less authentication such as Microsoft and Google authenticators. Almost all (n =136, 98.6%) reported about use of MFA. More than half (52.2%) of the respondents indicated having implemented CSPM solutions. The number and percentages of respondents who indicated using the various different ZTA practices are presented in Table 20.

*Table 18. Participants' Use of Different Technical Controls*

| Question | Response Category | $n$ | % |
| --- | --- | --- | --- |
| **Which form of password-less authentication is enabled for your users?** | Phone call | 16 | 11.6% |
|  | Text message | 41 | 29.7% |
|  | OATH token | 32 | 23.2% |
|  | Authenticator-Microsoft/Google/Duo | 119 | 86.2% |
|  | Email | 6 | 4.3% |
| **Have you enabled multifactor authentication (MFA) for users?** | Yes | 136 | 98.6% |
|  | No | 2 | 1.4% |
| **Do you use a Cloud Access Security Broker (CASB) solution to provide visibility and control over cloud** | Yes | 82 | 59.4% |
|  | No | 56 | 40.6% |
| **applications and services in your Zero Trust environment?** | | | |
| **Have you implemented a Cloud Security Posture Management (CSPM) solution to provide visibility and control overshadow IT and unsanctioned cloud services in your Zero Trust Architecture?** | Yes | 72 | 52.2% |
| | No | 66 | 47.8% |
| **Have you deployed a Network Detection and Response (NDR) solution to provide visibility and analytics into network traffic and behavior as part of your Zero Trust implementation?** | Yes | 68 | 49.3% |
| | No | 70 | 50.7% |
| **Have you integrated a Virtual Desktop Infrastructure (VDI) or Remote Desktop Services (RDS) solution to enable secure access to applications and data in your Zero Trust Architecture?** | Yes | 86 | 62.3% |
| | No | 52 | 37.7% |

Figure 15 represents the IBM SPSS code used in conducting such an analysis for the mentioned variables.

```
154  FREQUENCIES VARIABLES=A18_1_Call A18_2_Text A18_3_Token A18_4_Auth A18_5_Email A19_Use_MFA
155      A20_Use_CASB A21_Use_CSPM A22_Use_NDR A23_Use_VDI
156    /ORDER=ANALYSIS.
```

*Figure 15. IBM SPSS Code Snippet No 3 Frequencies*

In this analysis section, the answers to variables A19 through A30 were Yes or No. These dichotomous variables laid the groundwork for determining the relationship between technical controls and cyberattack susceptibility. Figure 16 helps to visualize the content described earlier in Table 20.

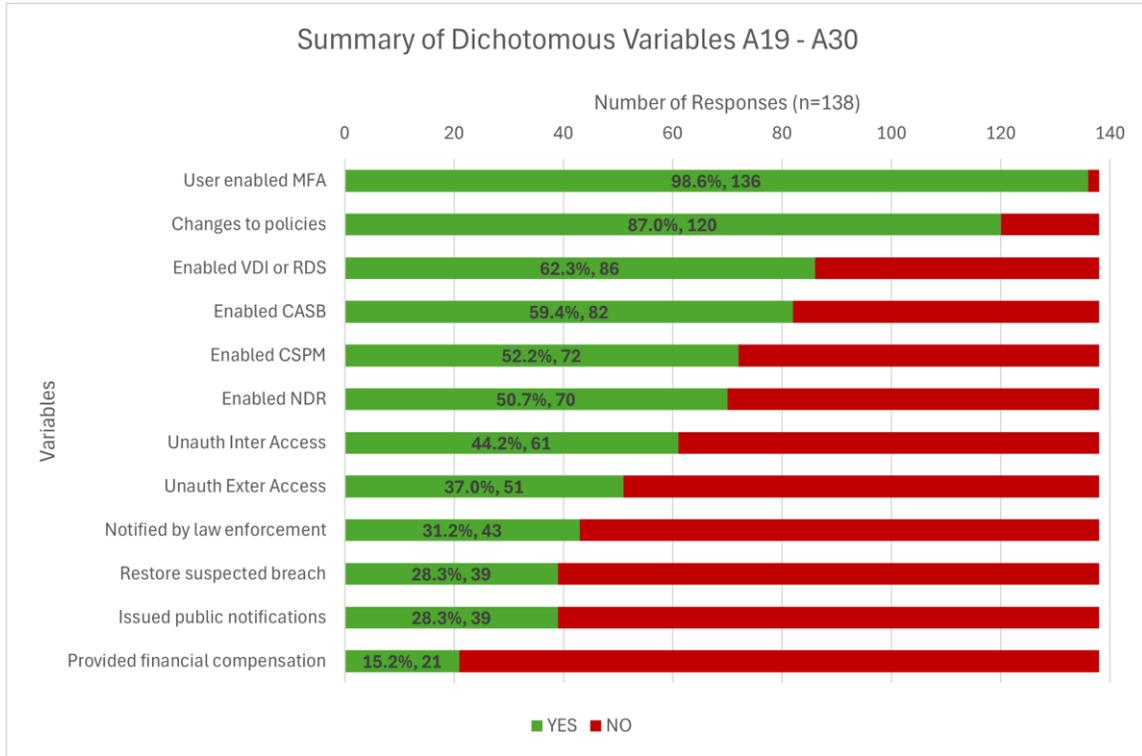

*Figure 16. Summary of Dichotomous Variables A19-A30*

**Research Question 3**

**What are the industry best practices for implementing a ZTA?**

**Zero Trust Architecture: Implementation Best Practices**

Based on the survey responses, the following are some best practices for implementing a Zero Trust Architecture that the survey corroborated. These technical controls and best practices feature significant enhancements in security posture and reduced risk of successful breaches in an increasingly complex threat landscape.

This research identifies the foundational controls for Initial and Developing. It also concludes that the technical controls associated with these posture levels are not sufficient to deter breaches.

**Foundational Elements of Zero Trust Architecture**

The implementation of Zero Trust Architecture has many facets since several components are considered critical. The foundation of ZTA is the Encryption of Data at Rest and in Transit; hence, it extends the core protection of sensitive information throughout all network segments.

SDPs are critical in building dynamic identity and application-centric perimeters that allow secure, location-agnostic resource access. A sound, pragmatic approach supported by comprehensive policy and risk management schemes forms the basis of all security in the ZTA environment.

### Least Privileged Access

Various respondents acknowledged the necessity of observing the least privileged access principle in their ZTA framework. This principle grants users and entities the least access necessary to perform a task, thus limiting the degree of exposure and the impact of such accounts if compromised.

### Comprehensive Identity and Access Management

The survey results reveal how essential comprehensive identity and access management is, including multi-factor authentication, privileged access management, Role-Based Access Control, and Single Sign-On. The understanding that these controls are the foundations for the best practice of the Zero Trust Architecture is profound.

### Dynamic Security Controls and Automation

Automation of policy enforcement and remediation systems is vital given the maintenance of security posture, ensuring quick response to potential threat scenarios and policy violations. ZTA heavily relies on microsegmentation and isolation techniques at the network architecture level to restrict lateral movement and keep a potential breach contained.

**Data and Device Management in ZTA**

Depending on the data's sensitivity, Data Classification and Labeling mechanisms ensure appropriate access control and protection. They also ensure device security through regular posture and compliance checks, blocking network access for non-compliant devices.

**Endpoint Security and Compliance**

Maximizing endpoint security and compliance is a standard best practice among the responders. This entails applications such as Endpoint Detection and Response, Unified Endpoint Management, Mobile Device Management, and Device Posture and Compliance Checks to ensure only clean devices are used within the ZTA circulation resources.

**Microsegmentation and Network Isolation**

Employing Microsegmentation and Network Isolation techniques such as software-defined perimeters, Zero Trust Network Access, virtual LANs, and Network Access Control is known to mitigate the lateral spread and potential challenges within the ZTA environment.

**Data Protection and Encryption**

The use of encryption, proper classification of information, implementation of data loss prevention, and usage of secure file-sharing systems are repeatedly outlined as best practices for ZTA deployment, making data secure and reducing the effects of successful compromises.

**Continuous Monitoring and Authentication**

Continuous Monitoring and Anomaly Detection Systems provide network surveillance activities in real-time and can respond promptly to potential security incidents. A Single Sign-On solution smooths the authentication processes without compromising on the most robust security controls.

### Continuous Monitoring and Analytics

Best practices utilized within the ZTA include security information and event management, user and entity behavior analytics (UEBA), real-time monitoring, and Threat Intelligence Platforms to achieve real-time threat detection, analysis, and response.

### Visibility and Reporting in ZTA

Through comprehensive reporting mechanisms and executive dashboards, stakeholders gain visibility into the organizational security posture and emerging issues.

### Breach Response and Incident Management

Breach response data analysis should be implemented as a predictive measure to tune the ZTA environment for potential threats. While only 44.20% of breaches were internally identified, an organization must improve its internal detection capabilities. At the same time, processes should be developed to respond to external notifications of a breach, as identified in 36.96% of cases.

### Post-Breach Practices and Policy Updates

The most crucial factor is robust backup and restoration processes, as 28.26% of responses related to breaches involved restoring data. Other major factors involve well-defined notification procedures, as incident response plans should be used to notify externally in 28.26% of these cases.

Although only 15.22% of the cases had financial compensation imposed, prevention should be emphasized since financial and reputational risks are considerably high. Besides, an incredibly high % of post-breach policy changes, 86.96%, underlines the step-up in improving security practices.

### Ongoing Assessment and Adaptation

Regular security assessments should be carried out, and all vulnerabilities detected. Given the quick adaptation of technology, any deployment will need frequent updates. However, following the principles of ZTA, the risks associated with this can be mitigated.

### Multi-Factor Authentication in ZTA

While no longer in the top-ranked controls, MFA is still a keystone in ZTA, and organizations should implement this on all systems and applications. It addresses technological, policy, and human factors in creating a comprehensive ZTA approach that adapts to ever-evolving threats.

### Integration and Automation

Many respondents acknowledged the need to consolidate multiple security products and workflows, including policy application, compliance management, and incident management. Therefore, they agreed that it makes managing the ZTA environment more efficient.

### Security Awareness and Training

The introduction of employee Security Awareness and Training Programs is viewed as one of the essential best practices that support the proper implementation and functioning of the ZTA framework and promote a proactive security environment among the employees.

## Cybersecurity Maturity: Advanced Elements for Breach Risk Reduction

Organizational security should be focused on several issues that enable them to advance beyond Initial and Developing to minimize the chances of a breach. First, organizations need to shift from ad hoc security measures to well-defined, structured security practices applied consistently throughout the organization. This formalization of processes and procedures should go hand in hand with a shift toward a more proactive approach to security. That means organizations can no longer afford to react to threats; they need to build in proactive security controls and risk management strategies to detect threats before they occur.

Similarly, developing a robust culture of security awareness among employees can further help advance security maturity. This includes regular training and education on cybersecurity best practices, spreading cybersecurity scenarios, and identifying potential threats. Organizations must also have appropriate resources for cybersecurity personnel and technology. This may involve special hires in cybersecurity or investing in special training for the existing IT staff. That means, as an organization moves further along with its security maturity, that set of core security controls will evolve to robust, well-defined security controls across every aspect of its operations. These security controls would be periodically reviewed for updates and the ever-changing emerging threats to stay ahead of the cybersecurity challenges that are to come. Meanwhile, the organization should identify and update detailed plans on how it can rapidly-detect, contain, and mitigate a security incident when it does happen.

Organizations should implement advanced security technologies and threat intelligence tools. Such technologies go a long way in enhancing the security posture of an organization. In addition, the integration of security practices into each fabric of an organization's operations should not be separated.

Organizations need to establish thorough security for all their systems, networks, and data repositories. This would involve a thorough search for and fixing of all weak spots in their cyber defenses to ensure that each part of their IT infrastructure is given equal protection with the same high level of security standards in each part of the organization.

Examples include organizations that move from Initial/Developing security to Advanced and then Optimized approaches. Such improvements better position organizations to identify cyber threats in a timely way, preventing any form of successful threat and limiting the damage in case such a breach occurs.

With this development, organizations reduce the risk of data breaches, financial loss, and other reputational consequences. This advanced level of protection is highly significant in the digital world, which, with time, becomes increasingly hazardous due to the continuous growth in the sophistication and complexity of cyber threats. By developing their security maturity, organizations protect not only their assets but also strengthen their resilience against complex and constantly changing landscapes of cybersecurity challenges.

**Supportive Evidence by External Research**

NIST is a federal agency under the U.S. Department of Commerce that fosters innovation and improves industrial competitiveness. Accordingly, NIST's mission has promoted the development of measurement science, standards, and technology for economic security and quality of life. It accomplishes this mission through four programs in laboratory research, development of technical standards and guidelines, providing measurement services, and collaboration efforts to drive innovation by partners in industry and academia. Work at NIST has touched a wide swath of sectors: cybersecurity, healthcare, manufacturing, and public safety, to name a few.

The NIST Cybersecurity Framework has gained widespread recognition as one of the leading standards for managing critical infrastructure cybersecurity risks. It helps set the scene for this study in its brief, state-of-the-art developments in cybersecurity risk management and their effect on business processes. The findings of the NIST study are important, especially in light of the new CSF 2.0, to imply stakeholder ideas and realities into what actually happens in this changing landscape. These findings speak directly to the research questions explored in the Prescriptive Zero Trust study about effective cybersecurity practices in the setting of modern organizations and their unique risk management strategies.

**Summary of External Findings**

On February 26, 2024, NIST published an updated version of the Cybersecurity Framework 2.0, ten years after the original publication was issued in 2014. It is designed to help organizations of all kinds understand the vivid landscape of cybersecurity and to better handle risks. The updated framework has five core functions: Identification, Protection, Detection, and Response and Recovery. Some of the major updates that this document emphasizes are in relation to 'cybersecurity governance,' a priority that places leadership at the center for minimizing cyber risks.

CSF 2.0 also includes expanded guidance on supply-chain risk mitigation in light of growing awareness of system interconnectedness and implied vulnerabilities. The scope has been expanded from cybersecurity-focused to other critical risks, such as privacy and fraud, resulting from an ever-sophisticated digital threat. Implementation Tiers and Profiles have also been updated for better alignment with organizations, permitting more effective integration of the Framework with specific business requirements and appetite for risk.

This update emphasizes measuring program effectiveness and provides updated guidance on determining cybersecurity program maturity levels. The focus on quantified results is one of the most pervasive challenges faced in the discipline: how to measure the effectiveness of security. Finally, CSF 2.0 enhances compatibility and alignment with other cybersecurity frameworks and standards to drive a more integrated approach to risk management across a diverse set of compliance requirements.

**Comparison with Research Findings**

Its results are closely aligned with the research on the Prescriptive Zero Trust study and technical measures, as it also follows a ***theme of multi-layering in cybersecurity*** due to no one security control being able to completely protect from threats that evolve continuously.

Accordingly, in the research on Prescriptive Zero-Trust technical controls, ***no single control is found to be highly correlated in preventing a breach.*** This would mean that it follows the point from NIST CSF 2.0, emphasizing an all-inclusive and integrated approach to cybersecurity. The streamlined core functions of the CSF are: Identify, Protect, Detect, Respond, and Recover. That again reflects this multi-layering approach and leads to a similar conclusion that different security controls take part in it.

The findings underlined how effective the postures of Adaptive Risk-Based Authentication were, which also corresponded to an increased focus on dynamic risk management by the NIST CSF 2.0. The research and update of CSF recognize the need for flexible, context-aware security measures that adjust to changing threat landscapes.

Controls identified as probably effective throughout the study included Privileged Account Management, Data Loss Prevention, and Centralized Logging and Auditing. This focus on enhanced governance and supply chain risk management is consistent with the emphasis of CSF 2.0 on managing access, protecting data, and maintaining visibility across complex, interconnected systems.

That research unveiled some interesting paradoxes between perceived importance and the effectiveness of differing controls in preventing breaches. This goes to the heart of what CSF 2.0 is trying to do: measure program effectiveness and determine maturity levels. Both point to the requirement for evidence-based cybersecurity, beyond traditional assumptions to data-driven strategies.

This conclusion that security must be assessed and adjusted constantly is supplemented by the emphasis the NIST CSF 2.0 places on cybersecurity aligned with business objectives. Both the research and the update of the CSF underline that ***cybersecurity strategies should be adapted*** to emerging threats and organizational needs.

Finally, this call for more research on the effectiveness of security controls within various organizational contexts aligns well with the recognition in CSF 2.0 of the need for adaptable implementation tiers and profiles. Both pieces of research note that cybersecurity strategies have to be very adaptable based on organizational characteristics and risk profiles.

While this research, in summary, focuses more in-depth on technical controls within the ZTNA framework and the NIST CSF 2.0 has a broader organizational framework for cybersecurity, it would seem that these key points do converge on a commonality: ***the need for multi-layered adaptive continuously evaluated cybersecurity strategies*** meeting organizational objectives, references to an evolving threat landscape.

**Implications for the Study**

The findings from the NIST CSF 2.0 have several key implications for cybersecurity research. First, the framework's framing around aligning cybersecurity with business objectives suggests, as a first avenue of possible research, one that examines how best CSF 2.0 can be implemented in support of an organization's strategic priorities. This may involve examinations of how cybersecurity risk management is integrated into broader enterprise risk management practices.

As such, attention within the existing context of changing risks – such as those posed by cloud computing or the Internet of Things – should focus on the position within CSF 2.0 that requires research in emerging technologies. This includes, but is not limited to, what changes an organization should make to the framework so that new technologies do not pose any new security risks.

Including CSF 2.0 in relation to other standards and standards consolidations fully provides opportunities for research that will advance a more cross-cutting approach to the management of cybersecurity risk. Investigators seek to find how an organization can satisfactorily implement adequate framework integrations within its entire security program in compliance with regulatory and industry standards.

Last but not least, the nature of the framework, as directed toward assessing the performance and maturity of cybersecurity programs, would invite studies concerned with metrics and methodologies that could help organizations understand the impact and effectiveness of their implementation of CSF 2.0. Examples can include proposing new assessment tools or improving existing tools.

## 5.6    Summary

The NIST Cybersecurity Framework (CSF) 2.0 also provides equal corroborative evidence concerning the Prescriptive Zero Trust study on cybersecurity, underlining themes of governance and adaptability to new emerging threats and providing measurable outcomes. Given the framework's broad adoption and continuous refinement through stakeholder contributions, it serves as a solid foundation for validating and informing cybersecurity research.

The corroborating evidence drawn from NIST CSF 2.0 provides a deeper focus on why it is important to study cybersecurity as one aspect of organizational strategy and not an end to itself as a technical issue. It also gives dynamic, adaptable approaches to cybersecurity that keep pace with evolving threats and technologies. CSF 2.0 is highly valuable to researchers, as it enables them to explore how best organizations could protect their digital assets within an increasingly complex and interconnected world by providing a thorough and flexible framework for managing cybersecurity risks.

# CHAPTER 6: FOUR-TIERED TECHNICAL CONTROLS MODEL

## 6.1    A Four-Tiered Posture Model

The research maintains that implementing a Zero Trust Architecture follows the paradigm of a progressive maturity model with four posture levels: Initial (Low), Developing (Moderate), Advanced (High), and Optimized (Very High). Each of these postures stands for a discrete set of cybersecurity capabilities and controls an organization must master toward the goal of advancing its Zero Trust transformation from basic device management and network segregation to advanced identity verification and threat detection.

The hierarchical model builds upon itself with each successive level incorporating and complementing the security posture of the preceding stages. While the Initial posture establishes the basic device attributes, like basic device management and processes in vendor assessment, the journey culminates in the Optimized posture with multi-factor authentication, Endpoint Detection and Response (EDR), and continuous monitoring systems implementing the two core Zero Trust principles: assume breach and maintain least privilege access. Not all controls are required for all organizations, and individual sectors should tailor the model to meet their specific requirements.

## 6.2    Understanding the Posture Levels

For organizations to understand how to reach a Zero Trust security posture, they must have an orderly framework that lays out not just the steps to get there but also the logic behind those steps and the dependencies among them. Posture levels, namely Initial, Developing, Advanced, and Optimized, are the categories in which we place controls and capabilities. The posture-level description of those controls and capabilities helps organizations identify the basic foundational practices for hitting the marks in the Initial level and then allows them to see how to make a sequence of moves that get them to the next level and then on to the next and the next.

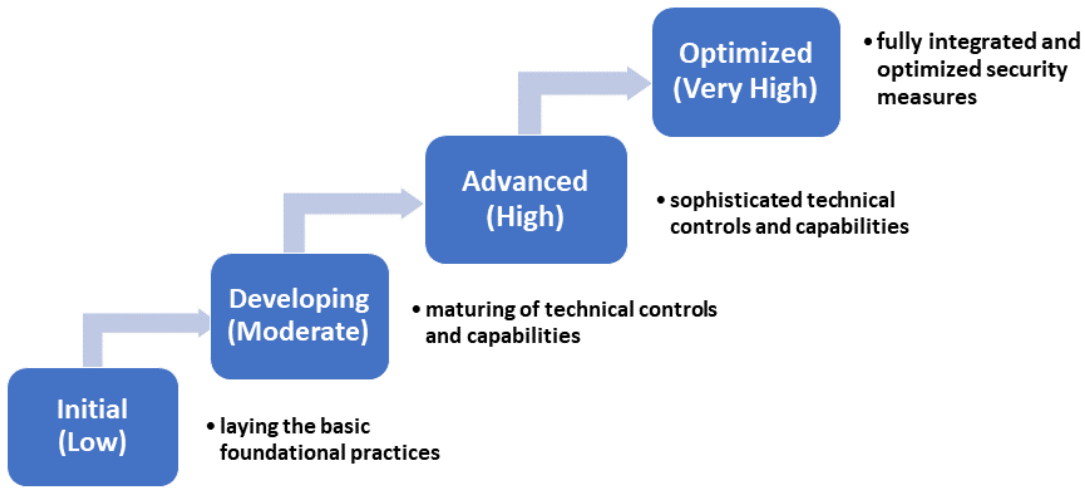

*Figure 17. Zero Trust Security Posture Levels*

When organizations reach the Developing level, they are moving beyond the Initial level and working to mature their security controls and practices. Developing-level organizations have much of Zero Trust's foundational security in place. They have implemented security architecture segmented into foundations (devices and users), a security perimeter, and an interior fortress. Most everything that is attempted in these environments is subject to multi-factor authentication and abundant attempts at social engineering around the MFA that never work. These organizations use advanced security for endpoint protection and layered defenses both inside and outside the security perimeter.

The conceptual framework serves as a solid foundation for the ZTA that not only yields better security outcomes but also improves the efficiency of maintenance and operations within the Zero Trust Architecture for today and for the future. The framework itself reflects the holistic nature of the organizational approach that is absolutely necessary to pursue ZTA across common information environments and various technical domains. The ZTA is capable of doing e2e at a better security level for access and with the boost of fundamental security principles for maintaining information and the robustness of processes against threats and hazardous events.

The Least Privilege Access Control (LPAC) principle, for instance, might be judged on how well it implements default deny permissions and achieves granular access control. The LPAC principle gets point values assigned to it based on how significant we think it is for a robust ZT posture. The total point value for all the criteria is what we use to judge the posture level of the organization. This is a multi-method assessment, and most of the methods we use are very similar to the ones we used in the last two reports. The first method is a questionnaire.

The main advantage of this assessment is that it provides a measure of network maturity. This measure can be easily compared between organizations and helps track trends over time, helping organizations measure their efforts to achieve ZT. Additionally, the review focuses on ZT's principles to ensure compatibility with current security measures and promote rapid methods of preventing cyberattacks.

However, the assessment also has limitations. Interpretation of questionnaire responses and data analysis might involve a degree of subjectivity requiring expert judgment. Furthermore, the effectiveness of the assessment hinges on the chosen maturity model and its alignment with ZT principles. Finally, the assessment necessitates continuous refinement to reflect the ever-evolving cybersecurity landscape and emerging threats.

This assessment presents a novel methodology for measuring cybersecurity maturity, specifically Zero Trust principles. By quantifying the implementation of ZT principles, organizations gain valuable insights into their security posture and can prioritize improvement efforts to prevent cyber attacks. Remember, this assessment serves as a starting point, and continuous adaptation is essential to maintain its effectiveness in the face of evolving threats.

**Dataset Analysis Supporting a Four-Tiered Posture Model**

A unique approach is not just beneficial but essential when it comes to cybersecurity maturity models. Therefore, the nature of the data and the context of cybersecurity maturity brought about a method based on the mean and standard deviation of the "Average Ranking" in this approach. By handling outliers better and considering the overall distribution of the data, this method is a significant step forward in identifying realistic posture levels.

The analysis begins by calculating each respondent's Average Ranking for each variable. The resultant stack rank is shown in Table 21.

*Table 19. Stack Rank Percentage and Count of Respondents*

| Variable | Name | Average Ranking | Overall Rank | Posture Description |
|---|---|---|---|---|
| Encryption of Data at Rest and in Transit | A13_1 | 1.891 | 1 | Initial (Low) |
| Software-defined perimeter (SDP) | A10_1 | 2.225 | 2 | Initial (Low) |
| Policy and Risk Management Frameworks | A16_1 | 2.290 | 3 | Initial (Low) |
| Automated Policy Enforcement and Remediation | A15_4 | 2.428 | 4 | Initial (Low) |
| Microsegmentation and Network Isolation | A11_1 | 2.435 | 5 | Initial (Low) |
| Data Classification and Labeling | A12_1 | 2.522 | 6 | Initial (Low) |
| Device Posture and Compliance Checks | A9_3 | 2.536 | 7 | Initial (Low) |
| Continuous Monitoring and Anomaly Detection | A15_5 | 2.587 | 8 | Initial (Low) |
| Identity Federation and Directory Services | A7_5 | 2.609 | 9 | Developing (Moderate) |
| Reporting and Executive Dashboards | A17_4 | 2.616 | 10 | Developing (Moderate) |
| Zero Trust Network Access (ZTNA) | A10_2 | 2.674 | 11 | Developing (Moderate) |
| API Gateways and Web Application Firewalls | A12_2 | 2.681 | 12 | Developing (Moderate) |
| Adaptive Risk-Based Authentication | A7_3 | 2.746 | 13 | Developing (Moderate) |
| Mobile Device Management (MDM) | A9_2 | 2.783 | 14.5 | Developing (Moderate) |
| Virtual LANs and Microsegmentation | A11_2 | 2.783 | 14.5 | Developing (Moderate) |
| Attribute-Based Access Control (ABA) | A8_3 | 2.797 | 16 | Developing (Moderate) |
| Compliance Monitoring and Reporting | A16_2 | 2.804 | 17 | Developing (Moderate) |
| Unified Endpoint Management (UEM) | A9_1 | 2.826 | 18.5 | Developing (Moderate) |

| Network Access Control 2 (NAC2) | A11_3 | 2.826 | 18.5 | Developing (Moderate) |
| Data Loss Prevention for Cloud Storage (DLPCS) | A14_1 | 2.841 | 20.5 | Developing (Moderate) |
| Security Awareness and Training Programs | A17_1 | 2.841 | 20.5 | Developing (Moderate) |
| Incident Response and Disaster Recovery Planning | A17_3 | 2.884 | 22 | Developing (Moderate) |
| Data Activity Monitoring and Analytics | A14_5 | 2.899 | 23 | Developing (Moderate) |
| Rights Management Services (RMS) | A13_4 | 2.913 | 24 | Developing (Moderate) |
| Privileged Access Management (PAM) | A8_4 | 2.935 | 25 | Developing (Moderate) |
| Role-Based Access Control (RBA) | A8_2 | 2.964 | 26 | Developing (Moderate) |
| Multi-Factor Authentication (MFA) | A7_1 | 2.971 | 27 | Developing (Moderate) |
| Security Orchestration Automation and Response (SOAR) | A14_4 | 2.986 | 28 | Developing (Moderate) |
| Single Sign-On (SSO) | A7_2 | 2.993 | 29 | Developing (Moderate) |
| User and Account Lifecycle Management | A8_5 | 3.000 | 30 | Advanced (High) |
| Secure Remote Access (VPN VDI RDP) | A12_4 | 3.036 | 31 | Advanced (High) |
| User and Entity Behavior Analytics (UEBA) | A14_3 | 3.043 | 32 | Advanced (High) |
| Integrated Dashboards and Reporting | A16_3 | 3.058 | 33 | Advanced (High) |
| Secure Boot and Hardware-Based Integrity | A9_4 | 3.065 | 34 | Advanced (High) |
| Network Traffic Analysis and Anomaly Detection | A12_3 | 3.094 | 35.5 | Advanced (High) |
| Change Management and Configuration Control | A17_2 | 3.094 | 35.5 | Advanced (High) |
| Application Control and Whitelisting | A10_3 | 3.145 | 37 | Advanced (High) |
| Digital Rights Management (DRM) | A13_2 | 3.181 | 38 | Advanced (High) |
| Centralized Logging and Auditing | A15_3 | 3.225 | 39 | Advanced (High) |
| Security Information and Event Management (SIEM) | A14_2 | 3.232 | 40 | Advanced (High) |
| Data Access Control and Granular Policies | A13_3 | 3.246 | 41 | Advanced (High) |
| Device Isolation and Quarantine | A10_4 | 3.254 | 42 | Advanced (High) |
| Security Analytics and Machine Learning | A15_2 | 3.261 | 43 | Advanced (High) |
| Endpoint Detection and Response (EDR) | A8_1 | 3.304 | 44 | Advanced (High) |
| Vendor and Third-Party Risk Assessments | A16_4 | 3.406 | 45 | Optimized (Very High) |
| Secure Web Gateways (SWG) | A11_4 | 3.435 | 46 | Optimized (Very High) |
| Automated Compliance Checks and Controls | A16_5 | 3.442 | 47 | Optimized (Very High) |
| Threat Intelligence Platforms (TIPS) | A15_1 | 3.500 | 48 | Optimized (Very High) |
| Cloud Access Security Brokers (CASB) | A11_5 | 3.522 | 49 | Optimized (Very High) |

| Device Endpoint Patching and Updates | A17_5 | 3.565 | 50 | Optimized (Very High) |
|---|---|---|---|---|
| Data Loss Prevention (DLP) | A12_5 | 3.667 | 51 | Optimized (Very High) |
| Biometric Authentication | A7_4 | 3.681 | 52 | Optimized (Very High) |
| Network Access Control 1 (NAC1) | A10_5 | 3.703 | 53 | Optimized (Very High) |
| Secure File Sharing and Collaboration | A13_5 | 3.768 | 54 | Optimized (Very High) |
| Micro-Agent or Agentless Endpoint Security | A9_5 | 3.790 | 55 | Optimized (Very High) |

The "Logic and Analysis" section substantiates the calculations, yielding a 3.0 median and a standard deviation .40761. These statistics form the basis for classifying the posture levels.

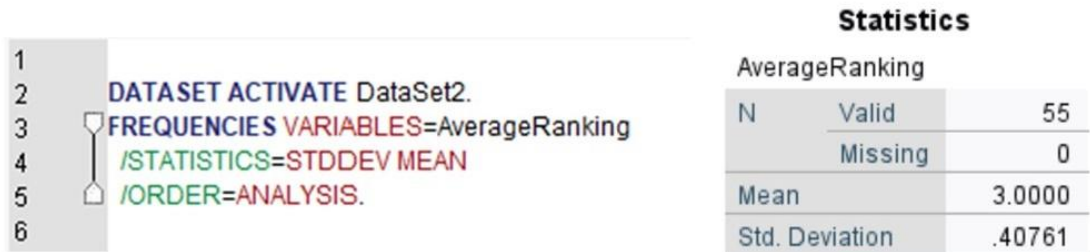

*Figure 18. IBM SPSS Code Snippet to Generate Mean & Std Dev*

These statistics were clearly and systematically established to determine cut-offs for four posture levels: Optimized (Very High), Advanced (High), Developing (Moderate), and Initial (Low). The Optimized level encompasses those variables whose percentages are greater than or equal to the mean of average percentage and one and half times the standard deviation. The Advanced level comprises those variables that are greater than or equal to the mean and half times the standard deviation and the mean. The Developing level incorporates the variables whose values are in excess of the mean but are less than or equal to even half of the standard deviation subtracted from the mean. Lastly, the Initial level incorporates the variables obtained from the mean subtracting 0.5 times the standard deviation up to the mean value.

The application of thresholds to the data provided results that are not only interesting but also practical. The study has revealed 11 variables grouped as Optimized which demonstrated astonishing results well above average. Within the Advanced category, there are 15 variables that exceed the average results but are still more than those within the Optimized group. The Developing category, which is biggest in size, encompasses 21 variables which tend to be concentrated around the average performance level. In the last category, which is the Initial, there are 8 variables that show potential areas that need the most improvement.

It is this advantage that most classification methods have over simpler methods such as servant methods containing quartiles. It adds a degree of complexity which captures the extent graphs allow for directional relationship between the measures. With regards to quartile based method, its advantage is that there are very few extreme cases and therefore the levels are very even when the ratio is assessed. As it enables such risks, variable and extreme outlier biases are restrained, meaning there is less balance in the classification one wishes to achieve. In addition, it makes it possible to have such distribution between the groups which are generally not the case in real practice.

Figure 19 and Figure 20 visualize the findings. Figure 13 is a scatter plot that indicates the distribution of the Average Rankings relating to the classification thresholds as marked. This chart shows the distribution of the variables across the various posture levels.

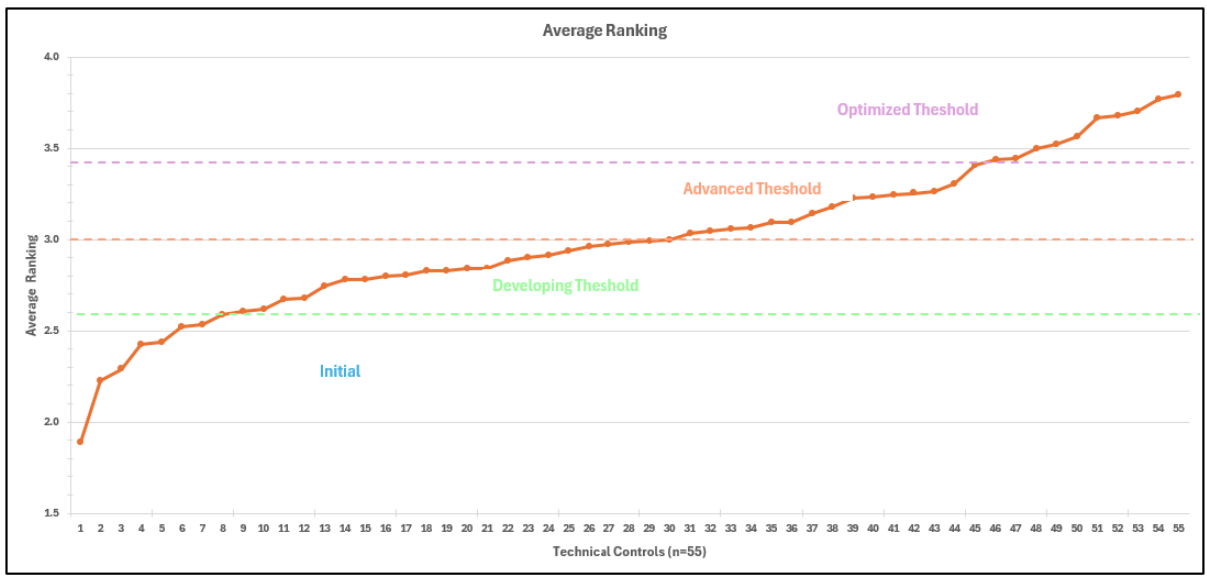

*Figure 19. Scatter Plot of Distribution and Threshold Levels*

Figure 20 includes a bar chart that displays the number of variables in each posture level, providing a quick overview of the distribution.

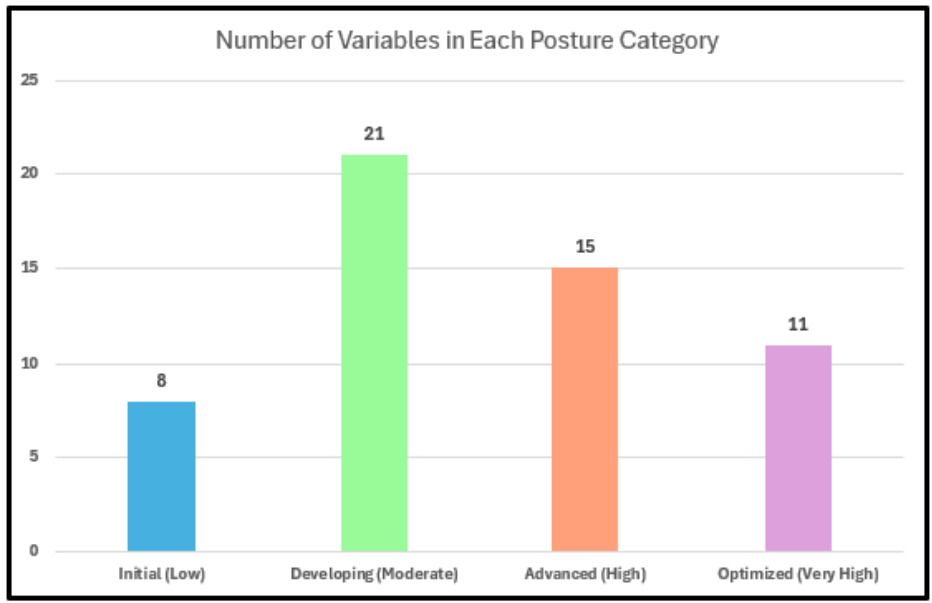

*Figure 20. Number of Variables in Each Posture Level Category*

The approach uses the average (mean) value of all the average rankings as a central point. Then, the standard deviation, which measures how spread out the numbers are from this average, is used to set classification thresholds. This technique is effective in identifying the frequency of very high-performing and low-performing variables; it is also reliable in cases when the dataset contains outliers.

This classification methodology, based on the variables´ mean and standard deviation, also allows for more advanced and statistically meaningful assessment and understanding of the variables' developmental stages. This process detects frequently identified variables, indicating the most widely deployed technical controls. Conversely, the least frequently identified controls can be assumed to be higher-level maturity variables. These insights are in great demand by organizations that want to improve their cybersecurity posture since they offer benchmarking to help them know where to put most of the effort.

*Initial (Low) Posture*

This group outlines the minimum cyber security attributes necessary to begin on a Zero Trust transformation journey. Basic device management, endpoint management systems, secure booting and minimal extreme device and network segregation such as VLANs are enough for simple devices in a normal state.

Technological capabilities like DLP for the cloud, threat intelligence integration, Security Analytics and Machine Learning, Digital Rights Management, and file sharing are advanced components that go well beyond risk or data protection as they enable a more nuanced approach to risk management, though this may at first be disparate in nature. Vendor risk assessment processes, functional reporting/dashboards and change control processes comprise basic operational disciplines.

Even though these factors display an early stage of low Zero Trust posture, they still constitute the basic cyber security characteristics that organizations can progressively enhance over time to advance the Zero Trust posture Implementation over the child phase.

*Developing (Moderate) Posture*

This group represents core operational processes, procedures, and cultural aspects that enable and support advanced technical controls. Secure remote access, like VPN, complements the functionalities of ZTNA. CASB extends the visibility and control to cloud environments.

The governance guardrails are made up of policy/risk frameworks, while compliance monitoring/reporting puts the aforementioned into practice. Automation of policy enforcement together with security awareness training fosters an environment of continuous compliance. Cyber resilience is maintained through regular patching and updates, along with appropriate incident response and recovery processes. These operational elements in place successfully form an environment for Zero Trust implementation.

*Advanced (High) Posture*

Based on what appears to be a strong identity, endpoint and data build-up, this group looks to implement risks mitigation strategies by controlling access precisely as necessary, constructing access control granularity. Microsegmentation coupled with RBAC, PAM, and attribute/context-based policies restrict lateral movement and enforce least privilege. Comprehensive validation checks are undertaken to control network access also further restricting access.

MDM extends endpoint hardening to mobile devices, while application control and API gateways reduce application/API vectored risks. Using a virtual desktop infrastructure application can be used securely with data. Just like SDN, the software-defined perimeter enables dark network microsegmentation, which facilitates the implementation of ZTNA. Taken together, these interlocking capabilities enhance the ability to maintain a posture of tight, well defined, continuously validated and highly controlled access.

*Optimized (Very High) Posture*

This group carries the core building blocks and the fundamental capabilities of a mature implementation of the Zero Trust Architecture. Factors like multi-factor authentication are the best way to ensure identity risks, whereby responding to incidents involving compromised endpoint with analytics related to endpoint detection and the response EDR systems can support 'identity to protect' model. Zone of trust access network and Data classification/labeling are core tenets of any Zero Trust principle regarding least privilege access and protection of data assets.

Certain facilities, for example, encryption of information, continual surveillance, SIEM/ UEBA, etc., also act as force multipliers, enabling organizations to maintain chronic surveillance of operations, quickly detect and contain threats, and take appropriate mitigation steps within a short period. Such interlock controls provide an optimal preventive, detective, and response posture that is consistent to Zero Trust principle that always seek to minimize risk and consider a breach to have happened.

Figure 21 gives a mind map illustrating a hierarchical arrangement of different technical controls' attributes and capabilities with respect to their varying posture levels in the context of fostering Zero Trust Architecture. Four specific posture levels are set out as follows: Initial (Low), Developing (Moderate), Advanced (High) and Optimized (Very High).

Each posture level consists of a cluster of logically related technical security controls and capabilities, which an organization would usually adopt or enhance at that stage in its Zero Trust implementation process. The levels build upon one another in succession, with the Initial level being the most basic level of security practices and the Optimized level being the most mature in terms of Zero Trust security capability.

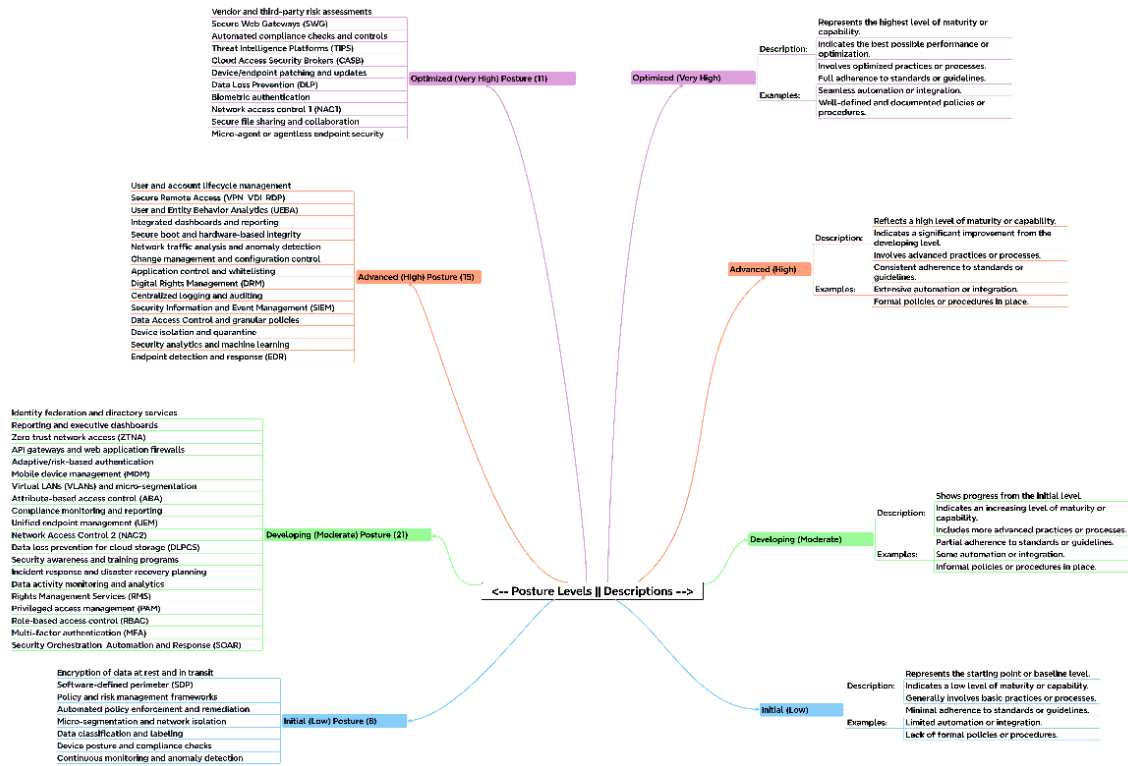

*Figure 21. A Four-Tiered Posture Model*

The graphic provides definitions and illustrations to demonstrate the level or the degree of compliance to the requirements in each level of posture, using generic situations as examples. It intends to help the organizations about the logical relationships and the changes that are required over time for the organization to progressively improve its Zero Trust security posture from one level to another until full optimization is achieved.

In short, this visual framework provides organizations with a systematic approach to strategizing, executing, and evolving critical technical controls and processes in various operations domains (identity, endpoints, network, data, etc.) to create an effective society of Zero Trust Architecture in accordance with the appropriate practices in the field.

When contextualized within different posture levels, technical controls and capabilities can help the organizational efforts to support security improvements over time through more effective planning and implementation of the necessary actions. This systematization offers organizations a progressive pathway in adjusting to best industry practices concerning Zero Trust security advancement.

The implementation of the Zero Trust Architecture is not bound to a specific area, but rather encompasses identity, endpoints, network, and data among others. In each of these areas, there is a need to work on an implementable and progressive plan for the requisite technical controls and processes. For example, using such a framework, organizations over time can be able to improve their security block from the Initial level to Optimize level. Such a way of improvement is consistent with the existing regulations and policies of the industry and allows organizations to effectively deploy Zero Trust Architecture with no risks.

Along with the technical controls and capabilities that every organization is equipped with, it is also very critical for organizations to grasp the logic in the steps and the relationships in enhancing their Zero Trust security posture. The posture levels, namely Initial, Developing, Advanced, and Optimized, provide a comprehension of the different activities to be carried out for organizations to evolve in their quest of improving their security posture.

This is the stage where the required processes, approaches, and policies are implemented within an organization's structure. This assists in implementing a strong Zero Trust Architecture which integrates with industry best standards in improvement.

The framework emphasizes that a perfect Zero Trust Architecture can only be achieved as a process, which compels organizations to level up steadily as each level and its associated technical controls and required operational processes are built upon sequentially. Organizations can develop strategies to assess their current conditions, identify shortfalls, and outline how to implement Zero Trust to accommodate logical interdependencies and optimal practices.

## 6.3	Technical Controls Model

The Technical Controls model outlines a structured and concise review and enhancement process for improving any organization's cybersecurity posture. In prioritizing scoping by impact and criticality, this model is purportedly designed to support a structured approach to cybersecurity. It offers a structured approach to cybersecurity, categorizing and ranking technical controls based on impact and importance. It considers eleven categories of cybersecurity technical controls, giving holistic security coverage.

The model ranks five technologies in each category, with a total of fifty-five technical controls. The controls are stack-ranked against criteria like deployment and relevance to the current threat landscapes. This hierarchical methodology allows the organization to focus first on the most impactful controls in its efforts to prioritize its own cybersecurity environment.

It provides help on several levels: First, it is an assessment tool because it allows any given organization to assess its position regarding its current cybersecurity and identify the gap that it needs to fill while prioritizing improvements. It enables informed decisions on resource deployment and risk mitigation through a comprehensive and ranked list of controls. It is also a development guide in as much as it helps organizations develop their security measures to ensure that they conform with industry best practices and regulatory requirements.

The technical controls model is dynamic and can be updated with new threats and technological changes. This assures that the matrix will remain useable in organizations as a guiding tool for maintaining stringent cybersecurity. The comprehensiveness of the model ensures that organizations get to touch on all the critical aspects of cybersecurity, starting from access management down to identity management.

The structured approach of the model also allows scalability to organizations of all sizes-from small and medium-sized enterprises to large corporations by being able to offer a detailed and prioritized list of technical controls that will let organizations make informed, data-driven decisions that enhance their security posture and more effectively fight today's cyber threats.

The model is not a check-the-box in nature but rather a strategic mechanism to facilitate organizations' prioritizing of efforts to posture their security systematically. It also enables them to make a fully informed decision based on a comprehensive evaluation of their cybersecurity posture.

## 6.4	Practical Insights from Chi-Square Tests

This section will show the results and discuss the implications of a detailed statistical analysis of the relationships among the different practices concerning security and incident response measures in organizations. Chi-square tests were carried out to explore the relationship of eleven categorical variables (A7_1 to A17_5) with the twelve specific dichotomous security management and incident handling variables (A18 to A30). These eleven focus points give the applicable overview of the association between different security practices with adopting security technologies, incident identification, and response actions and their respective outcomes.

The following analysis covers these twelve key areas:
1. Multi-Factor Authentication Usage
2. Cloud Access Security Broker Usage
3. Cloud Security Posture Management Usage
4. Network Detection and Response (NDR) Usage
5. Virtual Desktop Infrastructure Usage
6. Internal Identification of Security Issues

      7. External Identification of Security Issues

      8. Law Enforcement-Identified Security Issues

      9. Need to Restore After a Security Incident

      10. Issuance of Notifications After a Security Incident

      11. Financial Compensation After Security Incidents

      12. Significant Changes to Cybersecurity Policies

**Cyber Security Policy Approach Arising as a Result of Change(s)**

    This analysis highlights the number of significant associations for each area, the associations, and their issues for organizational security practice that are relevant to the scope of the analysis. Both cause-effect and enumerative approaches are essential for applying the suggested method to practical applications. The synthesis of information enabled practitioners to understand what could be done and what actions they needed to be prepared for, given the security concerns they had in mind.

**Multi-Factor Authentication Usage**

    The analysis helps demystify why the use of MFA was limited to three associations out of the 55 assessed variables. The evidence seems to imply that the sequential arrangement of particular variables is a function of the respondents' use of MFA. However, it is worth noting that several variables did not present any significant association, therefore suggesting that the adoption of MFA may be explained by other factors that were not examined in the survey.

**Cloud Access Security Broker Usage**

    Examination of CASB usage resulted in the identification of four significant associations. The implications of the findings are that organizations that use CASB may have different focus regarding network security and data protection. Quite interestingly, network security and access control variables that are very much associated with the use of CASB

appear to suggest that there are network security concerns that necessitate the need to adopt a CASB.

**Cloud Security Posture Management Usage**

The description of CSPM usage analysis is that out of 66 variables, only one lacked association with a greater than 0.05 probability level discrimination. This kind of low association infers that CSPM adoption could be reasoned more along particular cloud security needs than the general security posture. The positive correlation with a variable on a network security services, in which some organizations use a certain level of network security, means that those organizations are more likely to use CSPM. This means, CSPM adoption does not seem to follow organizational security practices and infrastructure but rather correlates with the cloud strategy of the organization and its particular cloud security requirements.

**Network Detection and Response Usage**

The analysis of the use of Network Detection and Response did not reveal any significant correlations, which was surprising. This could point to the fact that either the factors relevant for the implementation of NDR are not captured by the survey questionnaire, or that it is a very unsophisticated measure of security that is uniformly present, or absent, in organizations regardless of the state of security. The lack of correlations could be that organizations view NDR as a commodity security function that is independently enacted from other security controls.

Alternatively, this may reflect that NDR is an inexpensive networking solution that does not necessarily align with or complement other core security practices adopted by those organizations. Resulting from this, this unplanned outcome underlined the complexity of security technology adoption and called for further investigation of factors driving NDR implementation across diverse organizational contexts.

**Virtual Desktop Infrastructure Usage**

A single association was found, which falls into the category of security measures regarding access control and endpoint protection. This could be attributed to that VDI is adopted by an organization in order to extend or enhance particular security requirements.

The small number of associations would suggest that these adoptions of VDI were driven by pointed operational concerns or security issues rather than part of a general security strategy. That is, the implementation of VDI is to accomplish certain objectives of the organizations, such as access control or the reinforcement of endpoint protection, not because it is a comprehensive security solution.

**Internal Identification of Security Issues**

This analysis found two important associations to the factors concerning what measures to protect the data and the factors of governance, risk and compliance processes. The findings bring out the fact that the organizations with well-developed and defined security processes and the elements of governance can prevent and find security issues internally more effectively. It is therefore emphasized that the internal detection of security issues is more related to the organization's process and governance than the security instruments or measures employed.

**External Identification of Security Issues**

External identification revealed only one significant association related to data protection and network security measures. This indicates that exercising internal or external security measures contributes to issue identification regardless of its source. The low number of associations indicates that external identification might be subject to more factors that cannot be captured directly by the survey, such as the external threat landscape or the organization's image.

**Circumstances Witnessed by Law Enforcement- Identified Security Issues**

This analysis identified five significant associations encompassing identity and access management, access control, governance and compliance, and network security. The wide encompassing spread implies that adopting a spirit of collaboration on serious security positions among organizations and law enforcement agencies might be more effective than previously envisaged. Organizations with these practices might be able to intervene earlier and involve law enforcement after the detection of problems rather than these measures causing more security problems.

**Need to Restore Following a Security Incident**

This assessment identified three key linkages: network security, security surveillance and data protection. These findings imply that a multi-dimensional approach may raise additional risks of incurring needs for solution after an incident has already occurred. This does not mean that these procedures result in more risk of the security threats but assist in the better management of the situations that are likely to require restoration efforts.

**Issuance of Notifications After a Security Incident**

This analysis postulated four significant network security, governance and data protection associations. 28 % of the respondents answered that during the past year they had made public announcements or notifications to the public with respect to the data breach. The results imply that organizations lacking these best practices and other areas perfunctory notification are most likely to be recommended supplementary notification and possible punishment or closer supervision by the regulatory authorities. This could suggest the availability or the need for better reporting on serving the stakeholders.

**Compensation and Restitution After Security Incidents**

The analysis of financial compensation after security incidents identified seven strong associations with the strongest being linked to governance, risk management, and compliance practices. This means that companies which do not have or whose GRC practices are infant are likely to offer monetary compensation such as the ones financiers would require. 15% of the respondents had indicated that they were able to provide some form of financial restitution. Other associations also include information security monitoring, information security, and networks, which indicates that an organizational mixture of the theory provisions help an organization protect itself from incurring losses in case of an event.

**Significant Changes to Cybersecurity Policies**

Results from the conducted Significant Changes to Cybersecurity Policies could be associated with only 4 out of the 55 factors, hence allowing us to forecast a medium level of correlation between some practices after the incidents in question and changed practices in relation to the policies.

It was determined that the main correlations were in the segments including network security, segmentation, security monitoring and threat detection, governance, risk and compliance, and data protection and network security. These observations imply that the policy change practice should be advanced in these four particular areas and, therefore, increase the chances of the organization being able to deal with the emerging security threats easily through policy changes.

It is also noteworthy that in the course of such analyses, actions are described which seem to correlate with less, say, incidents being detected, or notified, or momentary compensatory actions being undertaken – but in most cases this does not mean that there are going to be more security incidents. Rather it would be the case that these activities are more useful in ensuring that security incidents are detected, reported and addressed properly. In this regard, the results underscore the need for a comprehensive strategy in addressing security, which combines technical, organizational and governance aspects.

## 6.5      Comparative Analysis of Technical Controls

**Evaluating Efficacy in Preventing Cybersecurity Breaches**

This section provides insights into the effectiveness of multiple technical controls in relation to cybersecurity breaches within the context of Zero Trust Network Access. ZTNA data was collected from 138 respondents of which 44.20% had self-identified with breaches. Using comparisons as the basis, the study examines eleven different categories of technical controls, their ranking, and how they relate with occurrence of breaches.

The approach taken consists in computing average ranks for the specified security controls for cases when there is a breach and when there is no breach, comparing such normalized scores with actual breaches, and ascertaining whether there are significant differences between the two case scores. It is crucial to mention that the barrier to entry controls, privileged access management, data leakage prevention, and Centralized Logging and Auditing demonstrate stronger negative relationships with breaches, thus can be seen by their effectiveness in the prevention of the breach.

Counterintuitive results paradigm has also been witnessed in this study for example where multi-factor authentication has performed abysmally compared to adaptive/risk based authentication despite the high reliance. It also reveals the inconsistency between the expectation of some controls that are continuously monitored and the relationship of these controls to breach effectuation.

The emphasis of the studies highlights that cybersecurity is very complicated and stresses the necessity of using multilayered security approaches. It makes clear that while cross tabulations are a very helpful way of looking at data, they do not equate to an explanation for the results and should be taken with caution. In support of this, this research offers evidence on the effectiveness of certain technical controls, adding value to the development of prevention policies and strategies regarding ZTNA in an organization.

This study applies to analyzing the breach incident these two approaches yielded: descriptively examining the technical controls in response to the existence of an internally identified implications of data analyses. The analysis of the findings suggests that 44.20% of respondents admitted to an external breach, while 61 respondents self-identified an internal breach, and 77 did not report any breach.

**Technical Control Group #1 Identity and Access Management**

*Table 20. Technical Control Group #1 Identity and Access Management*

| Technical Control | Average Rank Breached | Average Rank Not Breached | Difference | Breach Occurrence Correlation | Measure Effectiveness |
|---|---|---|---|---|---|
| Multi-factor authentication (MFA) | 2.11 | 2.14 | -0.03 | -0.01 | Negative |
| Single sign-on (SSO) | 3.25 | 3.60 | -0.35 | -0.15 | More Negative |
| Adaptive/risk-based authentication (RBA) | 2.69 | 2.61 | +0.08 | +0.03 | Positive |
| Biometric authentication | 4.08 | 4.04 | +0.04 | +0.02 | Positive |
| Identity federation and direct | 2.87 | 2.61 | +0.26 | +0.08 | More Positive |

*\* lower values equate to higher rankings*

The formal analysis of Technical Control Group number one, described in Table 22, produced several interesting results. Single Sign-On was emphatically agreed upon as the top priority in all breach and no breach, with a slight assumption that the occurrence rate of breach goes down. Organizations who reported breaches ranked SSO lower than the non-reporting ones implying that SSO may be considered less effective. However, those who seemed to experience a breach ranked it a bit higher indicating its broad perception as being critical.

Single Sign-On was the only control that had the strongest negative correlation with breaches, which was also greater than that of Multi-Factor Authentication. This somewhat surprising result illustrates relatively high efficiency of Single Sign-On which is remarkable since MFA is regularly regarded as gold standard for securing authentication.

Multi-Factor Authentication argued its case for weak relative performance which is rather unexpected considering its own ranking among the security community as quite an effective control. Correlating to the -0.03 level of breach occurrence, the effectiveness of a breach occurrence does not realistically demonstrate the anticipated outcome.

Remarkably, Biometric Authentication remained neutral regarding the research questions notwithstanding its increase in the device features. One might expect a stronger correlation (positive or negative) with breach occurrence. The neutral position suggests it is not viewed as a crucial factor, possibly due to difficulties in practical implementation or limited usefulness in minimizing breaches.

It was found that Identity Federation and Directory Services constituted the highest positive correlation with breach occurrence, which is quite surprising because the reason for adopting such advanced identity management solutions, in particular, is to lower the incidence of breaches. This is an area where more research is needed.

**Technical Control Group #2 Access Control and Endpoint Security**

A similar trend in Table 23 was obtained in the case of Technical Controls Group 1 as well. In both breach and no breach cases, two-factor (2FA) or multi-factor authentication emerged as most crucial for security planning with minimal risks of commitment failure. Organizations that reported breaches ranked MFA lower in their critical success factors than those that did not report breaches. In the case of those that experienced a breach, they ranked it lower than the no-breach case; however, it was highly ranked due to how important it is viewed to be.

*Table 21. Technical Control Group #2 Access Control and Endpoint Security*

| Technical Control | Average Rank Breached | Not Breached | Difference | Breach Occurrence Correlation | Measure Effectiveness |
|---|---|---|---|---|---|
| Endpoint detection and response (EDR) | 3.23 | 3.06 | +0.17 | +0.06 | Positive |
| Role-based access control (RBA) | 2.84 | 2.81 | +0.03 | +0.01 | Neutral |
| Attribute-based access control (ABA) | 3.54 | 3.32 | +0.22 | +0.08 | Positive |
| Privileged access management (PAM) | 2.62 | 2.92 | -0.30 | -0.11 | More Negative |
| User and account lifecycle management | 2.77 | 2.88 | -0.11 | -0.04 | Negative |

*\* lower values equate to higher rankings*

Organizations that suffered breaches rated Access Control based on attribute and Endpoint detection & response lower. The EDR is rated low; however, this is somewhat disheartening, as it is largely seen as being central to breach detection and response. This may imply that organizations only put EDR systems in place after a breach incident instead of being proactive.

For "User and Account Lifecycle Management", a slight negative difference in ranking was observed, which means that user and account management is given a higher priority in breach countries than in no breach countries. This indicates that engaging in effective user and account management may assist in minimizing the risk of a security breach, even though the impact appears limited.

Role-Based Access Control does not tend to change the ranking between the two groups and showed no relationship to breach occurrence whatsoever. The implication is that this would hold even for the reasons related to RBA control given its significance in the scope of defeating breaches.

**Technical Control Group #3 Endpoint Security and Management**

The analysis of Technical Controls Group 3 (Table 24) revealed that Device Posture and Compliance Checks, Unified Endpoint Management (UEM), and Mobile Device Management (MDM) might be the most effective controls in this group for preventing breaches. Organizations that have experienced breaches seem to recognize their importance more than those who have not.

*Table 22. Technical Control Group #3 Endpoint Security and Management*

| Technical Control | Average Rank Breached | Not Breached | Difference | Breach Occurrence Correlation | Measure Effectiveness |
|---|---|---|---|---|---|
| Unified endpoint management (UEM) | 2.64 | 2.75 | -0.11 | -0.05 | Negative |
| Mobile device management (MDM) | 2.97 | 3.08 | -0.11 | -0.04 | Negative |
| Device posture and compliance checks | 2.1 | 2.27 | -0.17 | -0.07 | Negative |
| Secure boot and hardware-based integrity | 3.61 | 3.42 | +0.19 | +0.07 | Positive |
| Micro-agent or agentless endpoint security | 3.69 | 3.48 | +0.21 | +0.07 | Positive |

*\* lower values equate to higher rankings*

Organizations that have been breached rated Secure Boot and Hardware-Based Integrity, as well as Micro-Agent or Agentless Endpoint Security, lower. This might suggest that such controls, though important to a comprehensive protection strategy, may be less effective against the types of breaches captured in this survey. Alternatively, it could mean that these controls are being adopted after-breach reactively, or are more difficult to implement effectively.

Because these correlations are relatively weak across all controls in this group, no control is very strongly related to breach occurrence. This points out the need for a comprehensive, layered approach to security rather than relying on any single control.

**Technical Control Group #4 Network Security and Access Control**

Zero Trust Network Access, according to Table 25, was the most critical control in preventing the breach of the Technical Controls Group 4. It showed the highest degree of negative difference of rank and the most extreme negative correlation with occurrence of a breach. ZTNA is the highest ranked and prioritized control of this group by the two groups. The determination of this makes this control widely recognized as an important breach prevention control.

*Table 23. Technical Control Group #4 Network Security and Access Control*

| Technical Control | Average Rank | | Difference | Breach Occurrence Correlation | Measure Effectiveness |
|---|---|---|---|---|---|
| | Breached | Not Breached | | | |
| Software-defined perimeter (SDP) | 2.84 | 2.9 | -0.06 | -0.02 | Negative |
| Zero trust network access (ZTNA) | 1.52 | 1.65 | -0.13 | -0.06 | Negative |
| Application control and whitelisting | 3.21 | 3.14 | +0.07 | +0.03 | Positive |
| Device isolation and quarantine | 3.87 | 3.84 | +0.03 | +0.01 | Positive |
| Network access control (NAC) | 3.56 | 3.47 | +0.09 | +0.03 | Positive |
| *\* lower values equate to higher rankings* | | | | | |

SDP ranked slightly lower with a negative difference and showed a weak negative correlation with breach occurrence, indicating perceived importance in preventing breaches. However, the effect is less pronounced than ZTNA.

Application Control and Whitelisting and Network Access Control ranked a little lower for breached organizations. That would be somewhat surprising with respect to NAC, in particular, given its general elevation to the level of network security savior. This may indicate that NAC alone is insufficient to prevent security breaches or that organizations implement it only reactively.

The ranking between Device Isolation and Quarantine was very little different between the two groups, and they also correlated the least with breach occurrence. This may suggest that there is a place for them in general security, but as a differentiator in preventing breaches, compared to other controls within this group, it may not be.

This is indicative of current cybersecurity trends that put a high emphasis on Zero Trust Architectures, as defined by the clear prioritization of ZTNA across both groups but mainly from those who reported breaches. This suggests that organizations view ZTNA as integral to their strategy in preventing unauthorized access and potential breaches.

**Technical Control Group #5 Network Security and Segmentation**

The analysis of Technical Controls Group 5 (Table 26) revealed that Secure Web Gateways (SWG) might be the most effective control in this group for preventing breaches. Organizations that have experienced breaches seem to recognize its importance more than those that have not, as evidenced by the most prominent negative difference in ranking and the strongest negative correlation with breach occurrence.

*Table 24. Technical Control Group #5 Network Security and Segmentation*

| Technical Control | Average Rank | | | Breach Occurrence Correlation | Measure Effectiveness |
|---|---|---|---|---|---|
| | Breached | Not Breached | Difference | | |
| Micro-segmentation and network isolation | 2.20 | 1.87 | +0.33 | +0.13 | More Positive |
| Virtual LANs (VLANs) and micro-segmentation | 3.10 | 3.05 | +0.05 | +0.02 | Positive |
| Network Access Control (NAC) | 3.34 | 3.42 | -0.08 | -0.03 | Negative |
| Secure Web Gateways (SWG) | 2.87 | 3.34 | -0.47 | -0.17 | More Negative |
| Cloud Access Security Brokers (CASB) | 3.49 | 3.32 | +0.17 | +0.06 | Positive |
| *lower values equate to higher rankings* | | | | | |

CASB and Microsegmentation and Network Isolation ranked lower for breached organizations. This could indicate either that these controls are less critical than SWG in preventing breaches or that organizations are implementing these controls reactively after having been breached.

On the other hand, NAC2 yielded only a slightly negative difference in ranking, which can be interpreted as being higher in importance for breached organizations. Therefore, it can be concluded that NAC2 is somewhat vital in ensuring the avoidance of breaches, although the effect here is not as clear-cut as that of SWG.

In turn, the ranking differences between VLANs and Microsegmentation were really very small among these two groups and very uncorrelated with the occurrence of a breach. This would suggest that, though such controls are important in relation to overall security, they may not constitute a discriminating factor in preventing breaches compared with the rest of the controls in this group.

**Technical Control Group #6 Data Protection and Network Security**

Of the controls in Technical Controls Group 6 in Table 27, DLP emerged as potentially the most effective at preventing breaches. Organizations that had a breach ranked it significantly higher in priority, and it showed the strongest negative correlation with breach occurrence.

*Table 25. Technical Control Group #6 Data Protection and Network Security*

| Technical Control | Average Rank Breached | Not Breached | Difference | Breach Occurrence Correlation | Measure Effectiveness |
|---|---|---|---|---|---|
| Data classification and labeling | 2.28 | 2.05 | +0.23 | +0.09 | Positive |
| API gateways and web application firewalls | 3.08 | 3.13 | -0.05 | -0.02 | Negative |
| Network traffic analysis and anomaly detection | 3.41 | 3.04 | +0.37 | +0.15 | More Positive |
| Secure remote access (VPN, VDI, RDP) | 2.98 | 3.04 | -0.06 | -0.02 | Negative |
| Data Loss Prevention (DLP) | 3.25 | 3.74 | -0.49 | -0.18 | More Negative |

*\* lower values equate to higher rankings*

Only minor negative deviations were observed for Secure Remote Access VPN, VDI, RDP), API Gateways, and Web Application Firewalls, meaning these controls are of higher priority among breached organizations. This would indicate these controls are key in preventing breaches; however, the effect is much weaker than DLP.

Organizations that have been breached rated Network Traffic Analysis, Anomaly Detection, and Data Classification and Labeling lower. This may indicate these controls are less effective in preventing a breach from occurring than DLP is, or that they are implemented reactively after an organization has been breached.

That is, with a few minor exceptions, the correlations are weak in all controls of this group, indicating that no one control is strongly related to the occurrence of a breach. This points to the importance of an overall and multi-layered approach to security rather than dependence on any one control.

**Technical Control Group #7 Data Protection and Information Security**

The analysis of Technical Controls Group 7 Analysis (Table 28) suggests RMS and Data Access Control and Granular Policies may be the best controls in this group. Indeed, breached organizations appear to understand their criticality more than their non-breached counterparts.

*Table 26. Technical Control Group #7 Data Protection and Information Security*

| Technical Control | Average Rank Breached | Not Breached | Difference | Breach Occurrence Correlation | Measure Effectiveness |
|---|---|---|---|---|---|
| Encryption of data at rest and in transit | 1.87 | 1.77 | +0.10 | +0.04 | Positive |
| Digital Rights Management (DRM) | 3.69 | 3.68 | +0.01 | +0.01 | Positive |
| Data access control and granular policies | 2.07 | 2.13 | -0.06 | -0.03 | Negative |
| Rights Management Services (RMS) | 3.97 | 4.06 | -0.09 | -0.05 | Negative |
| Secure file sharing and collaboration | 3.41 | 3.36 | +0.05 | +0.02 | Positive |
| *lower values equate to higher rankings* | | | | | |

Somewhat counterintuitively, Data at Rest and in Transit Encryption ranked lower for breached organizations. This could suggest that while encryption is paramount, it may not be enough on its own to prevent breach, or that organizations are doing it after the fact once they have been breached.

Secure File Sharing and Collaboration and Digital Rights Management were relatively close between the two groups and very little associated with breach incidence. This would indicate that they are indeed important to overall security, but not distinguishing variables in preventing breaches relative to other controls in this group.

Overall ranking indicates that this encryption of data at rest and in-transit is considered the top priority by both groups, second only to Data Access Control and Granular Policies. This follows general best practices in data protection.

**Technical Control Group #8 Security Monitoring and Incident Response**

Of the Technical Controls Group 8 in Table 29, SIEM became the most effective control to avoid breaches. Those organizations that had experienced a breach ranked it higher in priority, and it demonstrated a weak negative correlation with the occurrence of a breach.

*Table 27. Technical Control Group #8 Security Monitoring and Incident Response*

| Technical Control | Average Rank Breached | Average Rank Not Breached | Difference | Breach Occurrence Correlation | Measure Effectiveness |
|---|---|---|---|---|---|
| Data loss prevention for cloud storage | 3.56 | 3.60 | -0.04 | -0.01 | Negative |
| Security Information and Event Management (SIEM) | 2.39 | 2.58 | -0.19 | -0.07 | Negative |
| User and Entity Behavior Analytics (UEBA) | 2.66 | 2.64 | +0.02 | +0.01 | Positive |
| Security Orchestration, Automation, and Response (SOAR) | 3.05 | 3.00 | +0.05 | +0.02 | Positive |
| Data activity monitoring and analytics | 3.34 | 3.18 | +0.16 | +0.06 | Positive |

*\* lower values equate to higher rankings*

While there was a small negative difference in ranking for DLPCS, suggesting that Data Loss Prevention for Cloud Storage was considered a higher priority by organizations that had experienced breaches, the correlation with breach occurrence was close to zero, indicating again that though this is an important activity, the actual impact on breach occurrence is likely modest.

User and Entity Behavior Analytics (UEBA) and Security Orchestration, Automation, and Response (SOAR) showed very small positive differences in ranking, meaning organizations that experienced a breach ranked them slightly lower in priority. This suggests that while these controls are important, they might not be as effective in preventing breaches as other controls in this group.

Data Activity Monitoring and Analytics ranked lower for breached organizations and showed a weak positive correlation with the occurrence of the breach. This might show that while important, it is either not as effective against the prevention of breaches when compared with other controls in this group, or it is implemented reactively.

The overall ranking levels indicate that SIEM is considered of the highest priority importance for both groups, followed by UEBA. That's consistent with general best practices in security monitoring and incident response.

**Technical Control Group #9 Security Monitoring and Threat Detection**

Centralized Logging and Auditing appears to be the most efficient control in Technical Controls Group 9, based on the analysis of Table 30. Organizations that experienced a breach ranked it significantly higher in priority, showing the strongest negative correlation with breach occurrence.

*Table 28. Technical Control Group #9 Security Monitoring and Threat Detection*

| Technical Control | Average Rank Breached | Average Rank Not Breached | Difference | Breach Occurrence Correlation | Measure Effectiveness |
|---|---|---|---|---|---|
| Threat Intelligence Platforms (TIPS) | 3.25 | 3.34 | -0.09 | -0.03 | Negative |
| Security analytics and machine learning | 3.70 | 3.57 | +0.13 | +0.06 | Positive |
| Centralized logging and auditing | 2.74 | 3.08 | -0.34 | -0.11 | Negative |
| Automated policy enforcement and remediation | 2.77 | 2.69 | +0.08 | +0.03 | Positive |
| Continuous monitoring and anomaly detection | 2.54 | 2.32 | +0.22 | +0.08 | Positive |

*\* lower values equate to higher rankings*

TIP solutions showed a slight negative difference in ranking, which, in turn, means that it could get a higher priority in breached organizations, therefore indicating that, whereas the TIPS is somewhat important for prevention, this effect is not as pronounced as for Centralized Logging and Auditing.

Breached organizations ranked Automated Policy Enforcement and Remediation, and Security Analytics and Machine Learning lower. This suggests either these are less-effective controls in the context of preventing breaches when compared with other controls in this group, or that organizations are implementing these controls reactively.

The largest positive difference in rank, however, was Continuous Monitoring and Anomaly Detection, which fell significantly lower in priority for organizations that had been breached. That is somewhat surprising, since continuous monitoring has generally been considered one of the most critical controls for preventing breaches. It may indicate that organizations are implementing this control reactively after they have experienced a breach, or that it is not as effective as expected at preventing breaches.

The overall ranking shows that Continuous Monitoring and Anomaly Detection ranked the highest priority among organizations without breaches, while Centralized Logging and Auditing was ranked as the highest priority among organizations that have experienced breaches. This difference in priorities could reflect lessons learned from experiencing a breach.

**Technical Control Group #10 Governance, Risk, and Compliance**

Among the technical controls in Group 10, as outlined in Table 31, Vendor and Third-Party Risk Assessments, Compliance Monitoring, and Reporting emerge as potentially the most effective controls in preventing breaches. Organizations that have experienced a breach ranked these controls much higher in priority, while they conditioned a moderate negative correlation with occurrence of a breach.

*Table 29. Technical Control Group #10 Governance, Risk, and Compliance*

| Technical Control | Average Rank Breached | Average Rank Not Breached | Difference | Breach Occurrence Correlation | Measure Effectiveness |
|---|---|---|---|---|---|
| Policy and risk management frameworks | 1.89 | 1.84 | +0.05 | +0.02 | Positive |
| Compliance monitoring and reporting | 2.66 | 3.00 | -0.34 | -0.15 | More Negative |
| Integrated dashboards and reporting | 3.90 | 3.66 | +0.24 | +0.10 | More Positive |
| Vendor and third-party risk assessments | 3.36 | 3.73 | -0.37 | -0.15 | More Negative |
| Automated compliance checks and controls | 3.20 | 2.77 | +0.43 | +0.15 | More Positive |

*\* lower values equate to higher rankings*

Policy and Risk Management Frameworks had little positive difference in rank, indicating that these are indeed critical but less clearly linked to the prevention of breaches in a comprehensive manner compared to the other controls in this category.

Rankings of controls by breached organizations were rated far lower than expected on things like Integrated Dashboards and Reporting, Automated Compliance Checks and Controls. Of course, this could be an indication that while these controls are great, they alone may not be good enough to stop the breach, or that the organization has implemented them after already going through a breach.

In the overall ranking, Policy and Risk Management Frameworks tops in both groups, which is consistent with general best practices in GRC. However, the significant variance in the ranking in other controls between breached versus non-breached institutions may indicate shifting priorities of security due to breach experience.

**Technical Control Group #11 Operational Security and Incident Management**

Technical Controls Group 11 Analysis suggests the technical controls in Table 32 may have the most impact across this group to prevent breaches: Device Endpoint Patching and Updates, Change Management and Configuration Control, and Incident Response and Disaster Recovery Planning. It seems that organizations that have been breached recognize these controls as important more often than those who have not been breached.

*Table 30. Technical Control Group #11 Operational Security and Incident Management*

| Technical Control | Average Rank Breached | Not Breached | Difference | Breach Occurrence Correlation | Measure Effectiveness |
|---|---|---|---|---|---|
| Security awareness and training programs | 2.84 | 2.81 | +0.03 | +0.01 | Positive |
| Change management and configuration control | 2.46 | 2.57 | -0.11 | -0.05 | Negative |
| Incident response and disaster recovery planning | 2.79 | 2.92 | -0.13 | -0.06 | Negative |
| Reporting and executive dashboards | 4.7 | 4.27 | +0.43 | +0.22 | More Positive |
| Device/endpoint patching and updates | 2.21 | 2.43 | -0.22 | -0.09 | Negative |

*\* lower values equate to higher rankings*

Reporting and Executive Dashboards had the largest positive difference of rank, meaning that organizations that were breached ranked it as a considerably lower priority. This is somewhat surprising- one would generally consider reporting and the executive dashboard important for maintaining security. This may indicate that organizations implement this control reactively, after experiencing a breach, or that it is not effectively serving its intended purpose in breach prevention.

Contrasting that, the ranking difference for the Security Awareness and Training Programs was extremely minimal, possibly indicating that while significant in nature, they may not be as directly related toward preventing a breach when compared with other controls within this group.

The overall ranking does indeed show that Device Endpoint Patching and Updates is the highest priority for both groups, and this would align with general best practices in operational security and incident management. However, there are significant ranking differences in many other controls between the breached and non-breached organizations, which do point to breach experience influencing security priorities.

Note that correlation does not imply causation, and such breaches could therefore be due to a set of factors including how organizations may respond to breaches by changing the priorities of their security measures, the nature of the breaches that have occurred, or other variables not considered within the dataset. These findings should, however, be taken as a starting point upon which more detailed inquiry is started, to be weighed against other intervening factors and general knowledge in the cybersecurity domain.

## 6.6    Summary

This detailed assessment of the technical measures of the ZTNA framework offers a thorough analysis of the complicated areas related to prevention methods of cybersecurity breaches. It examines the eleven groups of technical security controls, their effectiveness, the breaches, and their relationships with the aim of better validation of different security measures. The key findings of this research underpin a multilayered approach to cybersecurity. There is no single control that has an exceedingly strong correlation with breach prevention, which strengthens the argument for the use of several layers of security controls. Apart from the policies, the surprising observations noted also include how Adaptive/Risk based Authentication postures stood out, especially in comparison to the mainstream Multi-Factor Authentication implements.

The study also identifies controls like Privileged Account Management, Data Loss Prevention, and Centralized Logging & Auditing as likely to be effective in disrupting organizational security policy. Of course, this correlation is not causation, and implementation and results vary due to organizational settings and specific possible threats. In addition, the study reveals some of the interesting paradoxes between the impression of how important some of the controls are in relation to the actual prevention of a breach.

The discrepancy accounts for the persistent need to revise and adjust the cybersecurity strategic approach within the context of what is revealed rather than traditional views. These implications are significant for cybersecurity practitioners and researchers. For practitioners, this research provides a foundation for prioritizing and implementing data-driven technical controls within the ZTNA framework. It amplifies that security is both wide and broad and emphasizes that its evaluation and modification must be ongoing efforts. This study has therefore emerged as an eye opener to other researchers. Additional research might consider the missing link between certain controls and prevention of breach, and others' dependence on different controls and conditions including organization size, industry and type of threats for the effectiveness of corresponding security controls.

The study undertakes a very critical analysis of the effectiveness of various types of technical controls. In examining cyber intrusion, this may indeed raise serious questions about the cybersecurity picture. Threats keep evolving, and so do thoughts on how security would be provided given the seriousness and intricacies of the current threat actor and their advanced tactics.

This work consists of a sequence of foot-in-the-door studies leading towards extensive evidence-based cybersecurity policies. It stresses the point that there should be learning and constant adaptation, along with a commitment to broad security in a fluid cyberspace.

# CHAPTER 7: CONCLUSION AND FUTURE WORK

## 7.1    Research Findings

This analysis of the cybersecurity survey provides insights into how organizations are working to build their security infrastructure in today's digital times. Data encryption leads, being the centerpiece of any security strategy, ranking highest with an average of 1.89, while endpoint security measures have a central rank of 3.79. This contrast in prioritization reflects a fairly mature understanding that data protection must be given a higher priority compared to traditional methods of perimeter-based security. It would appear that organizations understand fully that the very foundation of their security posture is the encryption of information, since this ensures comprehensive security against the cyber threats that are becoming increasingly sophisticated.

The surveyed findings draw a picture of organizations moving toward an abstract view of Zero Trust Architecture: one in which security solutions are no longer compartmentalized, but an integrated system; of these, identity and access management controls become highly critical, with organizations focusing strongly on multi-factor authentication, privileged access management, and Biometric Authentication. Emphasis on identity verification and access control speaks volumes about how it realizes that, in the modern era, security will need to be enforced through granular controls on who can access what resources, under what circumstances, and at what time. The multi-layer approach cuts across different security domains -  from network segmentation to application security -  indicating a conscious awareness by organizations of the interconnected nature of security challenges and the need for coordinated solutions.

Most telling is the clear result of moving towards the cloud-friendly security framework, reflected in the high ranking that cloud-specific security controls received in survey results. The organizations continue to make conscious adaptations regarding their security strategies, given unique challenges presented by the cloud environments, including CASB and Cloud Security Posture Management solution implementations. This thus frames the adaptation of strategy in a foresighted approach - one that not only views the cloud computing revolution as irreversible but also supports robust standards for security. In particular, protection in multi-cloud environments highlights how organizations get ready for life in the long term, where complex, distributed systems are the rule rather than the exception. This evolution of the thinking in security is a big maturation in how organizations approach and secure their digital assets, hence bringing together traditional security principles into modern technological realities.

## 7.2     Contributions

This dissertation investigates the field of Zero Trust and how it can be applied in a data-driven manner to assess the context and the impact of underlying technical controls being used primarily throughout the U.S. federal government under the guidance of the NIST Cybersecurity Framework. This research is intended to provide a prescriptive framework that can serve as both a guide for narrowly assessing and a way of broadly ensuring that an entity can effectively assess its overall cybersecurity posture. The second chapter provides context around the "why now?" issue and delves deeper into the underlying technical controls.

This research enhances the understanding of the digital defense mechanisms in place at the enterprise level. Primarily informed by the author's far-reaching analysis and specific insights and recommendations, the research is comprehensive in its assessment. It utilizes a variety of technical and contextual measures to determine the robustness of an organization's cyber posture. By "robustness," this refers to an adversary's difficulty in executing an attack successfully on their enterprise. In this instance, the "measure" is indeed a "mystery" since if the measures were well known, the posture would be far less robust.

Researchers are increasingly interested in zero trust due to its vast application in fulfilling complex network security requirements. Gartner also points out that a quantitative risk analysis framework called the Open FAIR™ Body of Knowledge is necessary for measuring, analyzing, and discussing risks in quantified manners; therefore, it is especially suited for Zero Trust (T. O. G., 2021). The above insights merit the relevance and importance of developing a quantitative framework for assessing an organization's cybersecurity posture using a zero-trust approach.

These guidelines are particularly well-suited for use with Zero Trust because they are narrative-driven and allow clear communication of what is at stake for an organization. One would hypothesize that implementing a ZTA correlates positively with an improved posture against these cyberattacks. Additionally, this is relevant and timely research for cybersecurity professionals since it would have practical implications in terms of the Zero Trust approach being most robust for fortification against cyber threats.

## 7.3     Research Limitations

The ZTA research has several limitations, including the lack of a practical implementation process and a focus on project goals. This gap between academic understanding and practical implementation is important for organizations trying to implement ZTA programs.

Limitations of this research study involve a few key areas that have to be considered for further research studies. The major limitation involves the sampling methodology used. Convenience sampling, although effective to carry out the process, has its drawback in that it will inherently limit generalizability across the wider population of organizations implementing Zero Trust Architecture. This can be shown particularly with the concentration of geographic areas of response, 78.3% of whom were from North America. Second, self-selection bias due to voluntary participation and restriction to those already familiar with ZTA makes results skewed towards organizations with more mature security practices.

Other methodological limitations relate to the comprehensiveness of the study. Only quantitative methods were adopted, which lacked qualitative validations. The research had a cross-sectional nature, able to capture a snapshot in time rather than the evolutionary changes in the ZTA implementation. Without longitudinal data, it becomes quite impossible to comment on the long-term effectiveness of the ZTA deployments. Also, without any comparison with a control group for non-ZTA-implementing organizations, it becomes rather difficult to identify comparative advantages.

Documented experiences of major technology corporations in implementing ZTA have identified critical gaps that have demonstrated a lack of understanding both in terms of scale and adaptability. While Google's BeyondCorp (Ward & Beyer, 2014) implementation demonstrates enterprise-scale deployment and Microsoft's Zero Trust (Zero Trust Guidance Center, 2022) journey highlights cloud integration challenges, these case studies offer limited applicability to small and medium enterprises (SMEs). With no scalable frameworks to suit various organizational contexts, many organizations are finding it difficult to adapt such enterprise-level implementations to their needs. The absence of standardized criteria about evaluation, performance benchmarking, and metrics about security posture improvements further aggravates the problem.

Other areas may be concerning reliability during the collection of data. The study depended on self-reported data with no independent verification, which can result in social desirability bias in the reporting of security incidents. Limiting the sample to English-speaking respondents perhaps excluded many valuable perspectives from other regions. No formal validation of the respondents' claimed expertise or credentialing was performed. Response accuracy would depend upon validity and reliability of the survey instrument and perhaps ambiguity in the definition of technical controls across the organizations.

There could be some limitations related to the contextual scope of the research. Industry-specific challenges and regulatory compliance issues that are diversifying across regions were not taken into consideration, while the differences in organizational size and resource level received limited attention. The focus on technical controls has been of value, but the organizational factors and implementation challenges have been left relatively unexplored. Integration challenges with legacy systems and return on investment metrics are also not deeply analyzed, which is very relevant for any organization considering the adoption of ZTA.

## 7.4	Future Work

This research does not address how AI and machine learning could help innovate ZTA or how ZTA can be applied to supply chain security. Future work includes:

- AI and Machine Learning in ZTA: Assess the possibility and possible applications of innovative approaches concerning AI and machine learning for ZTA enhancement and deployment, mainly focusing on anomaly detection and policy enforcing systems. Examine the implementation of ZTA and its optimization using AI applications.

- ZTA in Supply Chain Security: Examine how the ZTA paradigms can be advanced to protect intricate networks of suppliers and third parties. Build models that describe how the ZTA could be executed when organizational silos are breached.

These suggestions for further research are expected to fill the gaps in existing literature, facilitate the investigation of areas of growing concern, and significantly aid in the continued development and deployment of Zero Trust Architecture in evolving technological environments.

# APPENDICES

# APPENDIX A: GLOSSARY OF ACRONYMS & DEFINITIONS

This section gives the reader a brief overview of the definitions and acronyms used in the entire project.

**API Gateways**: API Gateways are the most important elements in the communication between the client and the backend services. They are very important in maintaining the uninterrupted and secure traffic within the client and the API. API gateways have a variety of functionalities such as authentication, authorization, rate limiting, and logging. The main reason is these functional features help control and manage the volume of inbound and outbound traffic directed towards the backend services. API gateways also extend features such as request modification, protocol conversion, and response blending that enhance client interaction with multiple backing services through a single system.

**Application Control and Whitelisting in the costing of information security policy**: This is a form of a security procedure that Authorizes the organizations to control the operation of applications on their systems by stating which applications are to be executed. Whitelisting refers to compiling a list of applications which are safe and sanctioned for use while other applications are restricted from being executed. This preventive measure is able to prevent the use of unauthorized or malfunctioning programs and applications, thereby improving the security orientation of the institution. Whitelisting is a very acceptable procedure for curtailing malware threats and ensuring the operation of desirable and recognized programs only within the computer environments.

**Automated Policy Enforcement**: Automated policy enforcement consists of the use of technology and software in any of the systems and processes to ensure that the more policies or rules go onto the system, the greater the attempts at following them are. In most cases, this is done by using automated systems which can detect, monitor, and correct any policy contraventions. The aims of depending on automated policy enforcement include the reduction of individual mistakes, efficiency of the compliance activities, and the protection of a firm in a normal, regulated, and safe environment. By automating the policy enforcement process, businesses have the capability of ensuring consistency and mitigating the risks associated with non-compliance from human error.

**Change Management and Configuration Control**: Change Management and Configuration Control are fundamental both to the practice of project management and to the discipline of software engineering. Change management involves changing control procedures minimal to the objects, the project scope, timelines and available resources. It includes systematic identification, documentation, assessment and correction of changes in order to reduce the likelihood of variants and the amount of time taken to complete the project. On the other hand, configuration control is concerned with the management of the project items to be configured that can be in form of software, documents, hardware among others. The procedure comprises defining the baseline, tracking the changes and making sure that all the changes that were made were authorized and documented properly.

**Cloud Access Security Brokers**: Cloud Access Security Brokers (CASBs) are the third-party tools which basically facilitate the users of cloud services to adhere and avails compliance regarding restrictive policies. These CASBs control both the information which is available for other consumers and information which is used by the providers of the cloud services. In doing so, they can perform security measures, comply with legal requirements, and protect the privacy of information in the cloud. A number of features are provided by Casb together with the ability to oversee the occurrence of cloud, show data security, and manage risk and control. Also, they would also help companies to implement additional security 'in the cloud' accompanying the policy deployment in the desktop and cloud environments.

**Compliance Assessment and Reporting**: Overseeing and reporting compliance means carrying out review and analysis with respect to legal, regulatory, and ethical obligations and standards of an organization. It also involves an over time watching of business operations, activities, processes, and transactions with a view to observing the laws and legal requirements governing them. Further, compliance monitoring includes enabling the collection and examination of relevant information to give account as to the level of compliance of the organization. Such reports are often used to demonstrate compliance with rules and requirements, identify further development areas, and provide disclosure to stakeholders, regulators and other parties. Therefore, organizations need to ensure that effective Compliance Monitoring and Reporting mechanisms are in place in order to maintain integrity, trust and legality in their businesses.

**Always-On Monitoring And Anomaly Detection**: Always-on monitoring and anomaly detection makes significant contribution to the safety and the proper functioning of any such system or network. Continuous monitoring means constant observation of the IT environment to test whether the security controls or measures in place are functioning as intended. This enables the early detection of such risks as vulnerably, threats, and non-compliance to policies.

**Data Classification and Labeling:** As a rule, Data Classification and Labeling means the organization of the information in an orderly manner based on the proportion of risk associated with that information, and putting tags to certain data that denotes the level of protection that.

It is important to protect the data and use the appropriate countermeasures when it comes to the protected information. By classifying and labeling any data, organizations are able to control who has access to such data, apply encryption, and use different safety measures to secure their data. This supports compliance with any regards to data protection and further decreases the risks that are usually associated with data non-compliance.

**Data Loss Prevention for Cloud Services:** Data loss prevention for cloud services refers to the strategies and tools that are developed for the purpose of protecting important information in the cloud from being lost, unauthorized access and other attacks. Because more and more cloud services are on use in storing and processing data, these raises the need for the companies to apply the necessary techniques in loss of information preventing strategies. This is essential in preserving the integrity of information. Various approaches can be used for loss prevention for cloud services including but not limited to encryption, access restrictions as well as user activities regulations.

**Device Posture and Compliance Checks:** Device Posture and Compliance Checks verify and enforce the understanding that a given device is compliant with an organization's security and policy requirements. Device Posture and Compliance Check involves evaluating all or at least several operational aspects of the device. For example operating system and its software versions, security settings, patch levels in order to check for compliance with the security requirements of the organization. All the concern about protecting and controlling the network in a secure stable posture demands repetitive posture and compliance checks. Such checks would be purposefully targeted in addressing the issues of devices and vulnerabilities in a proactive approach in all circumstantial requirements. These types of activities can be gauged using automated tools or through physical assessment of certain risk factors and are an important part of access control to counter cyber challenges.

**Digital Rights Management (DRM):** A systematic approach adopted in an attempt to protect the copyright of electronic publications. Its main objective is to prevent the illegal distribution of digital publication and also to restrict the ways of how the people who bought the content can make copies. Through DRM technologies, content providers can control who can access their content, when, and for how long. Several strategies are used to implement DRM including the use of encryption, watermarking and access controls. In spite of the fact that DRM has generated debate over infringement on consumer right to access content and fair use of such content, it is a great content protection strategy by content developers and marketers.

**Encryption of Data at Rest/Transit:** The encryption of information of data at rest entails changing data to a form that cannot be read unless a key is possessed that can decrypt the information. Encryption provides protection for sensitive information that is kept on physical devices such as hard drives, servers, and others storage equipment. Data in transit encryption is the procedure of protection of any information its source and its destination from unauthorized access while it is being relocated from one geographic location to another across networks. This is usually done by means of SSL/TLS encryption for web traffic and VPN for secured internet data transportation.

**Electronic Endpoint Security and Foren:** Electronic Endpoint Detection and Response – This is a subset of security products and techniques, aimed squarely at the detection and management of advanced persistent threats on the network endpoints. Such EDR solutions are designed with the aim of providing real time surveillance of the activity of the endpoints in relation to the data and also the activities which are potentially malicious.

**Incident Response and Disaster Recovery:** Device Endpoint Patching and Updates consist of the processes and activities that make sure device and endpoint located within an organization are changed with the latest available patch, security, and system advancement available. It is paramount to lower exposure to security threats, fix problems within software, and enable devices to be used effectively and efficiently.

**Patching and Updates:** Deployment of new software releases, sanitorium measures against bugs, and other actions such as installing security updates to guard against intruders. It is essential to provide IT infrastructure which is secure and stable in order to conduct activities within the organization. The process of applying patching and updates is done regularly and helps in boosting the level of security within the organization by reducing the risk of any possible attacks and also ensuring that devices and endpoints meet the industry standards and best practices.

**Integrated Dashboards and Reporting:** Integrated Dashboards and Reporting present an overlapping structure, where patterns of content which have been rendered understandable visually can be absorbed with the help of various other means. Such essential measurements and performance indicators are stored where stakeholders can quickly measure them and display them to support proper decision-making. Integrated dashboards aggregate data for many departments and systems and provide a far-ranging view concerning performance of the organization. What are the opportunities these reports present, ending with customizing reports to effectively present the insights and trends. The business organizations can grasp the whole business operations and decision making is no longer guess work but is based on data analysis due to the addition of data and reporting services.

**Microsegmentation:** Microsegmentation is a strategy used to construct secure zones within data centers or the cloud infrastructure. It is intended to confine workloads and secure each one individually. Security policies can be applied to specific workloads making it easier to control threat propagation and damage from a security incident. Organizations can avoid the compromise of critical assets from advanced threats by applying microsegmentation which allows them to implement more granular and frequent security policies.

**Mobile Device Management:** It is a security software solution that assists organizations in managing, controlling and protecting the mobile devices that their employees use for work. Such devices include phones, tablets and computers. CJ Parker Mobile Device Management allows IT department to remotely manage these devices in order to protect data, implement control and restriction policies and roll out applications and content. With the increased use of mobile devices in workplaces, it is evident that Mobile Device Management MDM has now become a necessity in ensuring the integrity and confidentiality of business information and networks.

**Multi Factor Authentication:** The user is required to provide two or more proofs of identity for successful identification which is usually referred to as contemporary biometrics which is a form of MFA. These factors belong to three classes: knowledge (passwords), possession (security token), and biometrics (fingerprint or facial recognition)

**Network Access Control:** Network Access Control is intended to prevent individuals from gaining unauthorized access to networks. It guarantees the enforcement of policy controls, checks the compliance of devices that seek to connect into the network and troubleshoots non-compliant devices. Many components Technology called Network Access Control NAC has efficient methods to enforce the network security policy, using hardware and software, as authentication & authorization, also the assessment of the security posture of users. Network Access Control, also referred to as NAC helps companies lower their vulnerability to theft and loss of control over their network architecture by restricting network availability to only those users and devices that have been authenticated.

**Policy and Risk Management Frameworks:** Policy and Risk Management Frameworks include processes or procedures that are important for organizations to properly identify, analyze, and treat risks. These frameworks help manage risks in the whole organization by establishing, executing and measuring continuous policies and procedures risk control. Policy and Risk Management Frameworks help to avoid non-compliance with the law, lessen monetary losses, and protect the organization and its stakeholders, among other things. Policy and Risk Management Framework is accomplished via the development of appropriate rules and procedures. A well-structured framework not only remains limited to the hierarchal levels, but also integrates the concept of risk awareness 'ownership' at all levels in the organization, hence a timely and appropriate response to risks is possible.

**Privileged Access Management:** The measure that cares for monitoring and restricting the use of privileged accounts in an organization. Collectively this includes management and monitoring of people with high level access, for example, system administrators, IT managers and other high level users. The main aim of the Legal Employees is to micromanage who gets to access certain sensitive systems and information in order to prevent potential threats or attacks from taking place. Typically, implementation of PAM system includes the use of specific software and tools employed to implement the policies, control the access and keep an all-inclusive audit trail on activities related to privileged accounts. If any institution is able to administer privileged access, the chances of violating security protocols decrease, and they would also comply with regulatory bodies' demands.

**Rights Management Services (RMS):** Rights Management Services are a technological solution that enables the safeguarding and controlled distribution of confidential data within an organization or with external collaborators. Implementing RMS facilitates organizations in protecting their data from breaches through document and email controls and also usage policies on sensitive data. Even if the safeguarded information is disseminated outside of the network and to individuals who work for the organization, no one can access and use it who is not permitted to do so. Correspondingly, RMS allows the organizations to control and manage the internal usage of rather sensitive information resulting into improved security and governance. A Records Management System (RMS) can prove beneficial to institutions that wish to safeguard sensitive information and control its use and distribution.

**Role-Based Access Control:** Role-Based Access Control helps users manage networks by limiting network access according to the roles of users in the firm. RBAC makes certain that employees are assigned rights of access only to that information that is required to perform a given responsibility and not beyond that. For this reason, the organization can appoint individuals to perform specific functions and then limit those individuals' access to different categories of information based on those functions. RBAC has created efficient control over users' accessibility without increasing the complexity of control on permission levels.

**Secure Boot And Hardware-Based Controls:** Secure boot is an industry specification designed to protect PC systems from being booted to any other OS other than that of the OEM. Ensuring that the hardware is fully ready includes a prior checking of the signatures of all the components of the bootstrapping process, into the MD5, the OS, and all the drivers pertinent to its operation. Unlike logical security measures, hardware security features entail the protection of a device from unauthorized access or alteration through the physical parts and components of that device. Such controls may include the use of physical devices that provide data encryption, secure zones and bio-metrics which help in data retention and maintaining the integrity of the system against any external modification. By providing secure booting along with hardware secure elements, entities can greatly increase the security of their products and protect them against lots of adverse cyber aspects.

**Secure File Sharing and Collaboration:** Secure file sharing and collaborative work entails the secure transfer of documents and working on them inside some protected area. It comprises of the use of safe means and methods to elicit, change or control documents while ensuring that sensitive information is not accessed by unauthorized parties. Nowadays, with the advancement of technology, people and organizations have to work and share information from different places and devices, but the most important for any person and organization is to maintain the highest level of security. There are numerous tools and applications that offer encryption, restrict access and other security measures so that safe file sharing and collaboration is possible.

**Secure Remote Access (VPN):** A virtual private network is a technology that allows an authorized remote user to connect to a private network and conduct transactions over public networks which is successful because it is highly protected. The data that one exchanges with the private network is well protected due to encryption applied by much verified VPN. Companies avail themselves of such technology as they are able to let their staff access their company's facilities from anywhere in the world. Furthermore, people use this particular technology to protect their internet connections while surfing the web or seeking confidential information. VPN's provide users with a private and safe connection which protects the user from any form of invasion and any other security threats that may be faced while at public networks.

**Security Analytics and Machine Learning:** Security analytics has incorporated machine learning, which amplified effectiveness of combative measures against cyber security threats. The security analytics can apply machine learning techniques to detect and resolve problems more quickly as threats arise. Such an approach is advantageous as it helps the organization in foreseeing the possible security danger and being able to avert it prior to its occurrence. One of the notable benefits of the inclusion of machine learning in security analytics is its ability to sort large amounts of information for potential patterns and irregularities that give rise to a possibility of a security breach.

**Security Awareness and Training:** Security awareness and training deeply engrain in the fabric of every organization's security measures. One particular definition of security awareness is concerning the education level of employees regarding the importance of following any security related policies or authorized procedures, best practices in handling sensitive data, and different types of threats. Conversely, training consists of providing employees with the proper knowledge and skills to manage security incidents and to take preventive measures. They incorporate security awareness and training, which have proven invaluable to aiding organizations in promoting a culture of preventing data breaches, cyberattacks, and any other security incidents. Security Awareness and Training helps to create a safer and healthier working environment both for the workers and for the organization itself.

**Security Information and Event Management:** A strategic approach to security management which incorporates the functionalities of both SIM and SEM into one. The SIEM technology enables real-time reporting and analysis of security threats and alerts produced by the junctions of the network devices and applications. It also produces reports for regulatory requirements. Most organizations use SIEM as a first responder to security breaches, accidents or instances, and as a means for compliance. SIEM systems offer security teams an exact perspective of the security health of an organization by aggregating and cross-referencing security information from various channels.

**Software-Defined Perimeter:** A specific security framework that geographically restricts access to sensitive information and systems in a way requiring authorization in each circle. As such, SDP is different from traditional perimeter controls which rely on fixed, net-based firewalls in that there is no "trust" granted to users before authorizing them to access a specific application or service. This model of security is proactive whereby the attackable surfaces are reduced making it hard to breach sensitive information or grant unapproved users access to data. Various measures like user authentication, encryption, microsegmentation, and continuous monitoring are employed by SDP solutions in enforcing security policies and mitigating the damage from cyberattacks.

**Threat Intelligence Platforms:** Companies utilize Threat Intelligence Platforms, which are solutions and tools that offer services to gather and interpret information about a possible or actual attack. These platforms pull information from several sources such as OSINT, technical data, and the black web. They then undertake analysis and correlation and feed the security teams usable findings. The aim is to avoid potential threats to an organization through finding and addressing them beforehand. In addition, Threat Intelligence Platforms usually include cross-organizational threat intelligence sharing functionalities, allowing organizations to be updated regarding the latest cybersecurity issues and threats faced and cooperation.

**Unified Endpoint Management:** Device Endpoint Patching and Updates is the process of patching and updating the software and/or the firmware of devices and or endpoints in the organization's network in a bid to enhance the security of such devices or endorsement against various disclosed vulnerabilities. It is important in ensuring that every device installs the latest update to the installed systems eliminating chances of any forms of cyberattacks and maximizing efficiency. Patching and updates generally involve deploying current security patches, bug fixes, and newer features offered by the software and device vendors. Consistent patch management is one of the pre-requisites for a safe and stable IT environment.

**User/Entity Behavior Analytics (UEBA):** User/Entity Behavior Analytics (UEBA) is a software that enables the IT administrators to ensure an organization's security by preventing internal breaches and advanced attacks through the help of user behavior pattern recognition of the internal network. UEBA is a complex system that employs algorithms and statistical methods to monitor user and entity status and performance in order to develop a standard and accurate behavioral pattern for every user and entity.

**Vendor and Third-Party Risk Assessments:** Is the process of risk assessment and determining the risks involved in the process of using external vendors or third party providers. Such a procedure is essential to ensure operating safety and efficacy as third parties or vendors can pose people's concerns in uncountable ways like security breaches, compliance issues, and Business continuity concerns. Proper risk management is a structured approach which organizations can adopt to assess and respond to risk, determine and implement resource allocations for protective measures, and maintain conformity of vendors and third parties to the applicable laws and standards. These assessments often involve evaluation of the data and information security policies, implementation of compliance and risk management programs, and availability of resources among the vendors and other third party providers. Businesses must create an efficient risk assessment model to manage and mitigate risks related to vendors and third party relationships processes in a timely manner.

**Virtual Desktop Infrastructure:** Virtual desktop infrastructure is a technology that allows users to work from anywhere while still using their computer or workspace, as all the components are in a network and do not require a physical desk. The user's operating system along with the software and data is not stored in the user's physical endpoint but in a virtual instance created at a datacenter or on the cloud within a centrally accommodating device. Users can connect to their desktops through a broad spectrum of devices such as laptops, tablets and smartphones, all capable of being a conduit to their virtual desktops as opposed to being stationary work stations.

**Virtual LANs:** VLANs are a method for logically segmenting a physical network into multiple independent logical networks for instance, by using software switches within the same physical facilities. VLAN allows network traffic isolation and increasing security and network efficiency. It is customary to manage VLANs according to device groups such as departments, functions or security concerns. By logically partitioning the network in a number of segments, an administrator can more effectively control and manage the network traffic further turning the utilization of the available network resources to optimum levels.

**Zero Trust Network Access:** Zero Trust Network Access is a security principle that holds that institutions should not trust anything or anyone, whether inside their border or outside. Thus, all users, devices, and applications that want to connect to a network must be authenticated and authorized in order to be allowed in. ZTNA is a step in a different direction where perimeter security is no longer the fundamental focus. Rather it is predicated on the certainty that there will be a breach and therefore the focus is on ongoing authentication and granting of the least privilege necessary.

# APPENDIX B: SURVEY QUESTIONS

## Mapping Exercise_v8 30 Questions

| | RQ1 | RQ2 | RQ3 | Demo |
|---|---|---|---|---|
| The three research questions pertaining to this research are stated below: RQ1- What are the key technical controls of a Zero Trust Architecture (ZTA) in organizations? RQ2- What is the impact of ZTA on cyber attack prevention in organizations? RQ3- What are industry best practices for implementing a ZTA? | | | | |
| **Section 1** | | | | |
| **Opt Out** | | | | |
| 1. Has your organization, or an organization that you are familiar with, implemented a Zero Trust Architecture to prevent cybersecurity attacks? You can abort this survey now if you are not familiar with the architecture. | | | | |
|   Abort this survey | | | | |
|   Continue | | | | |
| | | | | |
| **Section 2** | | | | |
| **Demographics** | | | | |

| 2. How long have you been working in the field of IT/Cybersecurity? |  |  |  | X1 |
|---|---|---|---|---|
| Required to answer. Single choice. |  |  |  |  |
| 5 or less years |  |  |  |  |
| 6 - 10 years |  |  |  |  |
| 11 - 15 years |  |  |  |  |
| 16 - 20 years |  |  |  |  |
| More than 20 years |  |  |  |  |
|  |  |  |  |  |
| 3. What is your current job function? |  |  |  | X2 |
| Required to answer. Single choice. |  |  |  |  |
| Administrative/executive |  |  |  |  |
| Cybersecurity/IT staff |  |  |  |  |
| Engineer/Architect |  |  |  |  |
| Staff/Technology Manager |  |  |  |  |
| Professional Staff |  |  |  |  |
| Academics/professor/faculty member |  |  |  |  |
| Consultant |  |  |  |  |
| FT/PT Graduate Student |  |  |  |  |
| FT/PT Undergraduate Student |  |  |  |  |
|  |  |  |  |  |
| 4. What active information security certificates do you hold? (check all that apply) |  |  |  | X |

|  |  |  |  | 3 |
|---|---|---|---|---|
| Required to answer. Multiple choice. |  |  |  |  |
| CISSP - Certified Information Systems Security Professional |  |  |  |  |
| CISM - Certified Information Security Manager |  |  |  |  |
| CISA - Certified Information Security Auditor |  |  |  |  |
| CRISC - Certified in Risk and Information Systems Control |  |  |  |  |
| CompTIA Security+ |  |  |  |  |
| CCSK- Certificate of Cloud Security Knowledge |  |  |  |  |
| ISO27001/2 |  |  |  |  |
| CEH - Certified Ethical Hacker |  |  |  |  |
| PMP – Project Management Professional |  |  |  |  |
| Vendor Certification(s) |  |  |  |  |
| CIPP/US- Certified Information Privacy Professional/United States |  |  |  |  |
| None |  |  |  |  |
| Other |  |  |  |  |
| XXX TEXT |  |  |  |  |
|  |  |  |  |  |
| 5. What is the size of your company based on USD Annual Revenue? |  |  |  | X 4 |
| Required to answer. Single choice. |  |  |  |  |
| < $1M |  |  |  |  |
| 1.1M - $10M | | | | |
| 10.1M - $50M | | | | |
| 50.1M - $200M | | | | |
| 200.1M - $500M | | | | |
| 500.1M - $1B | | | | |
| >1B | | | | |
| | | | | |
| 6. Which geographical region of the world is your organization predominantly located in? | | | | X 5 |
| Required to answer. Single choice. | | | | |
| Africa: North Africa and Sub-Saharan | | | | |
| Asia | | | | |
| Europe | | | | |
| Latin America and the Caribbean | | | | |
| North America | | | | |
| Oceania | | | | |
| | | | | |
| **Section 3** | | | | |
| **Zero Trust Architectural Components** | | | | |
| | | | | |
| 7. Please prioritize the following key technical controls for a Zero Trust Architecture, arranging them from most important to least important. | X 1 | | | |
| Ranking. | | | | |
| Multi-Factor Authentication (MFA) | | | | |
| Single Sign-On (SSO) | | | | |
| Adaptive Risk-Based Authentication | | | | |
| Biometric Authentication | | | | |
| Identity Federation and Directory Services | | | | |
| | | | | |
| 8. Please prioritize the following key technical controls for a Zero Trust Architecture, arranging them from most important to least important. | X2 | | | |
| Ranking. | | | | |
| Endpoint Detection and Response (EDR) | | | | |
| Role-Based Access Control (RBA) | | | | |
| Attribute-Based Access Control (ABA) | | | | |
| Privileged Access Management (PAM) | | | | |
| User and Account Lifecycle Management | | | | |
| | | | | |
| 9. Please prioritize the following key technical controls for a Zero Trust Architecture, arranging them from most important to least important. | X3 | | | |
| Ranking. | | | | |
| Unified Endpoint Management (UEM) | | | | |
| Mobile Device Management (MDM) | | | | |
| Device Posture and Compliance Checks | | | | |
| Secure Boot and Hardware-Based Integrity | | | | |
| Micro-Agent or Agentless Endpoint Security | | | | |
| | | | | |
| 10. Please prioritize the following key technical controls for a Zero Trust Architecture, arranging them from most important to least important. | X4 | | | |
| Ranking. | | | | |
| Software-defined perimeter (SDP) | | | | |
| Zero Trust Network Access (ZTNA) | | | | |
| Application Control and Whitelisting | | | | |
| Device Isolation and Quarantine | | | | |
| Network Access Control (NAC) | | | | |
| | | | | |
| 11. Please prioritize the following key technical controls for a Zero Trust Architecture, arranging them from most important to least important. | X5 | | | |
| Ranking. | | | | |
| Microsegmentation and Network Isolation | | | | |
| Virtual LANs and Microsegmentation | | | | |
| Network Access Control (NAC) | | | | |
| Secure web gateways | | | | |
| Cloud Access Security Brokers (CASB) | | | | |
| | | | | |
| 12. Please prioritize the following key technical controls for a Zero Trust Architecture, arranging them from most important to least important. | X6 | | | |
| Ranking. | | | | |
| Data Classification and Labeling | | | | |
| API Gateways and Web Application Firewalls | | | | |
| Network Traffic Analysis and Anomaly Detection | | | | |
| Secure remote access (VPN, VDI, RDP) | | | | |
| Data loss prevention (DLP) | | | | |
| | | | | |
| 13. Please prioritize the following key technical controls for a Zero Trust Architecture, arranging them from most important to least important. | X7 | | | |
| Ranking. | | | | |
| Encryption of Data at Rest and in Transit | | | | |
| Digital Rights Management (DRM) | | | | |
| Data Access Control and Granular Policies | | | | |
| Rights Management Services | | | | |
| Secure File Sharing and Collaboration | | | | |
| | | | | |
| 14. Please prioritize the following key technical controls for a Zero Trust Architecture, arranging them from most important to least important. | X8 | | | |
| Ranking. | | | | |
| Data Loss Prevention for Cloud Storage | | | | |
| Security information and event management (SIEM) | | | | |
| User and entity behavior analytics (UEBA) | | | | |
| Security Orchestration, Automation, and Response (SOAR) | | | | |
| Data Activity Monitoring and Analytics | | | | |
| | | | | |
| 15. Please prioritize the following key technical controls for a Zero Trust Architecture, arranging them from most important to least important. | X 9 | | | |
| Ranking. | | | | |
|    Threat Intelligence Platforms | | | | |
|    Security Analytics and Machine Learning | | | | |
|    Centralized Logging and Auditing | | | | |
|    Automated Policy Enforcement and Remediation | | | | |
|    Continuous Monitoring and Anomaly Detection | | | | |
| | | | | |
| 16. Please prioritize the following key technical controls for a Zero Trust Architecture, arranging them from most important to least important. | X 10 | | | |
| Ranking. | | | | |
|    Policy and Risk Management Frameworks | | | | |
|    Compliance Monitoring and Reporting | | | | |
|    Integrated Dashboards and Reporting | | | | |
|    Vendor and Third-Party Risk Assessments | | | | |
|    Automated Compliance Checks and Controls | | | | |
| | | | | |
| 17. Please prioritize the following key technical controls for a Zero Trust Architecture, arranging them from most important to least important. | X 11 | | | |
| Ranking. | | | | |
|    Security Awareness and Training Programs | | | | |
| Change Management and Configuration Control | | | | |
| Incident Response and Disaster Recovery Planning | | | | |
| Reporting and Executive Dashboards | | | | |
| Device Endpoint Patching and Updates | | | | |
| | | | | |
| **Section 4** | | | | |
| **Security Environment** | | | | |
| | | | | |
| 18. Which form of passwordless authentication are enabled for your users? | | | X1 | |
| Single choice. | | | | |
| Phone call | | | | |
| Text message | | | | |
| OATH token | | | | |
| Authenticator- Microsoft/Google/Duo/Authy | | | | |
| Email | | | | |
| | | | | |
| 19. Have you enabled multifactor authentication (MFA) for users? | | | X2 | |
| Single choice. | | | | |
| Yes | | | | |
| No | | | | |

| | | | |
|---|---|---|---|
| 20. Do you use a Cloud Access Security Broker (CAS solution to provide visibility and control over cloud applications and services in your Zero Trust environment? | | X 3 | |
| Single choice. | | | |
| Yes | | | |
| No | | | |
| | | | |
| 21. Have you implemented a Cloud Security Posture Management (CSPM) solution to provide visibility and control overshadow IT and unsanctioned cloud services in your Zero Trust Architecture? | | X 4 | |
| Single choice. | | | |
| Yes | | | |
| No | | | |
| | | | |
| 22. Have you deployed a Network Detection and Response (NDR) solution to provide visibility and analytics into network traffic and behavior as part of your Zero Trust implementation? | | X 5 | |
| Single choice. | | | |
| Yes | | | |
| No | | | |
| | | | |

| | | | | |
|---|---|---|---|---|
| 23. Have you integrated a Virtual Desktop Infrastructure (VDI) or Remote Desktop Services (RDS) solution to enable secure access to applications and data in your Zero Trust architecture? | | | X 6 | |
| Single choice. | | | | |
| Yes | | | | |
| No | | | | |
| | | | | |
| **Section 5** | | | | |
| **Reportable Breaches** | | | | |
| | | | | |
| 24. In the past 12 months, has your organization identified any unauthorized access to your computer systems or data? | | X 1 | | |
| Single choice. | | | | |
| Yes | | | | |
| No | | | | |
| | | | | |
| 25. Has your organization received any external reports (e.g., from customers, partners) of a potential data breach within the past year? | | X 2 | | |
| Single choice. | | | | |
| Yes | | | | |
| No | | | | |

| | | | |
|---|---|---|---|
| 26. Has your organization been notified by law enforcement or regulatory bodies of a potential data breach in the past 12 months? | | X3 | |
| Single choice. | | | |
|   Yes | | | |
|   No | | | |
| | | | |
| 27. Has your organization had to restore any systems or data from backups due to a suspected breach in the past 12 months? | | X4 | |
| Single choice. | | | |
|   Yes | | | |
|   No | | | |
| | | | |
| 28. Has your organization issued any public statements or notifications regarding a data breach in the past year? | | X5 | |
| Single choice. | | | |
|   Yes | | | |
|   No | | | |
| | | | |
| 29. Has your organization provided any financial compensation or credit monitoring services to | | X6 | |

| | | | | |
|---|---|---|---|---|
| customers affected by a breach in the past 12 months? | | | | |
| Single choice. | | | | |
| Yes | | | | |
| No | | | | |
| | | | | |
| 30. Has your organization made any significant changes to its cybersecurity policies or procedures in the past year (e.g., increased employee training, enhanced security software)? | | X 7 | | |
| Single choice. | | | | |
| Yes | | | | |
| No | | | | |
| | RQ1 | RQ2 | RQ3 | Demo |
| Totals | 11 | 7 | 6 | 5 |

# APPENDIX C: NAMES TO LABELS MAPPING

| NAMES | LABELS |
|---|---|
| A2 | Years Experience |
| A3 | Job Function |
| A4_1 | CISSP |
| A4_2 | CISM |
| A4_3 | CISA |
| A4_4 | CRISC |
| A4_5 | CompTIA Security+ |
| A4_6 | CCSK |
| A4_7 | ISO 27001/2 |
| A4_8 | CEH |
| A4_9 | PMP |
| A4_10 | Vendor Cert(s) |
| A4_11 | CIPP |
| A4_12_NONE | No Certs |
| A4_13_OTHER | Other Certs |
| A4_13_TEXT | Other - Text |
| A5 | Company Revenue |
| A6 | Geo Region |
| A7_1 | Multi-Factor Authentication (MFA) |
| A7_2 | Single Sign-On (SSO) |
| A7_3 | Adaptive Risk-Based Authentication |
| A7_4 | Biometric Authentication |
| A7_5 | Identity Federation and Directory Services |
| A8_1 | Endpoint Detection and Response (EDR) |
| A8_2 | Role-Based Access Control (RBA) |
| A8_3 | Attribute-Based Access Control (ABA) |
| A8_4 | Privileged Access Management (PAM) |
| A8_5 | User and Account Lifecycle Management |
| A9_1 | Unified Endpoint Management (UEM) |
| A9_2 | Mobile Device Management (MDM) |
| A9_3 | Device Posture and Compliance Checks |
| A9_4 | Secure Boot and Hardware-Based Integrity |
| A9_5 | Micro-Agent or Agentless Endpoint Security |
| A10_1 | Software-defined perimeter (SDP) |
| A10_2 | Zero Trust Network Access (ZTNA) |
| A10_3 | Application Control and Whitelisting |
| A10_4 | Device Isolation and Quarantine |
| A10_5 | Network Access Control 1 (NAC1) |
| A11_1 | Microsegmentation and Network Isolation |
| A11_2 | Virtual LANs and Microsegmentation |
| A11_3 | Network Access Control 2 (NAC2) |
| A11_4 | Secure Web Gateways |
| A11_5 | Cloud Access Security Brokers (CASB) |
| A12_1 | Data Classification and Labeling |
| A12_2 | API Gateways and Web Application Firewalls |
| A12_3 | Network Traffic Analysis and Anomaly Detection |
| A12_4 | Secure Remote Access (VPN, VDI, RDP) |
| A12_5 | Data Loss Prevention (DLP) |
| A13_1 | Encryption of Data at Rest and in Transit |
| A13_2 | Digital Rights Management (DRM) |
| A13_3 | Data Access Control and Granular Policies |
| A13_4 | Rights Management Services (RMS) |

| A13_5 | Secure File Sharing and Collaboration |
|---|---|
| A14_1 | Data Loss Prevention for Cloud Storage (DLPCS) |
| A14_2 | Security Information and Event Management (SIEM) |
| A14_3 | User and Entity Behavior Analytics (UEBA) |
| A14_4 | Security Orchestration, Automation, and Response (SOAR) |
| A14_5 | Data Activity Monitoring and Analytics |
| A15_1 | Threat Intelligence Platforms (TIPS) |
| A15_2 | Security Analytics and Machine Learning |
| A15_3 | Centralized Logging and Auditing |
| A15_4 | Automated Policy Enforcement and Remediation |
| A15_5 | Continuous Monitoring and Anomaly Detection |
| A16_1 | Policy and Risk Management Frameworks |
| A16_2 | Compliance Monitoring and Reporting |
| A16_3 | Integrated Dashboards and Reporting |
| A16_4 | Vendor and Third-Party Risk Assessments |
| A16_5 | Automated Compliance Checks and Controls |
| A17_1 | Security Awareness and Training Programs |
| A17_2 | Change Management and Configuration Control |
| A17_3 | Incident Response and Disaster Recovery Planning |
| A17_4 | Reporting and Executive Dashboards |
| A17_5 | Device Endpoint Patching and Updates |
| A18_1_Call | Phone call |
| A18_2_Text | Text Message |
| A18_3_Token | OATH Token |
| A18_4_Auth | Authenticator |
| A18_5_Email | Email |
| A19_Use_MFA | User enabled MFA |
| A20_Use_CASB | Enabled CASB |
| A21_Use_CSPM | Enabled CSPM |
| A22_Use_NDR | Enabled NDR |
| A23_Use_VDI | Enabled VDI or RDS |
| A24_Internally_Identified | Unauth Inter Access |
| A25_Externally_Identified | Unauth Exter Access |
| A26_Law_Enforcement_Identified | Notified by law enforcement |
| A27_Needed_to_Restore | Restore suspected breach |
| A28_Issued_Notifications | Issued public notifications |
| A29_Financially_Compensated | Provided financial compensation |
| A30_Policy_Changes | Changes to policies |
| A4_All_Certs_in_group | All Certifications Selected |
| A7_All_in_IAM | Tech Ctrls Grp 1 - Identity and Access Management |
| A8_All_in_ACES | Tech Ctrls Grp 2- Access Control and Endpoint Security |
| A9_All_in_ESM | Tech Ctrls Grp 3- Endpoint Security and Management |
| A10_All_in_NSAC | Tech Ctrls Grp 4- Network Security and Access Control |
| A11_All_in_NSS | Tech Ctrls Grp 5- Network Security and Segmentation |
| A12_All_in_DPNS | Tech Ctrls Grp 6- Data Protection and Network Security |
| A13_All_in_DPIS | Tech Ctrls Grp 7- Data Protection and Information Security |
| A14_All_in_SMIR | Tech Ctrls Grp 8- Security Monitoring and Incident Response |
| A15_All_in_SMTD | Tech Ctrls Grp 9- Security Monitoring and Threat Detection |
| A16_All_in_GRC | Tech Ctrls Grp 10- Governance, Risk, and Compliance (GRC) |
| A17_All_in_OSIM | Tech Ctrls Grp 11- Operational Security and Incident Management |
| A18_All_in_Pass_Auth | PW Auth- All Selected Options |

# APPENDIX D: IRB APPROVAL LETTER

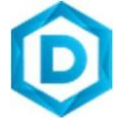

Institutional Review Board
DAKOTA STATE UNIVERSITY
820 N, Washington Ave
Madison, SD 57042

Permission to Modify Project

Date:              7/17/2024

To:               Dr. Varghese Vaidyan and Sam Aiello

Project Title:    Prescriptive Zero Trust: Assessing the impact of zero trust on cyber attack

Approval #:       20240612M

Type of Review:  Expedited

Dear Investigators:

The Dakota State University IRB has reviewed materials and statements you sent regarding the continuation and modification of your project, and granted permission to proceed.

It is noted you reduced the number of your survey questions from 60 to 30, utilizing preapproved questions.

If you have any questions regarding this permission, please contact us at 605-256-5100 or irb@dsu.edu. If you wish to make further changes to procedures, please contact us prior to implementation of any research-related activities. Thank you for your cooperation with the DSU IRB.

Best Regards,

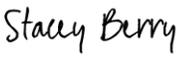

Stacey Berry,

IRB Chair